# Technical Report

# Design of Intra-body Nano-communication Network for Future Nano-medicine.

Mona Sayed Youssef
Fatima Ismail Ghanim
Noorulhuda Imad
Ayesha Ali Alqasim

**Project Supervisor:** Prof. Raed M. Shubair

# Table of Contents

# *Abstract:*

Intra-body communication (IBC) is a method that utilizes the human body as a broadcast biological medium for electromagnetic signals to inter-connect wireless body sensors. Study of the collaboration between electromagnetic waves and human cells has gained importance in recent years leading towards developing and establishing new novel concept which is the idea of nano-communications using nano-networks to form "*in vivo* communication" that are aimed to offer wireless communication between interior nano-sensors. The emergent of this advanced unprecedented prospective approach of deploying the *in vivo* communication concept in the health sector is considered as a key potential technology that enhances healthcare delivery and enables the progress of future applications and services. Various techniques of communication utilized for operating nano-devices is demonstrated in this report, the best applicable technique for data exchanging is utilizing wireless communication network (WCN) to operate using Electromagnetic waves in the Terahertz spectral band. By overcoming the Terahertz regime and developing sufficient minuscule transceivers and adequate nano-scaled antennas, wireless connection and incorporation of intra-body Nano-sensors will evolve to be permitted at both the Terahertz Band (0.1-10 THz) and optical spectral bands (400-750 THz). Consequently, this will promote the significance of investigating the electromagnetic waves propagation model. Furthermore, the analysis of the total path loss attained will be demonstrated by taking into consideration its various dependable factors such as the spreading loss due to the signal propagation, molecular absorption resulting from human tissues, and scattering due to both minor and outsized body constituent substances. The obtained analytical results verified that the path loss increases by increasing both the applied distance and frequency. In addition, this report also addresses the characteristics of *in vivo* wireless channel model and presents a comprehensive comparison between this desired channel with other parallel existing channels. Moreover, the main framework imposed for conducting link budget analysis of intra-body communication between intergrated bio-nano-sensors is expansively developed. Ultimately, to deliver a comprehensive and widely-broad intra-body communication model, the attenuation both of Transverse-Electrical (TE) and Transverse-Magnetic (TM) waves and cross-layer reflection inside human body taking into account . In conclusion, the challenges and future implications relevant to this evolving advanced technology are emphasized.

# *Chapter1:*

# *Introduction to Nano-scale communication:*

*I want to build a billion tiny factories, models of each other, which are manufacturing simultaneously...The principles of physics, as far as I can see, do not speak against the possibility of maneuvering things atom by atom. It is not an attempt to violate any laws; it is something, in principle that can be done; but in practice, it has not been done because we are too big.*

*Richard Feynman, 1959*

## 1.1 Introduction:

The last two decades have witnessed an exponential growth and tremendous developments in wireless technologies and systems, and their associated applications, such as those reported in [1-25]. The world today is witnessing a high rate of the role being played by nanotechnology in various industries in the world such as biomedical, military and environmental areas. Likewise, other sectors that have gained the role of nanotechnology are healthcare and bioengineering applications (Santagati & Melodia, 2015). Nanotechnology is now considered as one of the best alternatives to the traditional methods of treatments, the nanotechnology devices are the best in the industry as they offer reliable answers to medical problems such as spinal card as well as to gastrointestinal related problems. There has been a huge problem when it comes to the medical delays caused by the high dependency on using the traditional medical communication designs. This has become an issue over the years leading to huge problems in terms of the quality of healthcare that being offered at hospitals. (Malak & Akan, 2014). Linking and incorporating nano-devices together led to establish new unprecedented concept which is the idea of nano-communications using nano-networks, that are aimed at expanding the abilities of such devices in promotion and enhancement of future technologies. There are various techniques of communication utilized when operating nano-devices, the best applicable technique for data exchanging is utilizing wireless communication network (WCN) to operate using Electromagnetic waves in the Terahertz spectral band. The substantial contribution of this underutilized spectral band to potential upcoming medical technologies is its advantages of providing radiation that is less susceptible to propagation effects like spreading loss and scattering effect, more reliable in terms of data quality and time delay and harmless on biological tissues since it is considered a non-ionization radiation. This terahertz technique permits the use of bio-nano-sensors in future nano-medicine. As a result, various medical applications such as e-health monitoring system and e-drug delivery systems can be achieved. Ultimately, the main objective is to connect nano-devices wirelessly to the Internet, and accordingly e-medical operations and e-health applications can be accomplished. In current medical technologies, the study of body centric communication has been applied for extensive spectral bands. However, those prior examined bands lack the size reduction's characteristics unlike nano-scale technologies hence this makes such technologies an optimal desirable choice for upcoming applications of body-centric communication. Because of the short wavelength (λ), terahertz radiations, and due to the existence of molecular resonances at these frequencies, nano-devices can detect even minutes variations in water content and biomaterial body tissues. Therefore, one promising growing area in recent research is the analysis of terahertz electromagnetic propagation through human body tissues and layers. This is to come up with advanced medical diagnostic tools for early detection and treatment of some abnormalities that gives a hint or a sign to a serious condition such as skin cancer. In spite of effective solutions being offered by data processing, the multimedia nano-things though have to send a considerable amount of data in a reliable fashion just in its proper time. Luckily, graphene-based nano-transceivers and nano- antennas are expected to implicitly operate within the terahertz frequency band (0.1–10 THz). The advantage of operating in terahertz band is that it supports a very high bit rate information transmission, can be as many as terabits per second (Tbps). With the rise of these points of strength, yet it is considered one of the least explored and investigated frequency range among the full EM spectrum. The already existing channel models and prototypes is being designed to accommodate for lower frequency ranges and thus, not applicable to be used for terahertz band. This issue actually emerged since the available channels are unable to perceive a number of effects such as attenuation and noise introduced by molecular absorption, the scattering from particles which are comparable in size to the short wavelength of terahertz waves, or the scintillation of terahertz radiation.

## 1.2 Nano-networks applications:

Nanonetworks applications are very broad. They are mainly divided into; biological, environmental, industrial and military. [26] The table 1 below gives a detailed account of the nanonetwork application as it was intended to be. The networks are stable and large in size which gives them an advantage over other networks. In biomedical applications, these characteristics make them the optimum choice. [27] Nanodevices are considered as the main foundation behind Internet of Nano-Things (IoNT), that is the smallest part of the working network that can be used [28].

**Table 1:** Indication on the proposed future nano-network applications.

| **Field** | **Application** |
|---|---|
| Biomedical [4] | - Evaluation of Health<br>• Use of THz radiations and scans<br>• Use of remote detection of diseases<br>- Alternative methods<br>• Genetically modified cells<br>• Use of bio-hybrid tissues<br>• Delivery of drugs<br>• Minor surgeries |
| Environmental [5] | - Use of organic material<br>- Botanical methods of environmental control |
| Industrial [3] | - Quality of production<br>- Quality assurance department |
| Military [6] | - Use of biological and atomic weapons<br>- Nano-frictional equipment |

## 1.3 Motivation:

The persistent advancements in health monitoring solutions are strongly motivated by an ever-increasing demand for healthcare. The Body Area Networks (BAN) paradigm represents one such area of technological advancement. It consists of a set-up of continuously operating sensors designed to compute vital physiological and physical factors such as heart rate, the level of glucose and mobility. The integrated of wireless network connectivity capability is what specifically makes the BAN to be highly portable. The BAN mainly uses two types of hardware devices. These include a wearable device that people or patient wears on his body surface and a medical device such as Nano-sensors that are implemented in the human body. The wearable device contains sensors that provide a continuous health monitoring capability for the physicians. They also facilitate monitoring of a patient's diagnostic status as well as provides feedback to the patient sustain the most favorable health status. Consequently, it facilitates a transition in the health industry into health care that is more anticipatory and cost-efficient. Moreover, Radio frequency (RF) wireless technology has been used effectively and efficiently in the most (BAN) implementation. Even as the deployment of radio frequency (RF) wireless technology has met considerable successes in many BAN trial pieces of research, their major drawback has widely cited to be its high-energy consumption. This makes it particularly unreliable for health care monitoring processes that require a significant amount of electric power. Moreover they can be easily affected or influenced by electromagnetic interference. Another wireless communication technology is the Intrabody communication (IBC). This technology relies on the human body as a medium for transmitting signals. There is a consensus in the health technology industry that the intrabody wireless area networks along with their related technologies are spearheading the next phase of technological evolution in proving health solutions. They are both adaptable and cost-effective, particularly when integrated into wearable devices. As it can be implanted into the human body, the IBC is beneficial to a range of applications, including biomedical monitoring systems compared to BAN as it consumes low power because of its low power transmission. As this technology relies on implanted Nano-sensors in the human body to transmit data, it has a high energy-saving potential. Additionally, as it has a miniature transceiver, the IBC's communication capability tends to be robust and can overcome electromagnetic interferences. The IBC shows a great potential for developing the BANs. Their major advantage is that they can provide constant monitoring of health statuses of patients and reduce the need to undergo surgeries. They also facilitate the continuous collection of patient information for prolonged periods of time. This means that health practitioners assigned to a patient can monitor the health status of their subject as well as conduct consistent health analysis by leveraging the big data. It also implies that patients are freed from the trouble of visiting the hospital continually to have their health statuses monitored. It is, therefore, a big leap forward from the tradition of depending on health data collected at the hospitals during patient visits.

Another related technology called Vivo Nano-sensing system embodies current technological advancements that facilitate more precise diagnosis and treatment of patient's illnesses. It is designed to operate as an implant inside the patient's body and provides patient data in real time. The technology is widely suggested as a possible means for providing quicker and more precise diagnosis and treatment of diseases compared to the conventional Vitro medical technologies. Still, there is a concern that it has Nano sensor's that have narrow sensing range to the nano-environment. As a result, it is incapable of covering wide volumes or areas, unless it is installed with many Nano sensors. It also requires user interaction and external device to interpret the real measurement. On account of such improved communication capabilities, Nano-sensors can triumph over their drawbacks and realize increased application as viable medical solutions. Without a doubt, Nano-sensors will witness an improved capacity to transmit data in a multi-pronged manner to a gateway, and respond to commands faster.

Our main objective is to study the propagation of Nano-sensors for future healthcare applications. One of the ways to deploy Nano-sensors is to know how much the amount of power lost due to multi-layers reflection, because in wireless communications there are no multi-layers in free-space. However in free-space we have only the usual multipath, reflection, and path-loss. In the human body we have many different inhomogeneous medium such as the skin, blood, and fat. Many can wonder at this stage about why we should take into account studying the propagation aspects in which waves are travelling for short distance like human body. In fact, what we care about the most is not being the distance itself or the size of propagation medium, but the ration D/λ. For signals in the Terahertz band, they have a wavelength equals to $3 \times 10^{8}/1\,THZ$ = 3 mm, which is very small comparing to 1 cm. Therefore, and although the thickness is very small, the wave appears as if it's propagating in an open free spaces since D is much greater that λ and the ratio D/λ will be >>10. This criterion ensures us a successful intrabody communication model in which we can employ the previous concepts that were adopted in free space communication channels. Substantially, the need was raised for further analysis of intra-body wave propagation.

## 1.4 Problem Statement:

There are different channels for transmitting waves in an *in vivo* channel [26]. The effect of this transmission is that the wave speed is dispersed and diluted within each organ in the distribution channel. This absorption leads to low quality transmission of the *vivo* signal [33]. The waves from the antenna pass through a loss environment that reduces its wavelength. For this reason, the designed system should consider both the radial and the angular environment [34]. Waves produced travel differently in radiating fields that are closer to them as compared to those that are distant. This necessitates different kinds of links for transmitters [35]. Using the same link will affect the distribution of frequency which in turn makes nano communication over a larger area difficult. [30]. it is therefore essential to develop a customized system for each link putting into consideration the environment through which the signals are transmitted. By doing so, the WBAN devices functionalities and performances will be enhanced.

In recent years, body centric communication has been studied for a various wide range of frequencies [26]. However, this advanced technology enabled having small sized nano-devices which is considered as a preferred choice by many consumers. Their adaptability to different technologies and portability has particularly made them desirable among its users. Given that most of their use in the health sector, nanodevices have been engineered with special sensors that will transmit all the bio data and make communication easier. This will make it easier for the different health departments to communicate and exchange information over a large geographical region [27].

## 1.5 Challenges of Nanoscale Communication:

Just like any other technology, there is bound to be challenges that will occur from time to time. In the process of communication, these nano-devices too are no exemption. These devices are made to shrink in size and from a macro-scale which are challenging enough in the transmission of information. For this reason, there are various methods that can be used to tone down the challenges associated with them.

- Despite the advancement of technology, it has become increasingly difficult to manufacture nano-scaled devices that cover all the multiple disciplines of nano-communication. This has therefore made their performance very limited. The design required for this is very complex and we are not there yet.
- Besides this, it is still challenging to develop an interface that balances out the components of nano-scales and macro-scale networks.
- There is need for establishing a system in which the devices work together especially when the scope of the communication is wide.
- Challenges arise when using a nano-network due to its reliability. This arises as most systems are unable to process and diffuses molecular communication [28].

## 1.6 Other research challenges of this project:

Every project has its fair share of challenges. Establishing a project of such intricate details comes with many hitches along the path. Without proper handling, this technology can go wrong in so many aspects. In developing a working THZ technology that will facilitate communication over the scope of area is difficult due to the fact that different regions have different terrains. It is therefore hard to find a straight line wavelength that fits all and this delays the procedure.

- One of the major difficulties is brought on by the path loss occurred at the THz band frequencies. The path loss for a transmitted wave in the THz band ought to incorporate the increase of the spreading loss, the molecular absorption loss and additionally the dispersing loss. The spreading loss represents the attenuation because of the development of the wave as it proliferates through the medium. The absorption loss represents the weakening that a proliferating wave endures on account of sub-atomic retention and relies on upon the focus and the specific blend of particles experienced along the way. Molecule diffusing can come about into critical extra loss that should be considered with the loss subject to the size and thickness appropriation of the particles. In this way, the divert in the THz band is exceptionally recurrence particular [29].

- Another test is the nonattendance of the LOS transmission because of the nearness of snags. In particular, NLOS transmissions can be sorted into: specular reflected spread, diffusely scattered proliferation and diffracted engendering. To represent NLOS spread, it is fundamental to portray the coefficients for reflection, dissipating and diffraction of EM waves at THz frequencies. These coefficients rely on upon the material and geometry of the surface, and also on the recurrence and point of the occurrence EM wave [30].

- In connection to the ultra-wide transmission capacity, every recurrence part in the transmitted wave encounters diverse lessening and deferral. These recurrence scattering impacts, or proportional bending in the time space, are not caught in existing multipath channel models. Subsequently, new multipath channel models should be produced for correspondence THz band.

- There are a few clamor sources in the THz band. The surrounding clamor in the THz band channel is essentially contributed by the atomic retention commotion. Without a doubt, retention from atoms display in the medium does weaken the transmitted wave, as well as presents clamor [31]. Thus, a model produced for EM THz intra-body communication ought to represent the atomic absorption commotion that that a wave going in the THz band endures when spreading over short separations.

- The real bottleneck that existed inside THz frameworks is the absence of productive, room-temperature worked, high power THz band handsets. Be that as it may, new application regions have driven the THz look into towards less expensive, minimized, room temperature parts that prepare towards remote communication applications [32].

## 1.7 Research Objectives:

Planning proficient body driven frameworks working at the THz band depends on a far reaching investigation of the correspondence channel. Since the human body is a lossy and dispersive medium, the impacts of its reality on both the radio wires and proliferation channel must be appropriately examined and displayed. The underlying destinations of this exploration are condensed as takes after

- Explore the likelihood of utilizing the THz wave for body-driven nano-communication, particularly intra-body communication.

- Assess the intrabody channel transmission by recognizing the contrasts between this channel and the conventional ones. It is imperative to acknowledge how might we concentrate the channel and what sort of instruments are utilized to assess the channel execution. Really, a great learning of the channel conduct empowers the best possible estimation of the wave parameters. • Develop a spread model for EM THz intrabody communication among nanodevices. This model ought to represent the aggregate path loss and the atomic retention clamor that a wave in the THz band experiences when spreading over short separation.

- Perform connect spending counts to evaluate the execution of the proposed correspondence medium.

- Identify the impact of the multipath segments that exist because of the reflections that happen inside the human body.

- Compare EM THz intrabody with correspondence other recurrence groups including ultrasonic and infrared frequencies. This is finished by evaluating the achievability and execution of every recurrence band.

- Understand the operation system of plasmonic nanoantennas since it is the key empowering influence of the remote interconnection between bio-nanosensing applications.

## 1.8 Project Description:

The project generally illustrates the applicability of THz communication in the biological domain. This is illustrated in the diagram 1.1 below. The test project entails assessing the transmission of electromagnetic waves through the tissues. Part of the research will include designing a system that will balance the loss mechanism. Finally the process will conclude with formulation of different channels that will be used to present the architectural design of the nanodevice. Emphasis will include showcasing the different components of the nanodevices that play a crucial part in establishing communication. With time after the assessment of the efficiency of the system, additional units will be added to ensure it works at its best. As a maintenance procedure, case studies will be conducted to determine the advantages and the disadvantages of the system.

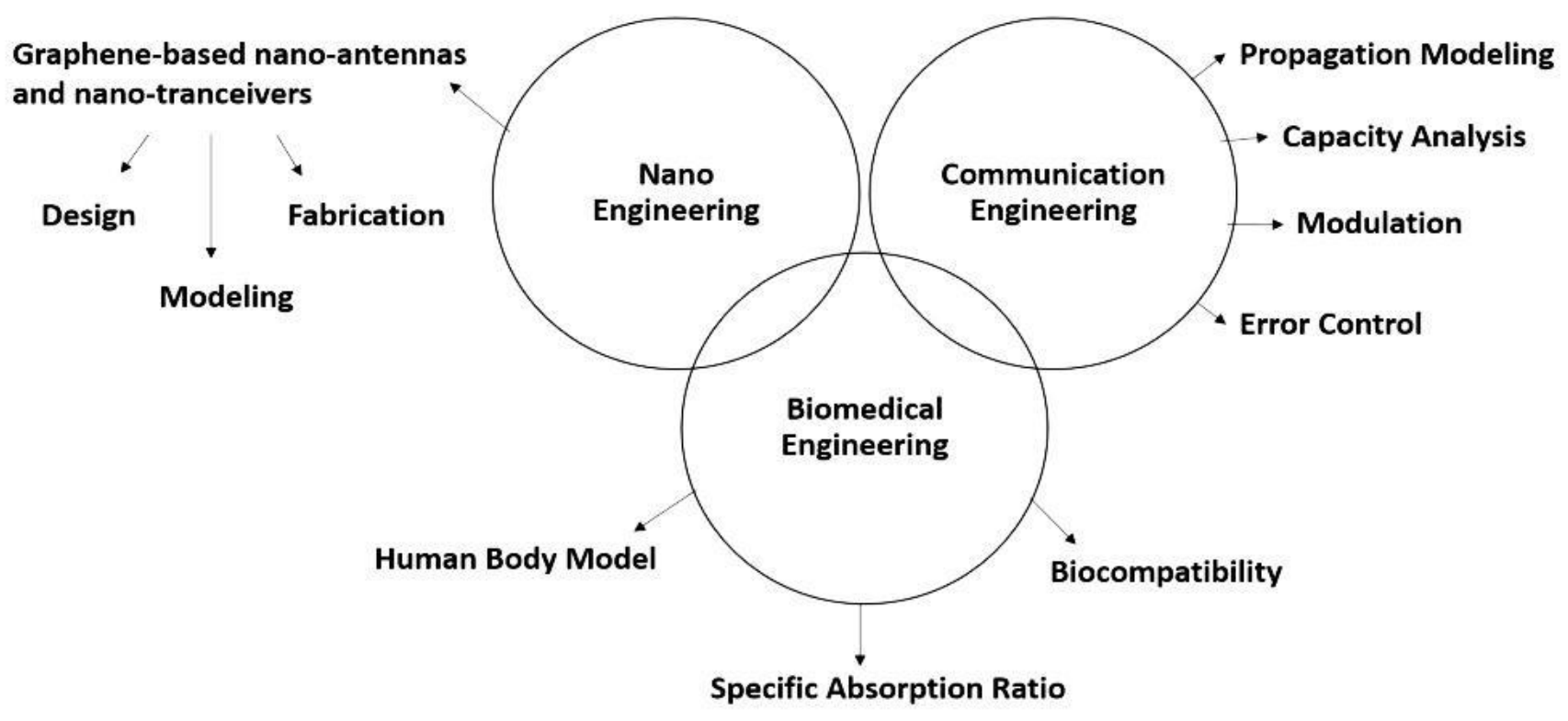

**Figure 1.1:** Project scope showing the interdisciplinary of the system

## 1.9 Project Plan:

This project underwent various stages and different phases that are needed to accomplish its outcomes throughout a period of two semesters (eight months). A Gantt chart of an eight months strategy has been formed as demonstrated in Figure 1.2.

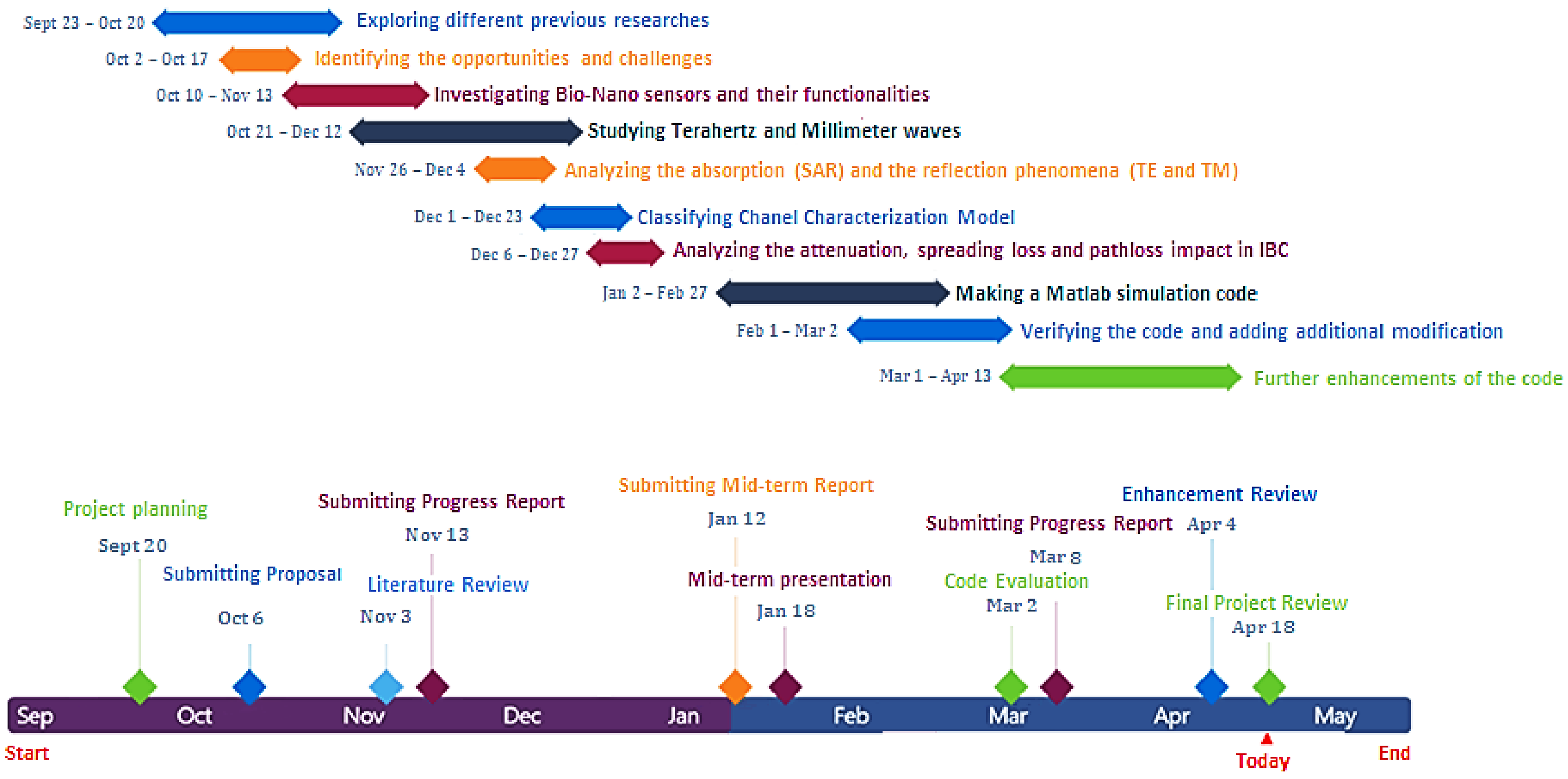

**Figure 1.2:** Gantt chart showing the various stages of this project.

# *Chapter2:*

# *Revolutionizing the Healthcare of the Future through Nano-medicine: Opportunities and Challenges:*

## 2.1 Background:

Nanotechnology is an astoundingly developing multidisciplinary logical field worried with the investigation of frameworks in view of their nuclear or atomic details. The importance of this innovation lies in its capacity to configuration, combine and portray materials and gadgets whose littlest practical association in no less than one measurement is on the nanometer scale (one-billionth of a meter) [26]. It gives unprecedented open doors to enhance material and material gadgets as well as to make new "savvy" gadgets and innovations where existing and more regular advancements might achieve their cutoff points. Essentially, nanotechnology includes the advancement of nanomachines, which are coordinated utilitarian gadgets fit for performing straightforward errands at the nano level. The interconnection of nano-machines in a system or "nanonetwork" is the basic usage utilized as a part of request to conquer the confinements of individual nano-gadgets [34]. Therefore, nanonetworks offer ascent to an unlimited cluster of utilizations in the medicinal, modern, natural and in addition military fields.

The way that natural species constitute of atomic structures at the nanoscale level makes nanotechnology crucial to the progressions in science. Really, these species hold onto various structures, for example, proteins, polymers, starches (sugars), and lipids, which change in their concoction, physical, and utilitarian properties [26]. Such basic assortment and framework flexibility have significant ramifications on the plan and advancement of new and counterfeit gatherings that are basic to organic and medicinal applications. By the methods for nanotechnology, these organic units will be better fathomed with the goal that they can be particularly guided or coordinated. Therefore, a careful comprehension of nanotechnology could unleash achievements in pharmaceutical, hereditary designing, genomics, and also proteomics in the coming decades [27].

In this report, current advances in present day prescription are displayed and examined. Bunch applications used in the finding and treatment of sicknesses are pondered including drug conveyance, quality conveyance, invulnerable framework bolster and additionally wellbeing observing. Moreover, the innovation of bio-nanosensors is exhibited showing their capacity to detect any biochemical and biophysical wave related with ailments at the sub-atomic or cell levels. Encourage, nanoparticles are exhibited as basic building pieces of nanotechnology in which different sorts of materials are introduced and portray. At last, the difficulties and future patterns related with this creating innovation are shown.

This chapter of this report is composed of the following categories. Segment 3.2 introduces the idea of nano-medicine which has upset the medicinal services framework. Area 3.3 portrays different uses of bio-nanosensors in the field of solution. In Section 3.4, nanoparticles are exhibited alongside their applications in indicative medication and clinical treatment. Segment 2.5 outlines the difficulties and future research headings connected with the field of nanotechnology.

## 2.2 Nano-medicine:

The meeting of late advances in nanotechnology with present day science and drug has brought about another exploration space of nanobiotechnology. In particular, nanomedicine is viewed as a branch of nanotechnology alluding to a very particular therapeutic mediation at the sub-atomic scale. The field of "Nanomedicine" is the science and innovation of diagnosing, treating, and averting illness and traumatic harm, of calming agony, and of saving and enhancing human wellbeing, utilizing nanoscale organized materials, biotechnology, hereditary designing, and in the end complex machine frameworks and nanorobots [27]. In-vitro diagnostics, nanomedicine could expand the effectiveness and unwavering quality of the conclusion utilizing human liquids, tissue tests, or particular nanodevices bringing about various examination at subcellular scale. In vivo diagnostics, nanomedicine could create gadgets ready to work inside the human body keeping in mind the end goal to distinguish the early nearness of an ailment, or to recognize and evaluate both dangerous particles and tumor cells [26]. In particular, one of the developing multidisciplinary fields in nanomedicine is called regenerative medication. Looks into in tissue regenerative medication points in creating inserts equipped for conveying drugs, development components, and hormones for tissue repair. They give maintained delivery of bio-active molecules to support survival, infiltration and expansion of cells for tissue building. The normal result of such treatment methodology is to have finish tissue substitution and utilitarian recuperation [28]. Among the utilizations of nano-medicine are medication and quality conveyance, invulnerable framework bolster and wellbeing observing which will be addressed in the resulting areas.

**Drug Delivery**

The special feature of nanotechnology in this regard, is that it is able to deliver the medicine into specific damaged cells with the aid of the nanoparticles. At the same time, neighboring cells will not be affected at all. One of the main applications for that, the smart insulin nano-acutators which constantly measure the sugar level in blood and hence, inject only the needed amount that is pre-calculated already. These highly accurate systems would lead to significant improvement in the absorption, stability and therapeutic concentration of the drug within the target tissue [30].

**Gene Delivery**

Gene therapy is lately introduced method utilized for the prevention and treatment of genetic disorders. This technique can repair defective genes which have the responsibility for the disease development by replacing the incorrect genes. One of the fundamental approaches is the insertion of normal genes any location within the genome which then will lead to replacing the nonworking gene.

**Immune Support**

Nanoparticles can be designed either to target or avoid interactions with the body immune system. The reason for that is that the interaction between immune system and nanoparticles is desired when it results in beneficial applications. For instance, vaccines or therapeutics for inflammatory and autoimmune disorders [31]. Moreover, special types of Nano-biosensor can detect any foreign bodies and pathogens within the human body [29].

**Health Monitoring**

Nano-sensors have the ability to offer highly precise data at the cellular level. The concentration of glucose, sodium and oxygen in blood as well as the level of cholesterol and other infectious agents can be measured and monitored by nonosensors. This information can be delivered with the aid of nano-networks [32].

## 2.3 Bio-Nanosensors:

Recently, the world has witnessed many dramatical changes and unexpected developments in all aspects, particularly the technological advances. These advances have totally changed the traditional ways by which diseases have been diagnosed, treated, and cured. For example, the nanotechnology can be considered as one of these amazing inventions. The medical field or health care sector owed the bio-Nanosensors greatly, because of its limitless pros whether in its high sensitivity or the variable tasks that can be carried out by using this technology. This invention has made diagnosing diseases in their early stages more possible and approachable since it can magnify and capture 3Ds images of tiniest molecules, cells or tissues. This process has eased the idea of detecting diseases and treating them. In addition to that the small size of these instruments or inventions gives more space and tasks for medical specialists whether in their labs, or surgical operation rooms. Finally, the most amazing or unbelievable invention is the Nano-robot which can be considered as the crown of these invention. This matchless instrument can perform many magical tasks in the human body through the vascular system. In addition to that all cells and tissues have become approachable and accessible through this invention.

## 2.4 Nano-particles:

Through nanotechnology, novel nano-scaled materials and particles are created. These materials have exceptional properties and superior abilities that conventional combination and assembling units can't make [32]. Such little particles, alluded to as nanoparticles, move through cell layers and diffuse through basic hindrances. Subsequently, they are imagined as nanoscaled ships, which transport high potential pharmaceutics absolutely to their goal . Nanoparticles go about as perfect bearers to convey anticancer medications and other remedial medications at the objective site with ideal effectiveness and least inadvertent blow-back to the neighboring sound tissues. The improvement procedures are intensely interwind with biotechnology and data innovation, making its degree wide [33]. Being the auxiliary components of nanotechnology, nanoparticles can be connected from multiple points of view including fluorescent natural markers, protein location markers, attractive reverberation picture enhancers, and also organic particles' separators and purifiers [33]. Carbon nanotubes, quantum dabs and dendrimers, outlined in Figure 2, are couple of cases of nanoparticles that will be talked about in the consequent subsections in this review. In addition, table 2.1 records various nanomaterials alongside their applications in the field of nanomedicine.

**Table 2.1:** Applications of Nano-Substances.

| **Nanomaterial** | **Uses** | **Ref.** |
|---|---|---|
| Graphene Oxide | Detect low level of cancer cells (3-5 malignancy cells/ml blood) | [34] |
| Single-Walled Carbon Nanotubes (SWNT) | Monitor blood nitric oxide level in incendiary illnesses | [35] |
| Magnetic Nanoparticles | Allows constant checking the micro-vesicles in the blood | [36] |
| Gold nanoparticles coated with influenza antibodies | Detect flu infection in test. | [37] |
| Gold/Bismuthbased nanoparticles | To amass radiation utilized as a part of radiation treatment to treat disease tumors. | [38] |
| Nanocrystalline Silver | Antimicrobial operator for treatment of wounds. | [39] |
| Fullerene nanoparticles | Reduce hypersensitive responses | [40] |

**Carbon Nanotubes:**

Carbon nanotubes have pulled in the consideration of numerous specialists since their disclosure a decade ago. These carbon particles are little tubes with distances across down to 0.4nm, while their lengths can grow up to a million times their breadth. Utilizing their astounding electrical properties, basic electronic rationale circuits have been constructed. These structures are promising for the semiconductor business which is driving the look for scaling down. They are great conductors, as well as have all the earmarks of being the yet discovered material with the greatest particular solidness, having a large portion of the thickness of aluminum [31]. Also, these tubes can change electromagnetic vitality into warmth, creating a temperature increment deadly to disease cells. This is accomplished either by expanding the attractive field or by illumination with an outside laser source at the area where nanoparticles are bound to tumor cells [33].

**Quantum dots**

Quantum specks are a focal topic in nanotechnology. They show properties that are halfway between those of mass semiconductors and those of discrete atoms. Their optoelectronic properties change as a component of both size and shape. Due to the irrelatively simple and inexpensive synthesis, quantum dots have already entered the market for experimental biomedical imaging applications [32]. Furthermore, they can be made to emit light at any wavelength in the visible and infrared extents. Quantum specks can be likewise joined to an assortment of surface ligands, and embedded into an assortment of living beings for in vivo examine [33].

**Dendrimers**

Dendrimers are macromolecular mixes comprised of a progression of branches around an inward center [28]. Dendritic polymers give a road to the conveyance of qualities. They can frame to a great degree little particles, on the request of nanometers, and have appeared to be successful as DNA conjugates [28]. Dendrimers have enhanced physical, compound, and natural properties in contrast with conventional polymers. Because of the tree like structure of dendrimers, an assortment of atoms including medications can be joined to these mixes. Under 5nm in distance across, dendrimers are sufficiently little to sneak past minor openings in cell layers and to pass vascular pores and tissues in a more proficient manner contrasted with bigger polymer particles [31].

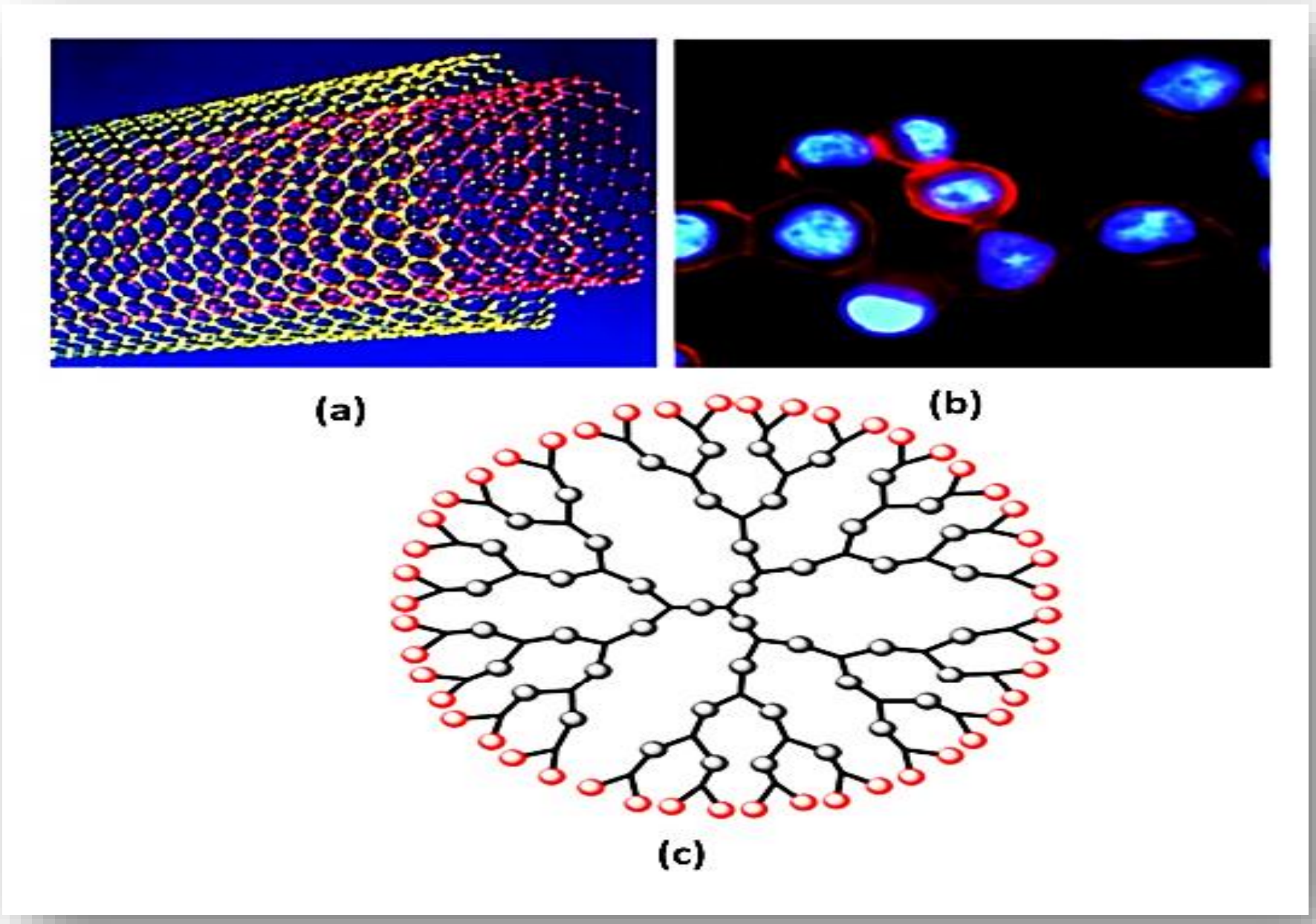

**Figure 2:** Representative types of nanopraticles: a) Carbon nanotube b) Quantum dots c) Dendrimer.

## 2.5 Challenges and Future Directions:

Nanomedicine represents one of the fastest growing research territories and is viewed as one of the promising instruments for frontier malady treatment. Table 2.2 gives a future point of view toward the open doors in the field of nanotechnology in the medicinal field. By and by, nanomaterial portrayal, security concerns, and regulatory issues hamper the widespread application of nanomedicine therapeutics. In this way, complete and reproducible portrayal of nanomedicine items to foresee their viability and security in people is justified. When all is proposed in achieved, the principle physico-substance elements of nanocarriers are structure, piece, measure, surface properties, porosity, charge, and collection conduct [26]. Fluctuation inside these properties makes it difficult to characterize nanomedicine items prior and then afterward organization. Quantitative logical techniques must have the capacity to screen all vital quality parts of the nano-sized mixes. Subsequently, in vitro and in vivo models precisely speaking to the clinical setting must be produced.

Besides, across the board utilization of nanoparticles makes it important to address poisonous quality issues for the human wellbeing and condition.The nano-scale measurement of nanomedicine items is like that of intracellular organelles or biomolecules required in cell signaling. A few reviews have exhibited that nanoparticles might be related with inconvenient natural cooperations. This has prompted the rise of nanotoxicology as a free field of research [27]. An expanding measure of information is getting to be plainly accessible with respect to the lethality of nanoparticles. In any case, it stays hard to contrast the poisonous quality of nanomaterials and that of macromaterials. At present utilized harmfulness appraisals for nanomaterials are the same as those utilized for traditional medications. Thusly, the assessment of nanoparticle poisonous quality might be lacking at present and the improvement of reciprocal lethality examines for nanomedicine mixes ought to be energized [28]. It ought to be noticed that various variables tweak the poisonous quality of nanomaterials. Properties, for example, measure, shape, surface territory, and surface charge influence the conduct and execution of nanomedicine medications at the nano-bio interface.

Additionally, without proof and direction, administrative choices concerning nanomedicine therapeutics depend on individual evaluation of advantages and dangers [29]. However, this procedure is tedious and may bring about administrative deferrals for nanomedicine items. Viable direction requires keeping up abnormal state aptitude in inventive innovations. For instance, the FDA teams up with the Nanotechnology Characterization Laboratory (NCL) to encourage the administrative survey and inside and out portrayal of nanomedicine items. As a piece of the Horizon 2020 venture, the European Technology Platform on Nanomedicine (ETPN) plans to set up an European Nano-Characterisation Laboratory (EU-NCL) [5]. Notwithstanding, it is important that the administrative endorsement of nanomedicine stays testing particularly that created nanomaterials with cutting edge properties are rising regular.

**Table 2.2:** The future of nanotechnology in the medical field [31]

| Desires of Nanotechnology in Medicine Year | **Year** |
|---|---|
| Widespread utilization of medications to cure viral liver sickness. - Widespread utilization of an oral insulin treatment technique. | 2014 |
| - Development of a cell treatment strategy for myocardial infraction.<br>- Widespread utilization of early analysis strategies in light of blood tests for all growths. | 2015 |
| -Development of treatment strategies prepared to do totally curing hypersensitivities, for example, atopic dermatitis | 2016 |
| - Widespread utilization of techniques to anticipate growth in light of hereditary conclusion | 2017 |
| - Improvement in the normal five-year survival rate for a wide range of disease to over 70%.<br>- Widespread utilization of regenerative treatment innovation for harmed organs utilizing embryonic undeveloped cells, of treatment strategies able to do totally curing Alzheimer's ailment. | 2020 |

# *Chapter3:*

# *In Vivo Wireless Body Communications: State-of-the-Art and Future Directions:*

## 3.1 Background:

Remote Body Area Networks (WBANs) is an energizing innovation that guarantees to convey medicinal services observing applications to another level of personalization. The point of these applications is to guarantee nonstop checking of the patients' key parameters, while giving them the flexibility of moving in this way bringing about an improved nature of human services [32]. Truth be told, a WBAN is a system of wearable registering gadgets working on, in, or around the body. It comprises of a gathering of small hubs that are outfitted with biomedical sensors, movement indicators, and remote specialized gadgets [33]. Really, propelled medicinal services conveyance depends on both body surface and inner sensors since they lessen the intrusiveness of various therapeutic systems [34]. Sensors, for example, those appeared in Fig. 1 transmit information to checking gadgets of the clinic Information Technology (IT) foundation. Electrocardiogram (ECG), electroencephalography (EEG), body temperature, beat oximetry (SpO2), and pulse are developing as long haul checking sensors for crisis and hazard patients [30]. Propelled sensors for synthetic, physical and visual applications will become some portion of future monitoring platforms to check, for instance, insulin or hemoglobin. The advantage gave by WBAN is clear to the patients' solace particularly for long term observing and additionally complex checking amid surgery and restorative examinations [35]. One appealing element of the rising Internet of Things is to consider *in vivo* organizing for WBANs as an imperative application stage that encourages nonstop remotely empowered social insurance [36]. *In vivo* communication, otherwise called Intra Body Communication (IBC), utilizes the human body to transmit electrical signs, where the emanated vitality is mostly confined within the body [37]. Internal health monitoring [38], inward medication organization [39], and insignificantly obtrusive surgery [40] are cases of the pool of uses that require communication from *in vivo* sensors/actuators to body worn/surface hubs. Be that as it may, the investigation of *in vivo* remote transmission, from inside the body to outside handsets is still at its initial stages.

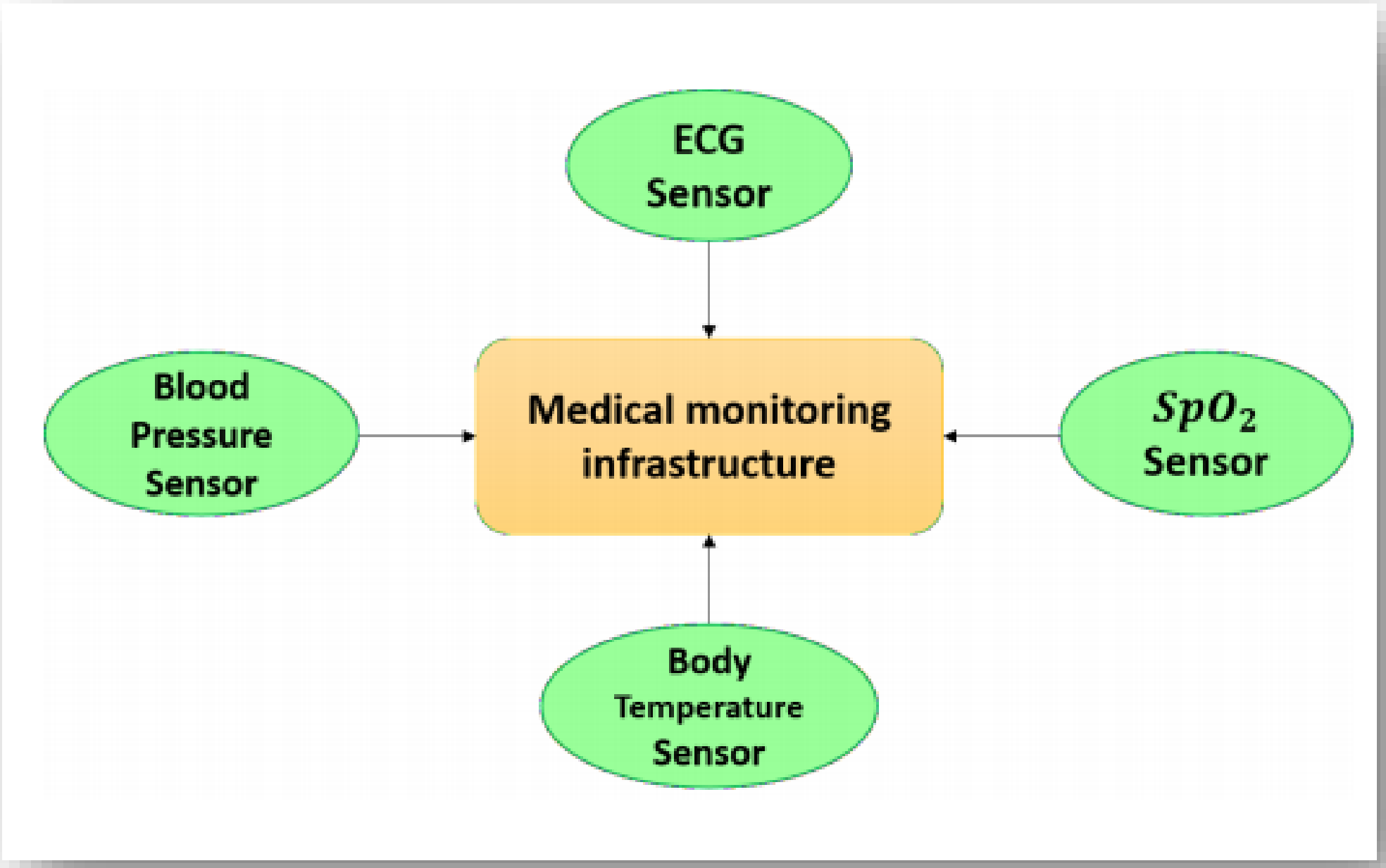

**Figure 3.1.1:**.Sensor system of biomedical observing applications.

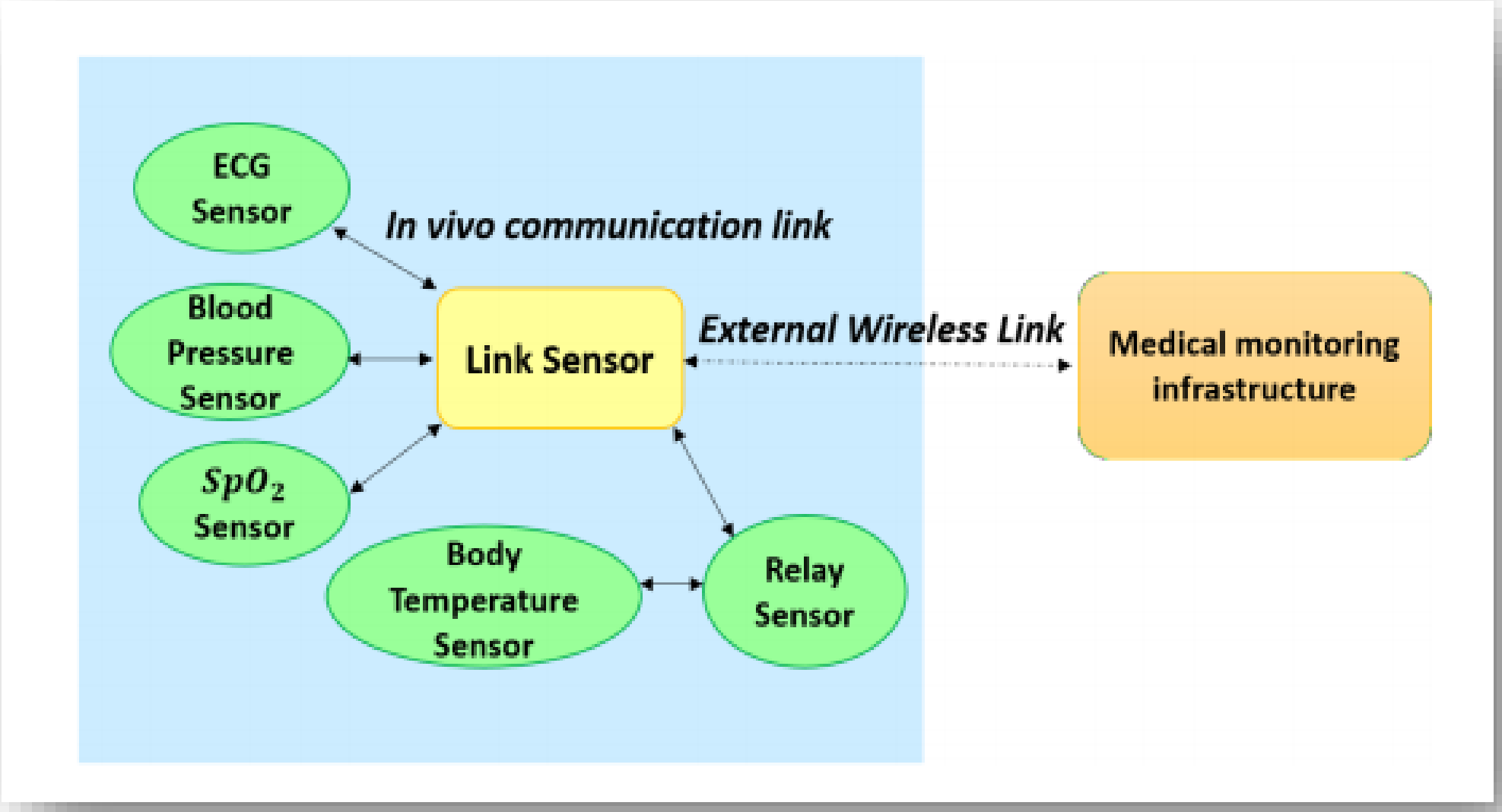

**Figure 3.1.2:** Disentangled diagram of the *in vivo* communication system.

Figure. 3.1.2 demonstrates the altered system association for interconnecting the biomedical sensors. The information is not exchanged specifically from the biomedical sensors to the healing center foundation as in Figure. 3.1.1 For sure, the sensors send their information by means of a reasonable low-power and low-rate *in vivo* communication connection to the focal connection sensor (situated on the body like every single other sensor). Any of the sensors may go about as a transfer sensor between the coveted sensor and the focal connection sensor if an immediate association is restricted. An outside remote connection empowers the information trade between the focal connection sensor and the outer healing center framework [35].

This report studies the current research which examines the state of art of the *in vivo* communication. It likewise concentrates on describing and displaying the *in vivo* remote channel and standing out this channel from the other commonplace channels. MIMO *in vivo* is likewise respected in this diagram since it altogether improves the execution pick up and information rates. Besides, this report proposes some future research territories. Whatever remains of the report is composed as takes after. In Section 2.2, we show the condition of specialty of *in vivo* communication. Directed research on *in vivo* divert portrayal is given in Section 2.3. The MIMO *in vivo* framework is depicted in Section 2.4. In Section 2.5, future research ranges are introduced.

## 3.2 State of art of "*In Vivo Communication*":

*In vivo* communication is a demonstrable wave transmission field which uses the human body as a transmission medium for electrical signs [37]. The body turns into an essential part of the transmission framework. Electrical current enlistment into the human tissue is empowered through refined handsets while savvy information transmission is given by cutting edge encoding and compression. Figure.2.2 below shows the main components of an *in vivo* communication connection.

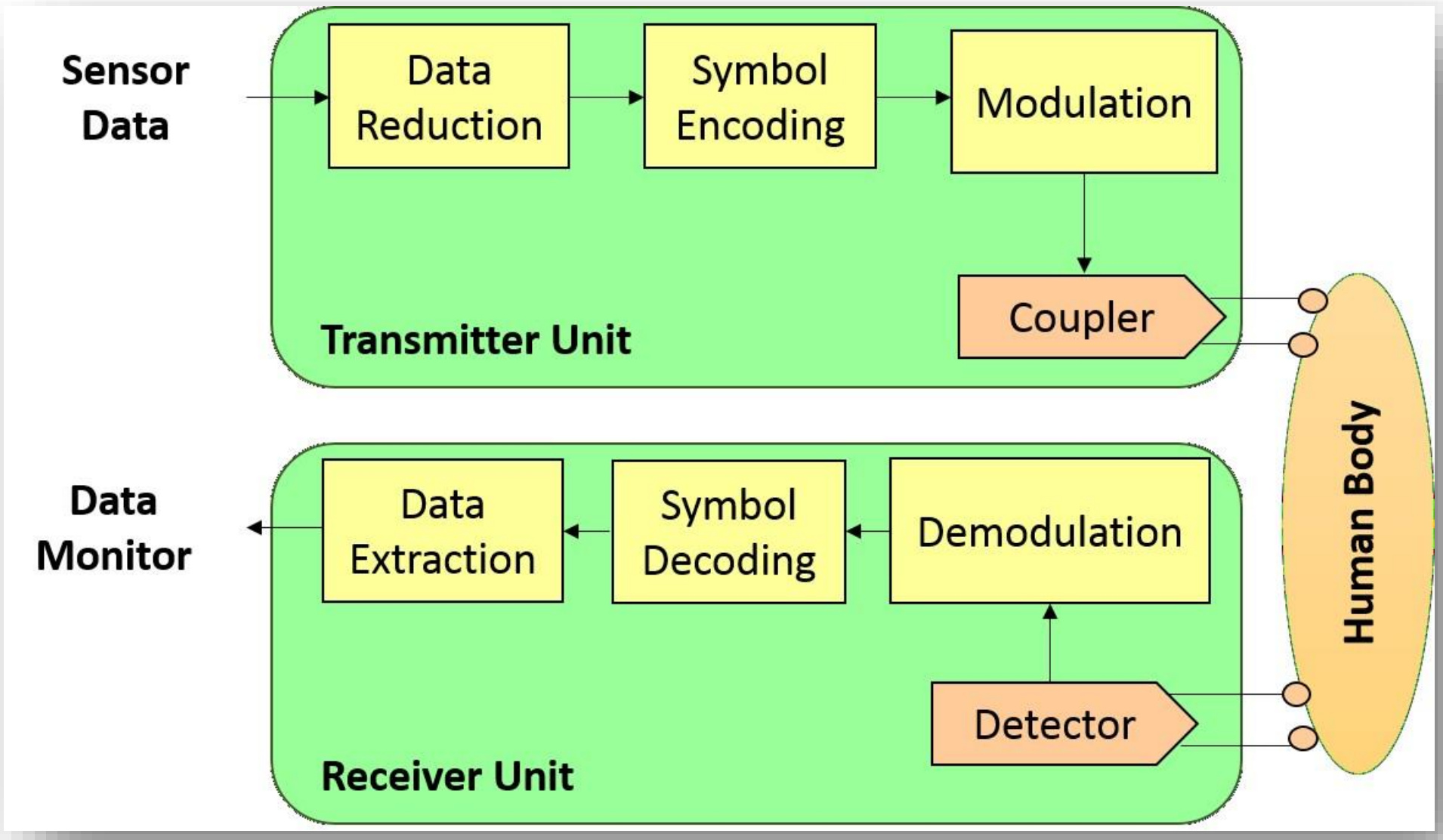

**Figure 3.2:** *In vivo* communication for information transmission between sensors empowered by transmitter and collector units.

A transmitter unit grants sensor information to be packed and encoded. It then passes on the information by a current-controlled couple unit. The human body acts as the transmission channel. Electrical signs are coupled into the human tissue and dispersed over various body locales. Then again, the collector unit is made out of a simple finder unit that intensifies the actuated wave and computerized substances for information demodulation, translating, and extraction [35]. Creating body transmission frameworks have demonstrated the feasibility of transmitting electrical waves through the human body. Regardless, point by point qualities of the human body are deficient with regards to up until now. Not a ton is thought about the effect of human tissue on electrical wave transmission. Really, for cutting edge handset outlines, the impacts and points of confinement of the tissue must be warily mulled over [35] [32].

## 3.3 *In Vivo* channel modeling and characterization:

An *in vivo* channel is portrayed as being both inhomogeneous and extremely lossy [58]. Fundamentally, in an *in vivo* channel, the electromagnetic wave goes through different disparate media that have diverse electrical properties, as showed in Figure. 3.3. This prompts the diminishment in the wave engendering speed in a few organs and the incitement of noteworthy time scattering that varies with every organ and body tissue [34]. The past impact combined with weakening due ingestion by the diverse layers result in the debasement of the nature of the transmitted wave in the *in vivo* channel.

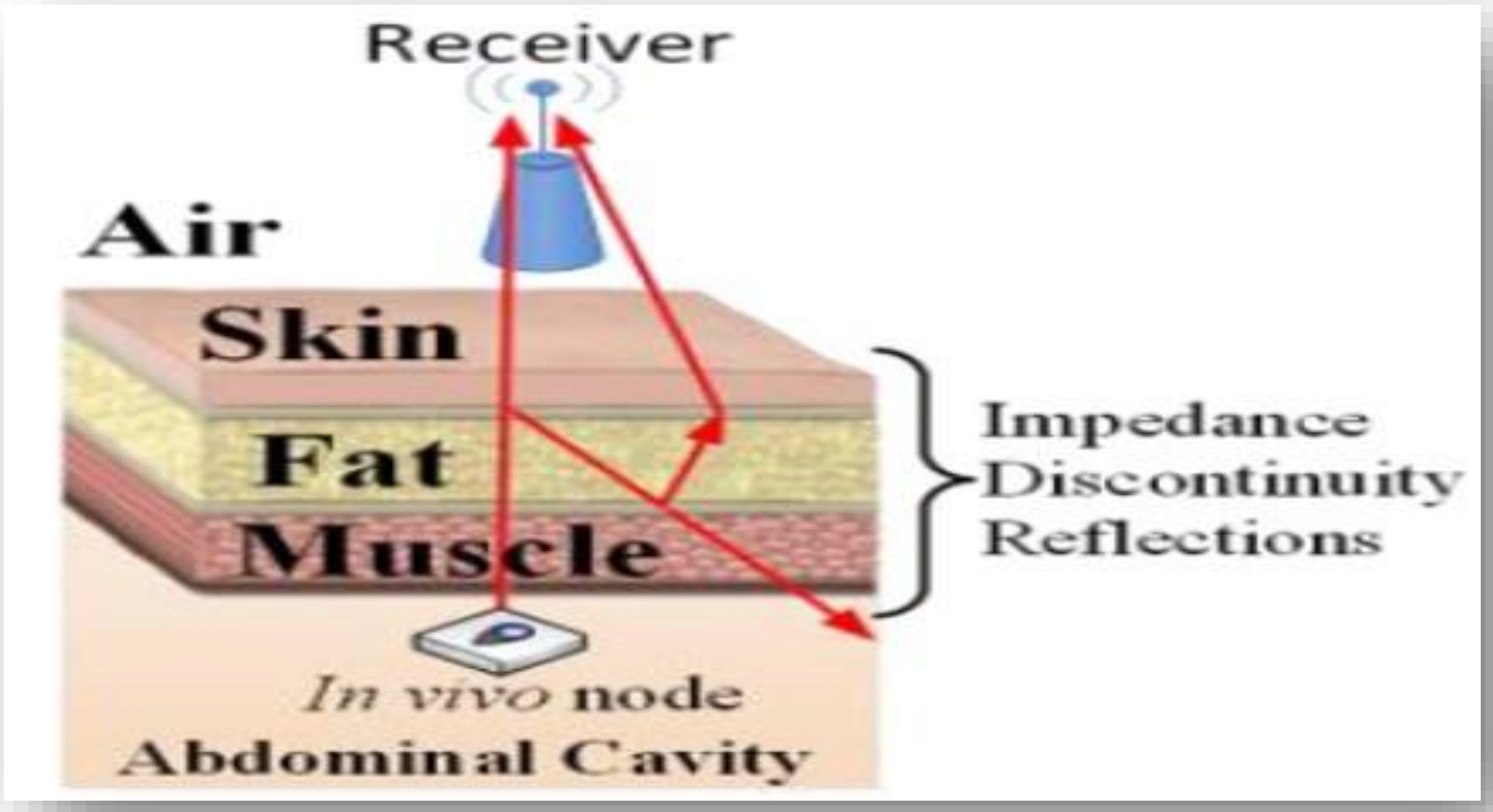

**Figure 3.3:** *In vivo* multi-path channel [9].

Moreover, since the *in vivo* reception apparatuses are transmitting into a complex lossy medium, the emanating close fields will firmly couple to the lossy environment. This signifies that the radiated control depends on both the spiral and precise positions; subsequently, the close field impact must be constantly considered when working in an *in vivo* condition [35]. The electric and attractive fields carry on distinctively in the transmitting close field contrasted with the far field. In this way, the remote channel inside the body requires diverse connection conditions [36]. It must be noted too that both the postpone spread and multipath dispersing of a cell system are not specifically relevant to close handle channels inside the body. The purpose for this is the way that the wavelength of the wave is any longer than the spread condition in the close field [33]. The researchers in [34] utilized an exact human body to explore the variety in signal loss at various radio frequencies as a component of position around the body. They saw huge varieties in the Received Signal Strength (RSS) which happen with changing places of the outside get receiving wire at a settled position from the interior radio wire [34]. By and by, their examination did not consider the essential portrayal of the *in vivo* channel. In [37], the researchers utilized an immersive perception condition to describe RF spread from restorative inserts. In light of 3-D electromagnetic recreations, an experimental path loss (PL) model is created in [38] to distinguish losses in homogeneous human tissues. Further, numerical and trial examinations of biotelemetry radio stations and wave weakening in human subjects with ingested remote inserts are introduced in [39]. Demonstrating the *in vivo* remote channel including building a phenomenological path loss model is one of the significant research objectives in this field. A significant comprehension of the channel qualities is required for characterizing the channel limitations and the ensuing frameworks' imperatives of a handset plan [35].

### 3.3.1 Path Loss in an *in vivo* channel:

Attenuation *in vivo* channels can be explored utilizing either a Hertzian-Dipole receiving wire or a monopole reception apparatus. The previous case is introduced in [43] in which path loss is inspected with negligible radio wire impacts. The length of the Hertzian-Dipole is so little bringing about little collaboration with its encompassing condition. The path loss can be figured as

$$Path\ loss\ (\ r,\theta,\Phi\ ) = 10 \times log_{10}(\frac{|E|^2_{r=0}}{|E|^2_{r\ \theta,\Phi}}) \quad (1)$$

where $r$ expresses to the separation from the beginning, i.e. the sweep in round directions, $\theta$ is the polar edge and $\Phi$ is the azimuth edge. $|E|^2_{r\ \theta,\Phi}$ is the square of the greatness of the electric field at the measuring point and $|E|^2_{r=0}$ is the square of the extent of E field at the source. Because of the way that the *in vivo* condition is an inhomogeneous medium, it is obligatory to estimate the path loss in the circular framework [43]. The setup of this approach can be found in Figure. 3.3.1.a in which it incorporates the truncated human body, the Hertzian-Dipole and the circular framework.

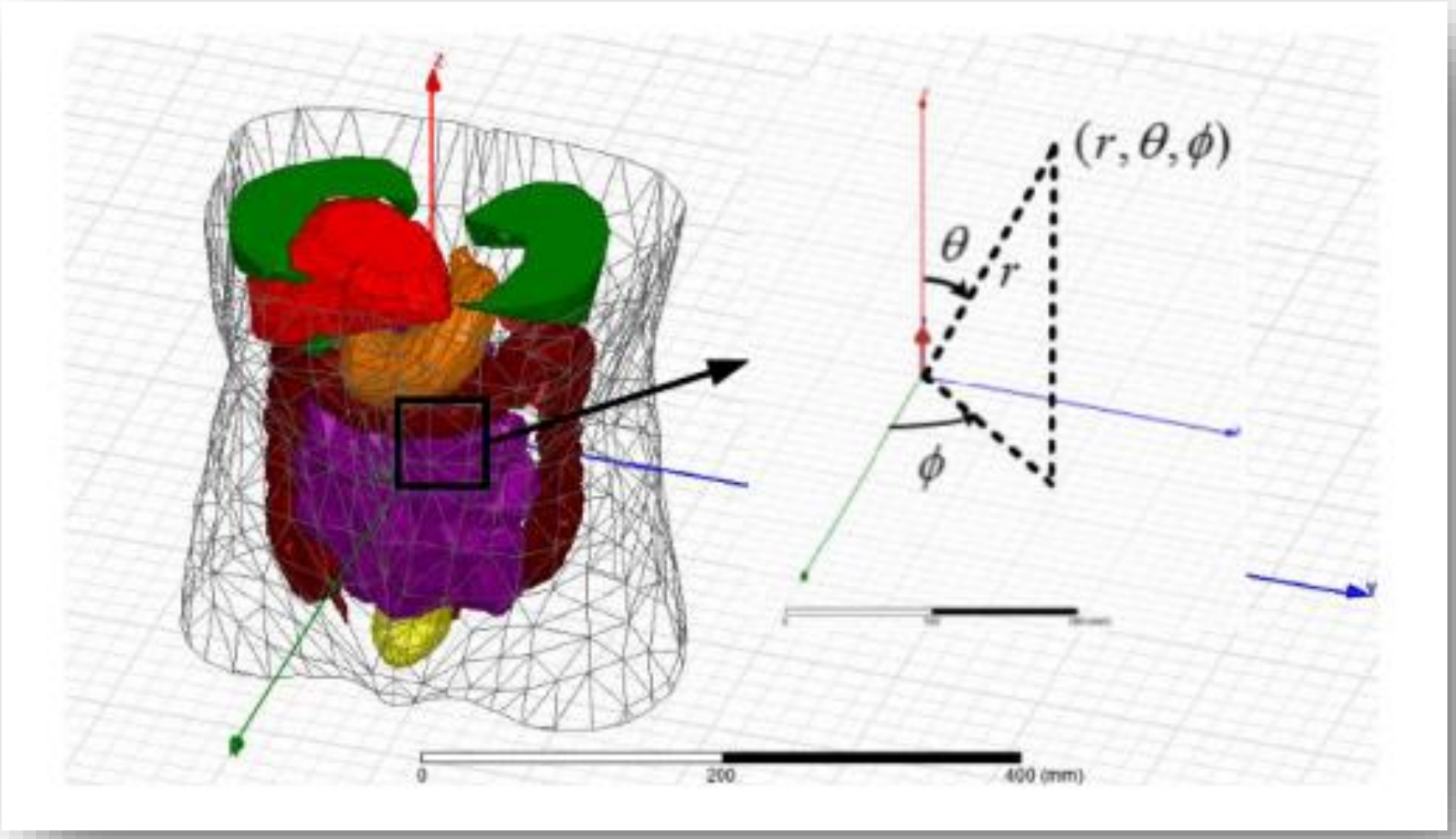

**Figure 3.3.1.a:** Truncated human body with Hertzian-Dipole at the root in spericl framework [58].

The last case is introduced in [44]. Really, monopoles are great decision of commonsense receiving wires since they are little in size, basic and omnidirectional. The path loss can be measured by dispersing parameters (S parameters) that depict the input output connection between ports (or terminals) in an electrical framework [44] As per Figure. 2.3.1.b, on the off chance that we set Port 1 on transmit reception apparatus and Port 2 on get receiving wire, then $S_{21}$ expresses to the power pick up of Port 1 to Port 2, that is

$$|S_{21}|^2 = \frac{P_r}{P_t} \tag{2}$$

Where $P_r$ is the received power and $P_t$ is the transmitted power. In this manner, we figure the path loss by the equation underneath,

$$Pathloss(\,dB) = -20\ \times log_{10}(|S_{21}|) \tag{3}$$

In light of the representation introduced in [43], it can be perceived that there is a significant distinction in the practices of the path loss between the *in vivo* and free space condition. Indeed, noteworthy weakening happens inside the body bringing about an *in vivo* path loss that can be up to 45 dB more noteworthy than the free space way attenuation. Vacillations in the out-of-body locale are experienced by the *in vivo* path loss. Then again, the free space attenuation rises. The inhomogeneous medium outcomes also in precise ward path lost [43].

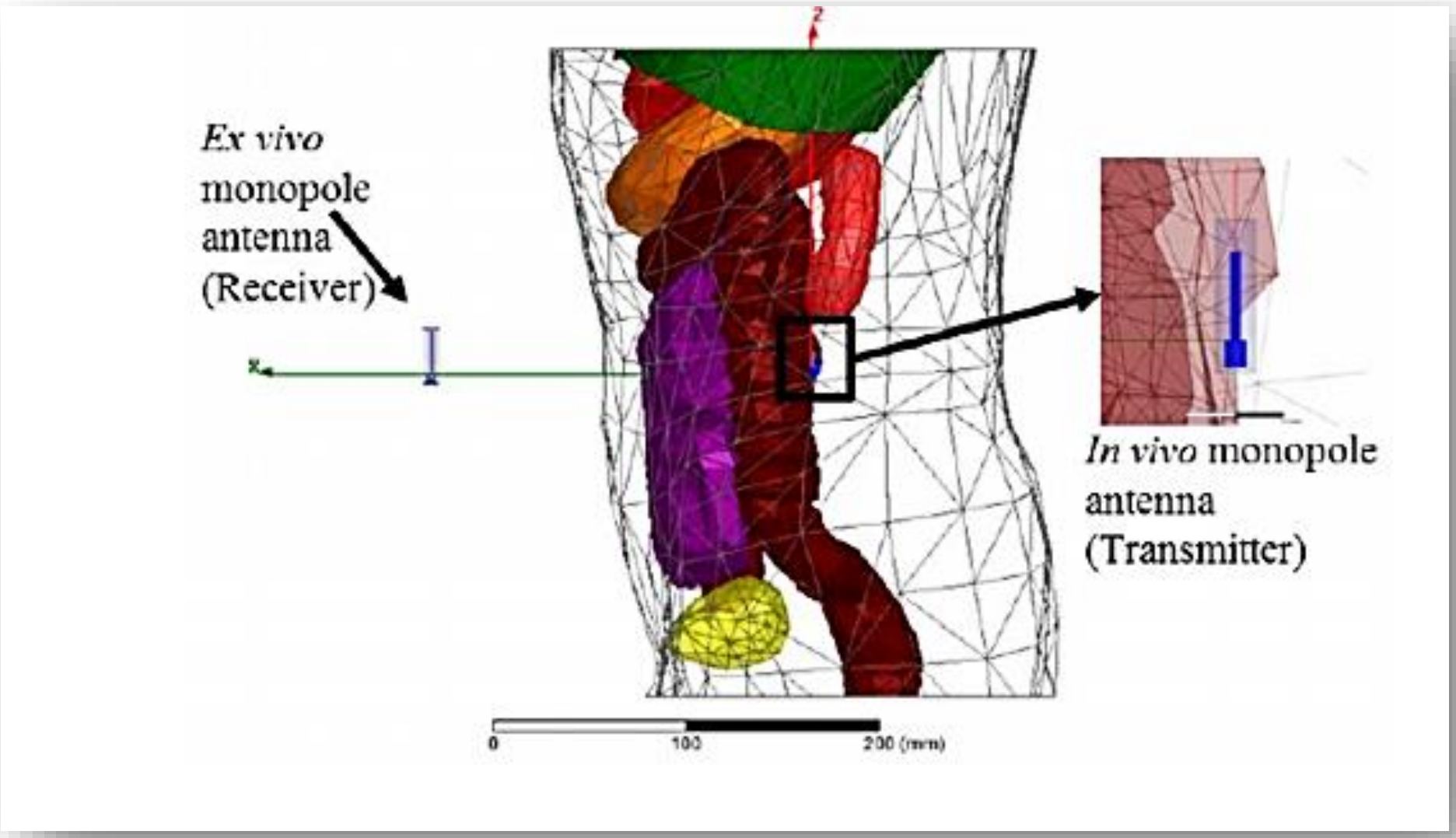

**Figure 3.3.1.b:** Representation setup by utilizing monopoles to estimate the attenuation [43].

### 3.3.2 Examination of *Ex Vivo* and *In Vivo* Channels characteristics:

The distinctive qualities between *ex vivo* and *in vivo* channels are abridged in [44] as illustrated in Table 3.

**Table 3:** Comparison of *Ex Vivo* and *In Vivo* channels [44].

| **Features** | ***Ex vivo*** | ***In vivo*** |
|---|---|---|
| Physical Wave Propagation | Constant speed<br>Multipath- reflection, scattering and diffraction | Variable speed<br>Multipath and penetration |
| Attenuation and Path Loss | Lossless medium<br>Decreases inversely with distance | Very lossy medium<br>Angular (directional) dependent |
| Dispersion | Multipath delays-time dispersion | Multipath delays of variable speed - frequency dependency-time dispersion |
| Directionality | Propagation essentially uniform | Propagation varies with direction<br>Directionality of antennas changes with position |
| Near Field<br>Communications | Deterministic near-field region around the antenna | Inhomogeneous medium near field region changes with angles and position inside the body |
| Power<br>Limitations | Average and Peak | Plus specific absorption rate (SAR) |
| Shadowing | Follows a *log normal* distribution | To be determined |
| Multipath Fading | Flat fading and frequency selective fading | To be determined |
| Antenna Gains | Constant | Angular and positional dependent<br>Gains highly attenuated |
| Wavelength | The speed of light in free space divided by frequency | $\lambda = \frac{c}{\sqrt{\epsilon_r} f}$<br>at 2.4GHz, average dielectric constant $\epsilon_r = 35$, which is roughly 6 times smaller than the wavelength in free space. |

## 3.4 Multi-input multi-output (MIMO) *in vivo* model:

Due to the lossy path of the *in vivo* medium, achieving high information rates with dependable execution is viewed as a test [51]. The explanation for this is the *in vivo* reception apparatus execution might be influenced by close field coupling as said before and the signs level will be restricted by a predefined Specific Absorption Rate (SAR) levels. The SAR is an estimation of how much power is retained per unit mass of conductive material, for our situation, the human organs [43].

This estimation is constrained by the Federal Communications Commission (FCC) which in turns restrains the transmission control [40]. The researchers in [44] examined the Bit Error Rate for a MIMO *in vivo* framework. Really, by contrasting their outcomes with a 2× 2 SISO *in vivo*, it was apparent that critical execution additions can be accomplished when utilizing a 2× 2 MIMO *in vivo*. This setup permits greatest SAR levels to be met which brings about the likelihood of accomplishing target information rates up to 100 Mbps if the separation between the transmit (Tx) and get (Rx) reception apparatuses is inside 9.5 cm [40].

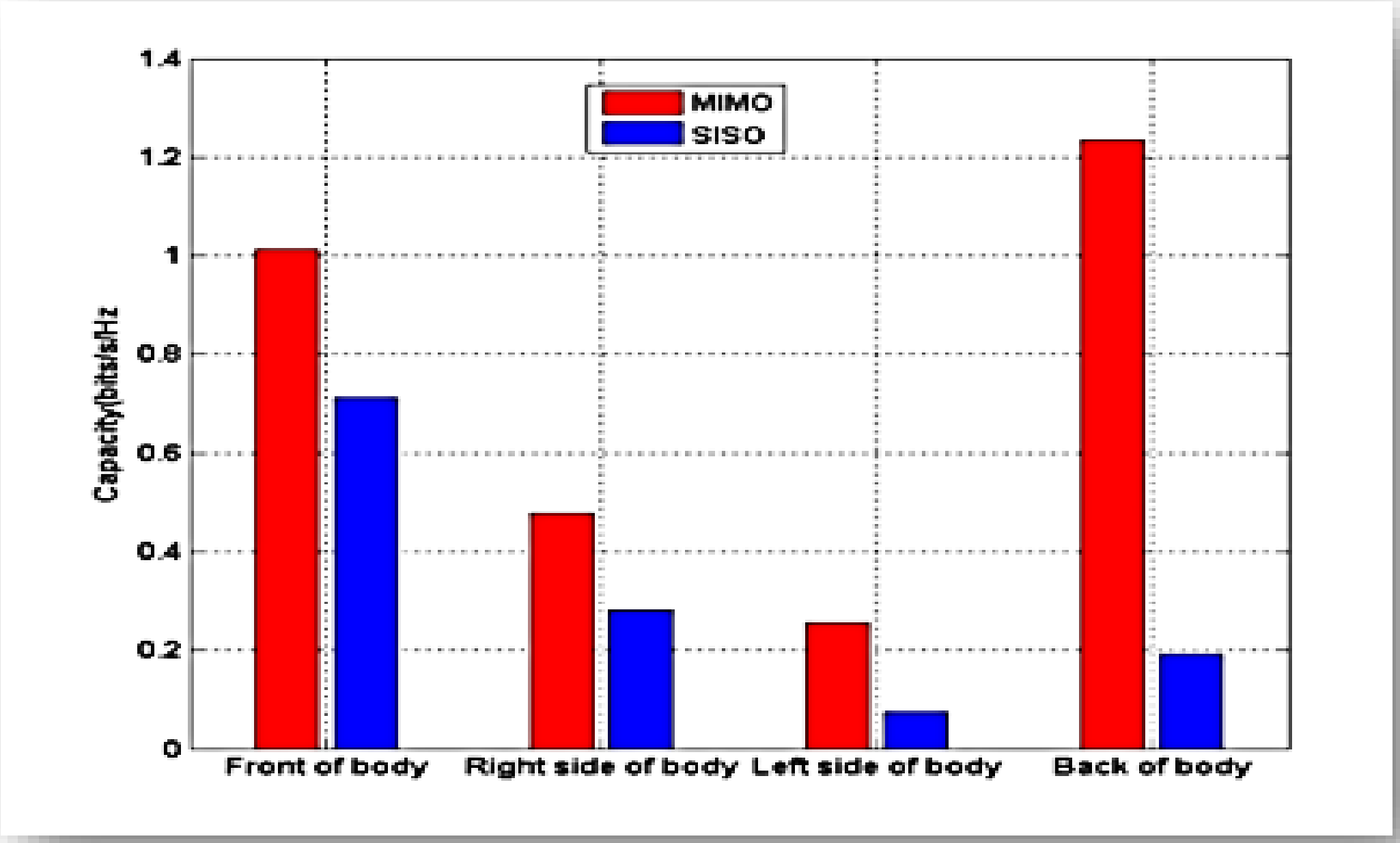

**Figure. 3.4:** 2 × 2 MIMO and SISO *in vivo* system capacity comparison [36].

Additionally, in [36], it was demonstrated that MIMO *in vivo* can achieve better performance in comparison to SISO systems as well as impressively better framework limit can be watched when Rx radio wires are put along the edge of the body. Figure 2.4 presents the *in vivo* framework limit with respect to front, right side, left side, and back of the body. Also, it was seen that keeping in mind the end goal to meet high information rate prerequisites of up to 100 Mbps with a separation between the Tx and Rx radio wires more noteworthy than 12 cm for a 20 MHz channel, transfer or other comparative helpful arranged correspondences are important to be brought into the WBAN organize [36].

## 3.5 Future Directions:

The real future research course concentrates on exploring high throughput, proficient and strong novel systems administration innovations that empower solid data transmission between gadgets. So as to accomplish a framework with such particulars, awesome concentrate ought to be spent on architecting, acknowledging and organizing a group of remotely controlled *in vivo* gadgets. Also, electronic and mechanical scaling down of complex frameworks are required. Fundamentally, movement control, video zoom, auto center and LED light are a portion of the different capacities that require watchful control of the *in vivo* remote gadgets. The principal viewpoint that ought to be respected keeping in mind the end goal to guarantee enhanced future *in vivo* communication is recurrence. In reality, the recurrence extend chose for such communication assumes a critical part in the outline and execution of the framework.

In fact, there is an immediate connection amongst recurrence and tissue warming in which the higher the recurrence of the electromagnetic wave, the higher is its ingestion by the tissue, in this manner the more noteworthy the tissue warming. Accordingly, it is great to utilize bring down frequencies for communications. However, the lower the recurrence, the bigger the radio wires measurements. Therefore, a tradeoff between reception apparatus measurements and more prominent tissue warming ought to be accomplished [16]. Additionally, confinements exist when transmitting at high frequencies from *in vivo* gadgets to *ex vivo* handsets since the greatest transmit power is limited by the SAR wellbeing rules [40]. Notwithstanding when working under low clamor conditions with direct BER necessities, solid information transmission to an outer beneficiary must be accomplished when found near the body. Along these lines, when clamor levels increment or the BER turns out to be more stringent, a hand-off system or utilization of various reception apparatuses is fundamental to accomplish high information rates [40]. In view of the reproductions directed by [40] , the most extreme SAR levels happen at focuses nearest to the transmit reception apparatus. Thus, it could be inferred that by putting the transmitter advance from organs, the power levels could be expanded to acquire higher wave levels at the outer collector.

Moreover, the base transmission control levels required to build up a dependable remote information association must be recorded with the inserts put in various areas inside the body. The principal point of this information is to contrast the power levels and global security controls, restricting the human tissues presentation to radio recurrence signals. The second goal is to give reference levels for power utilization amid information transmission that can be valuable to gauge battery lifetime, since this errand is a standout amongst the most vitality requesting for the embed, especially, if contrasted and sensor information obtaining.

Besides, a considerable element in future in vivo research is the improvement of parametric models for the in vivo channel reaction which can beneficially affect the streamlining of cutting edge correspondence systems. This includes the factual portrayal of the in vivo channel and the use of MIMO innovation for enhanced transmission dependability and execution. Channel models are required to accomplish exact connection spending which helps in finding streamlined areas for setting the transmitter and collector on and inside the human body [42]. Then again, MIMO in vivo limit ought to be hypothetically considered consolidating both the close and far field impacts [44].

It must be noted also that the divert at the head in vivo communication is of a specific enthusiasm since most human correspondence organs, for example, mouth, ears and eyes are situated there. Along these lines, it is proposed to consider the ear to ear interface in reproductions, for it is the direst outcome imaginable at the head as it does not have the viewable pathway segment [43].

Subsequently of the assorted qualities found in the human body including liquids, fats, bones, and muscles, variety in vitality absorption exists combined with the likelihood of tissue harm because of warming by radiation. Such issues bring about muddling the plan of biosensors and biosensor systems; in this way, cautious consideration must be paid in future work to beat such confinements [41]. Another specific concern is biocompatibility. Essentially, any sensor must be organically inactive, as well as it ought not annoy the typical science its investigating [44].

Nanoscale in vivo gadgets are another significant part of organically motivated system frameworks since they open up astonishing open doors in human services. Undoubtedly, these frameworks can be brought into the human body and could mediate intimately with organs and cells. Nanodevices mediating at the atomic level will permit novel medicinal services answers for be created which will be more proficient as well as cost effective as a result of the likelihood of extensive scale generation [35].

Ultimately, testing utilizing genuine creatures to approve the reproduction results is one of the major research headings [42] notwithstanding hardware imitation with human body apparitions.

# *Chapter4:*

# *Terahertz Intra-body Communication:*

## 4.1 Trends of Terahertz in Wireless and Wired line communication System:

Within the last decade, new unprecedented techniques in generating, exchanging and transferring data information have emerged. This transformation implies that both wired and wireless traffic will significantly increase leading to a massive and extensive increase in response for the demanding ultra-high speed wireless communication. By relatively referring to Edholm's law of bandwidth, "Wireless data rates have doubled every 18 months over the last three decades". Therefore, as illustrated in below Figure, the drastically convergence of wireless data rates is in parallel with the Ethernet wired line systems. Subsequently, following this approach and by exploiting this new emerging technology, the applications of wireless Terahertz (THz) transmission channels are predicted to exist and be a reality within the upcoming decade [36]. Consequently, both novel physical structure characteristics and distinctive frequency bands are obligatory required to accommodate this significant increase in data rates. Subsequently, this implies that there is no other alternative apart from turning towards developing advanced extensively high carrier frequencies that utilizes the terahertz carrier frequencies to operate at various current modulation schemes such as Amplitude Shift Keying (ASK), Frequency Shift Keying (FSK) and Phase Shift Keying (PSK). In the view of satisfying such high demands, Terahertz (THz) band communication is proposed as a promising imperative wireless technology that has the capability of alleviating spectrum scarcity as well as capacity constraints of advanced highly anticipated applications in various industries and revolutionary fields [37].

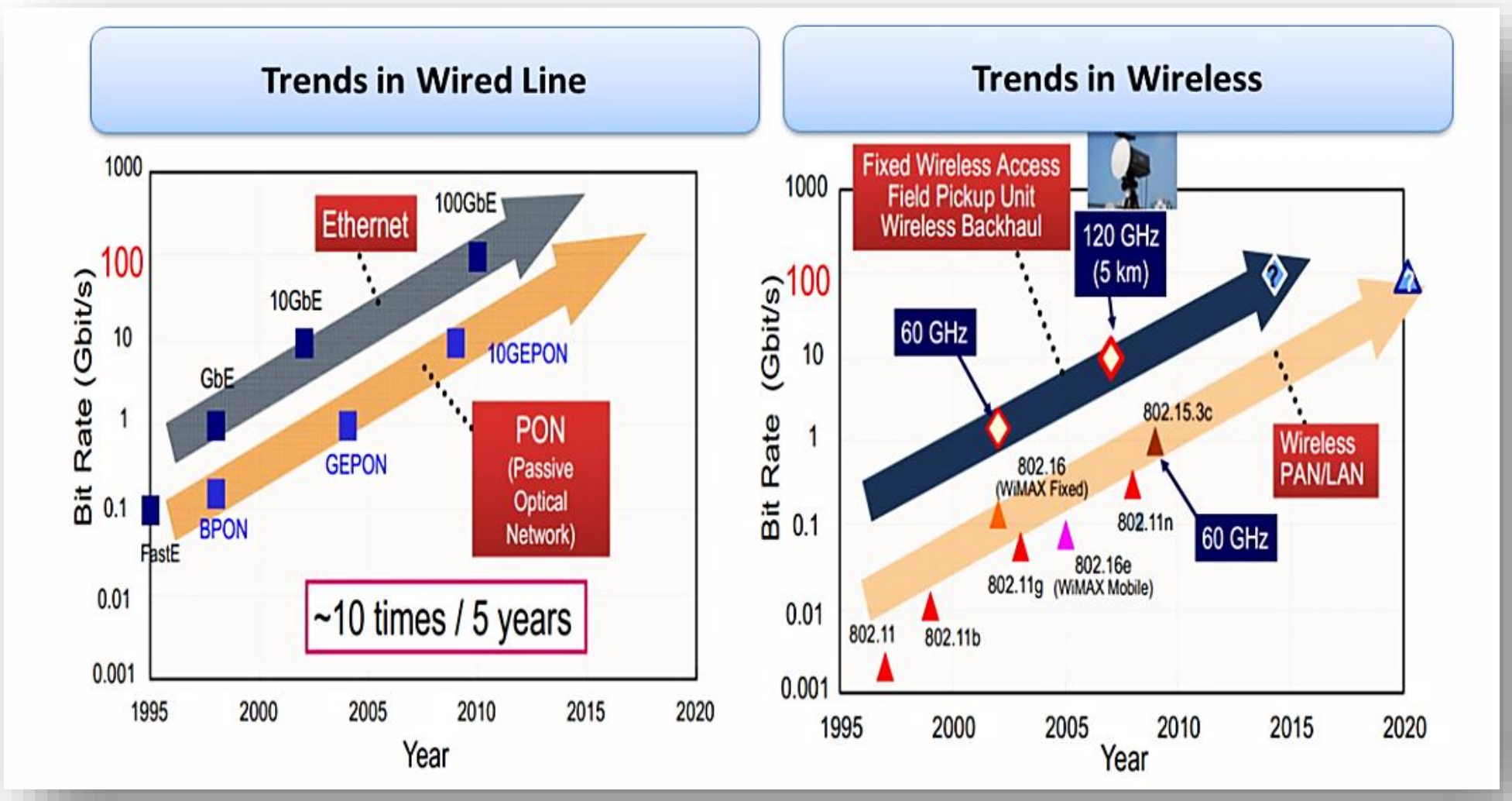

**Figure 4.1.1:** The convergence trends of both wireless data rates and wired line.

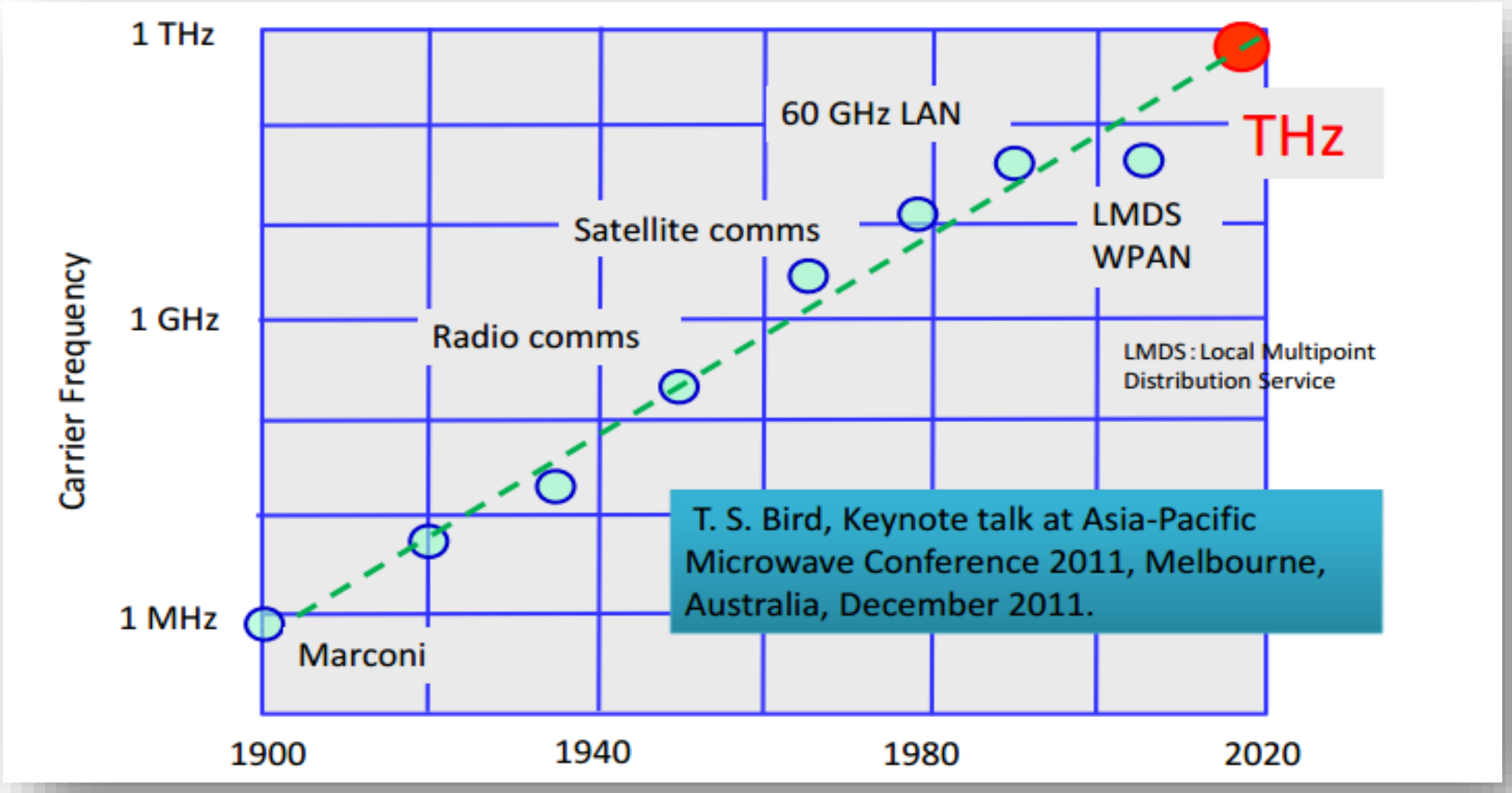

**Figure 4.1.2:** The convergence trends of Carrier frequency band.

## 4.2 Advantages of Utilizing Terahertz Communication Band for Intra-body Communication:

- ***Permits various integrated nano-devices utilized in intra-body communication of operating at the nano-scale level:***

Linking and incorporating nano-devices together led to establish new unprecedented concept which is the idea of nano-communications using nano-networks, which aimed at expanding the abilities of such devices in promotion and enhancement of future technologies. There are various techniques of communication utilized when operating nano-devices, the best applicable technique for data exchanging is utilizing wireless communication network (WCN) to operate using Electromagnetic waves in the Terahertz spectral band. By Terahertz communication bandwidth prior examined bands that lack the size reduction's characteristics unlike nano-scale technologies hence this makes such technologies an optimal desirable choice for upcoming applications of body-centric communication. Because of the short wavelength (λ), terahertz radiations, and due to the existence of molecular resonances at these frequencies, nano-devices can detect even minutes variations in water content and biomaterial body tissues. Therefore, one promising growing area in recent research is the analysis of terahertz electromagnetic propagation through human body tissues and layers. This is to come up with advanced diagnostic tools for early detection of some abnormalities that indicates to a serious condition such as skin cancer. Regardless of effective solutions being offered by data processing, the multimedia nano-things though have to send a considerable amount of data in a reliable fashion just in its proper time. Additional advantage of operating in terahertz band is that it supports a very high bit rate information transmission, can be as many as terabits per second (Tbps).

- ***Exploits the Non- ionizing Radiation characteristics of Terahertz communication band in intra-body communication:***

The substantial contribution of this underutilized spectral band to potential upcoming medical technologies is its advantages of providing radiation that is less susceptible to propagation effects like spreading loss and scattering effect, more reliable in terms of data quality and time delay and harmless on biological tissues since it is considered a non-ionization radiation. To elaborate, such radiation does not destruct the human tissues Unlike X-rays and many other radiations which are deployed in today medical field.

One of the main concerns of utilizing the terahertz radiation for intra-body communication is the thermal effect on human tissues caused by the absorption. However, many suggestions were proposed that this impact can be reduced by considering the terahertz exposure parameters. In particular, the exposure duration using an ON-OFF technique which means that the body we will be exposed to such radiation intermittently. Subsequently, ON-OFF keying modulation scheme were proposed since such technique is valid for short distances, in which molecular absorption does not drastically influence the transmission link.

## 4.3 Comparison between Terahertz and Photonics Utilized bands for Intra-body Communication technologies:

### A. Terahertz Transmitters

The feasibility of THz band communication networks, for many years, has been limited by the lack of methods efficient in generating THz band signals. The new advanced communication technologies' use has, however, lifted the stumbling blocks as regards to the limitations mentioned. Through the use of the advanced technologies, tremendous progress has been witnessed as regards to the use of nanoscale devices. For the devices to communicate, however, they need to be compact. That is, they should reach an area size of hundreds of square nanometers or micrometers at most. The aim is to ensure that they are fast for them to support gigahertz’s (GHz) modulation bandwidths. Additionally, they have to be the effective energy as well as tunable [38].

The several designs and performance features exhibited by electronic sources, as a result of the modern advanced technologies, imply that the sources could be feasible for biological research. An example is that the electronic sources can, at lower THz frequencies, provide high levels of average output power [39]. Their characteristics including the ability to generate narrow linewidth continuous THz radiation and operate at room temperature as well as being compact and ragged support the conclusions made above [40].

It is also important to note that, to enable multi-gigabit per second (Gbps) and Terabit per second (Tbps) links in the THz band, ultra-broadband antennas are needed as well. Consequently, the novel antennas' potential needs to be investigated as far as metamaterials and nanomaterials are concerned. Some past investigations have shown that, through the exploitation of the behavior of global oscillations of surface charges, graphene can be used to make plasmonic nano-antennas. Therefore, the Graphene-based nano-antennas can be, through material doping, tuned to be integrated with almost everything. The problem towards achieving the above depends on how to characterize as well as account for nearby antennas' coupling and interaction effects [41].

### B. Optical Transmitters

To generate frequencies covering around 400THz to 750THz of optical frequency window, photonic sources are needed. The sources must, in optical communications, possess particular features including being stable, compact, long lasting and monochromatic. In this context, the laser diodes and light-emitting diodes (LEDs) are the popular optical sources in use [42]. Scientists in many years have been exploring and developing many types of the mentioned sources thereby realizing significant success in enabling cost-effective optical transmissions. The solid, liquid and gas lasers are the various categories of the laser sources while those of LEDs include the broadband, ultraviolet, and infrared LEDs. Despite the progress, only the transmitters that are compact and can operate at room temperatures are used for biological purposes [43].

### C. Terahertz Receivers

There has been significant progress in the THz detector sensitivity for over 70 years with the noise equivalent power (NEP) value decreasing by a factor of $10^{11}$. The decrease corresponds with the improvement of a factor of two after every two years [44]. The detectors, to overcome the very high loss at THz frequencies, are needed to provide high detection sensitivity as well as exploit large bandwidth at THz band frequencies. The maximum received measurable power can be in the mW and µW range depending on detector type [45].

### D. Optical Receivers

The use of optical radiation detectors is essential to applications in the field of photonics. The thermal and photon detectors are the two primary types of the optical detectors. The thermal detectors enable the conversion of optical energy to heat energy, after which an electrical signal is generated. On the contrary, the photon detectors are able to create one electron for every incoming photon of optical energy where an electronic circuitry identifies the electron [46]. Avalanche photodiode (APD), the PIN photodiode, and quantum well and dot detectors, and Schottky photodiode are the various types of photon detectors. For many photonic applications within their spectral range, in practice, the silicon photodiodes have been the popular choice [47].

## *Chapter5:*

# *EM Channel Characterization accounting for the Path loss at In-Vivo Terahertz Nano-Communications in WBANs:*

## 5.1 Relationship of Optical Parameters to Electromagnetic Parameters:

In the optical frequencies, the data is usually referred in terms of the refractive index which is indicated by:

$$\tilde{n}(f) = n_r(f) - jn_i(f)$$

The real part which is $n_r$ is easily obtained by measurements while the imaginary one $n_i$ is found by calculation by knowing the absorption coefficient α.

$$n_i(f) = \frac{\alpha(f)\lambda_0}{4\pi}$$

Where $\lambda_0$ is the wavelength in free-space and it is derived from:

$$\lambda_0 = c/f$$

Knowing ˜n, the relative permittivity can be found by:

$$\epsilon(f) = \tilde{n}(f)^2$$

And accordingly,

$$\epsilon' = n_r(f)^2 - \kappa(f)^2$$

And,

$$\epsilon'' = 2n_r(f)\kappa(f)$$

Those values correspond to the real and imaginary part of the relative permittivity.

## 5.2 Electromagnetic Modeling of the Human Body:

In order to examine the in vivo wireless communication channel, it is essential to have an accurate body models as well as knowing the EM properties of the tissues. For that purpose, human autopsy materials in addition to animal tissues were measured over the range of frequencies between 35 GHz to 1 THz. The properties of tissues that are frequency dependent dielectric were modeled using the double Debye theory.

## 5.3 Path-loss :

There are three main frequency dependent parameters that contribute to the total path loss at both the Terahertz and the optical frequencies. Those parameters are: the spreading loss $L_{spr}(f)$, the molecular absorption loss $L_{abs}(f)$ and the scattering loss $L_{sca}(f)$. Whereas the total attenuation factor is found by:

$$L_{tot}(f) = L_{spr}(f) \times L_{abs}(f) \times L_{sca}(f)$$

In this research paper, the frequency range of 0.1-10 THz and 400-750 THz or equivalently their corresponding wavelengths will be analyzed. The wavelength ranges are: 30µm-3mm and 400nm-750nm. The value of the overall path loss for an EM wave that is in the frequency band of Terahertz in dB is found by:

$$PL_{[dB]} = PL_{Spread} + PL_{Absorption} + PL_{scattering}$$

### 5.3.1 Intra-body Path Loss Due to Wave Spreading in Human Tissue:

The spreading path loss is caused by the propagation of that wave in the medium while the absorption path loss is due to absorption of the medium to the wave. Those two different values of the path loss can be obtained by:

$$PL_{Spread} = -10 \log_{10} (\lambda_g/4\pi d)^2$$

$\lambda_g$ is the wavelength in the medium, and it is given by the ratio of free space wavelength $\lambda_0$ to the real part of refractive index n, this ratio is indicated by:

$$\lambda_g = \lambda_0/n'$$

Where n is; n = n' – jn''. It is good to know that $n=\sqrt{\epsilon_r \mu_r}$, $\epsilon_r$ is the relative permittivity and μr is the relative permeability that is considered to be equal to 1since the biological tissues are nonmagnetic, hence, we have $\epsilon_r = \epsilon'_r + \epsilon''_r = n2$, and $\epsilon'_r = n'^2 - n''^2$. The term directivity D, is the maximum gain of the nano-antenna, and it is given as:

$$D = \frac{P(\theta, \phi)_{max}}{P(\theta, \phi)_{av}}$$

Where P(θ, φ)max is the maximum power density in W/m$^2$ and P(θ, φ)av is the average value over a sphere and it can be finalized according to [39] as:

$$D = \frac{4\pi}{\iint_{4\pi} P_n(\theta, \phi) d\Omega} = \frac{4\pi}{\Omega_A}$$

Where Pn(θ, φ)dΩ = P(θ, φ)/P(θ, φ)max is called the normalized power pattern whereas $\Omega_A$ refers to the radiation solid angle that depends on the particular radiation pattern of the source and antenna that is used. For instance, for a narrow beam directional source that has a width Δθ, $\Omega_A$ is given as:

$$\Omega_A = \int_{\phi=0}^{2\pi} \int_{\theta=0}^{\Delta\theta} \sin\theta d\theta d\phi = 2\pi(1 - \cos\Delta\theta)$$

In case of THz and optical sources, it is good to consider a light source that has a Gaussian beam with a radiation pattern given by [40]:

$$E_\theta = \frac{1 + \cos\theta}{2}$$

The radiated power, P is proportional to Eθ2, the solid angle, ΩA, of a Gaussian beam of width Δθ is given as:

$$\begin{aligned} \Omega_A &= \int_{\phi=0}^{2\pi} \int_{\theta=0}^{\Delta\theta} \frac{1}{4}(1 + 2\cos\theta + \cos^2\theta)\sin\theta d\theta d\phi \\ &= \frac{\pi}{2}\left[\frac{8}{3} - (\cos\Delta\theta + \cos^2\Delta\theta + \frac{1}{3}\cos^3\Delta\theta)\right]. \end{aligned}$$

It is important to note that one of the main limitations of the Terahertz band is the large amount of the spreading loss. As a result, this would remarkably reduce the transmission range of the future Nano-devices. Even though it is considered as a major problem for the current applications in the Terahertz communications, this band suits the Nano and micro scales of communication where the transmission distance considerably very small, and it is about several millimeters.

### 5.3.2 Intra-body Path Loss Due to Molecular Absorption by Human Tissue:

The EM waves at particular frequencies within the Terahertz can excite the molecules of any medium. The internal movement of the atoms is vibration which reflects a periodic motion. While the whole molecule appears as it is in translation and rotation motions. Both the Terahertz and the optical waves are non-ionizing and inducing the vibrational motion. But at the same time, they will not break the molecules. This vibration will cause some of the propagating wave energy to be converted into a kinetic one. This amount of converted energy is referred to in communication as a loss. The molecular absorption is found by calculating the fraction of the EM radiation that is incident and have the ability to pass through the medium at specific frequency. With the aid of Beer-Lambert law [41], the amount of attenuation caused by the molecular absorption for an EM wave that travels at a distance, d, is given by:

$$L_{abs} = e^{-\mu_{abs} d}$$

Where $\mu_{abs}$ is called the molecular absorption coefficient and it depends on the medium composition. This coefficient can be calculated using the following formula:

$$\mu_{abs} = 4\pi (n'')/ \lambda$$

The imaginary term n'' contributes to the absorption and it is zero when the medium is non-conducting; ($\sigma = 0$) [44]. There are two different methodologies to find out the absorption coefficient. The first one is through the individual particles. The radiation absorption efficiency of a particle can be modeled using the absorption efficiency.

$$Q_{abs} = \sigma_{abs}/\sigma_g$$

Where $\sigma_{abs}$ is the molecular absorption and $\sigma_g = \pi r2$ is the geometric cross section. Hence, the absorption coefficient $\mu_{abs}$ can be derived from:

$$\mu_{abs} = \rho_v Q_{abs} \sigma_g$$

Where $\rho_v$ is the particle concentration and it is given by $\rho_v = \kappa/( 4/3\pi r^3)$ while $\kappa$ is the fraction volume of that particle. For the second method, since we are dealing with huge amount of molecules, it is good to consider the effective medium assumption. In case of pure materials, it is possible to consider the multiple Debye processes. This theory was used to model the Terahertz spectroscopy figures of the polar molecules, electrolytes and fully hydrated nucleotides.

The Debye parameter was deployed for analyzing the interaction of Terahertz radiation with the human tissues. These results showed that the tissue could be demonstrated as a semi-infinite homogeneous medium that has a dielectric constant that is similar to that of the liquid water. The complex permittivity is explained by [46] and it is given as:

$$\epsilon = \epsilon_\infty + \sum_{j=1}^{n} \frac{\Delta\epsilon}{1 + jw\tau_j}$$

$\epsilon_\infty$ is the permittivity at the high frequency limit and $\Delta\epsilon = \epsilon_j - \epsilon_{j+1}$.

In order to properly estimate the complex permittivity for polar liquids at frequencies up to 1 THz, the double Debye equations are used [45].

$$\epsilon(w) = \epsilon_\infty + \frac{\epsilon_s - \epsilon_2}{1 + iw\tau_1} + \frac{\epsilon_2 - \epsilon_\infty}{1 + iw\tau_2}$$

Where $\mathcal{E}_\infty$ is the limiting permittivity value at high frequency, $\mathcal{E}i$ is the epsilon value at middle frequency step, $\tau i$ represents the mechanical relaxation time at that frequency step, and $\omega$ is the angular frequency. Both the real and imaginary parts of the complex permittivity are separated as illustrated below:

$$\epsilon'(w) = \epsilon_\infty + \frac{\epsilon_s - \epsilon_2}{1 + (w\tau_1)^2} + \frac{\epsilon_2 - \epsilon_\infty}{1 + (w\tau_2)^2}$$

$$\epsilon''(w) = \frac{(\epsilon_s - \epsilon_2)(w\tau_1)}{1 + (w\tau_1)^2} + \frac{(\epsilon_2 - \epsilon_\infty)(w\tau_2)}{1 + (w\tau_2)^2}$$

Those double Debye relaxation coefficients are illustrated in the table below.

**Table 5.1:** Permittivity and relaxation time values

| Model | $\epsilon_\infty$ | $\epsilon_1$ | $\epsilon_2$ | $\tau_1$(ps) | $\tau_2$(ps) |
|---|---|---|---|---|---|
| Water | 3.3 | 78.8 | 4.5 | 8.4 | 0.1 |
| Whole Blood | 2.1 | 130 | 3.8 | 14.4 | 0.1 |
| Skin | 3.0 | 60.0 | 3.6 | 10.6 | 0.2 |

Approaching the near-infrared frequency, especially toward the optical window, the table below demonstrates the wavelengths and their corresponding permittivity.

**Table 5.2:** Relative permittivity vs. wavelength [37] [39]

| $\lambda(nm)$ | Fat | Hemoglobin | Water |
|---|---|---|---|
| 450 | 2.13-$j$6.68$\times10^{-7}$ | 2.04-$j$3.46$\times10^{-3}$ | 1.78-$j$2.72$\times10^{-9}$ |
| 500 | 2.13-$j$2.20$\times10^{-7}$ | 2.03-$j$1.26$\times10^{-3}$ | 1.78-$j$2.68$\times10^{-9}$ |
| 550 | 2.13-$j$9.89$\times10^{-8}$ | 2.01-$j$2.86$\times10^{-3}$ | 1.77-$j$5.35$\times10^{-9}$ |
| 600 | 2.13-$j$6.47$\times10^{-8}$ | 1.99-$j$2.50$\times10^{-4}$ | 1.77-$j$2.91$\times10^{-8}$ |
| 650 | 2.13-$j$7.12$\times10^{-8}$ | 1.99-$j$2.87$\times10^{-5}$ | 1.77-$j$4.36$\times10^{-8}$ |
| 700 | 2.13-$j$5.26$\times10^{-8}$ | 1.99-$j$2.43$\times10^{-5}$ | 1.77-$j$9.22$\times10^{-8}$ |
| 750 | 2.13-$j$1.70$\times10^{-7}$ | 1.99-$j$4.68$\times10^{-5}$ | 1.76-$j$4.14$\times10^{-7}$ |
| 800 | 2.13-$j$7.45$\times10^{-8}$ | 1.99-$j$7.83$\times10^{-5}$ | 1.76-$j$3.35$\times10^{-7}$ |
| 850 | 2.13-$j$1.26$\times10^{-7}$ | 1.99-$j$1.08$\times10^{-4}$ | 1.76-$j$7.81$\times10^{-7}$ |
| 900 | 2.13-$j$9.66$\times10^{-7}$ | 1.99-$j$1.29$\times10^{-4}$ | 1.76-$j$1.33$\times10^{-6}$ |
| 950 | 2.13-$j$8.69$\times10^{-7}$ | 1.99-$j$1.37$\times10^{-4}$ | 1.76-$j$7.79$\times10^{-6}$ |
| 1000 | 2.13-$j$6.17$\times10^{-7}$ | 1.99-$j$1.23$\times10^{-4}$ | 1.76-$j$7.67$\times10^{-6}$ |

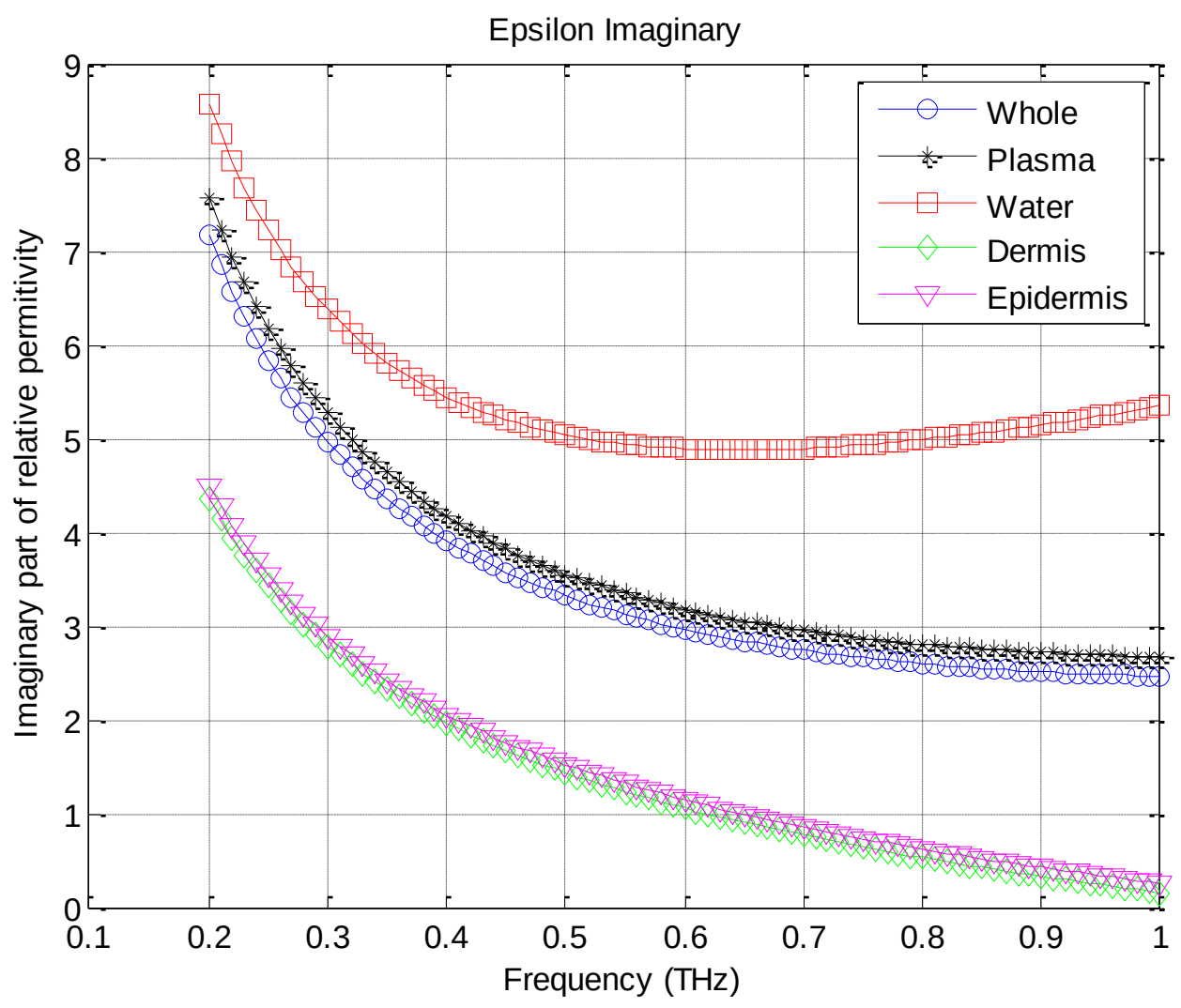

**Figure 5.1:** Imaginary part of the relative permittivity.

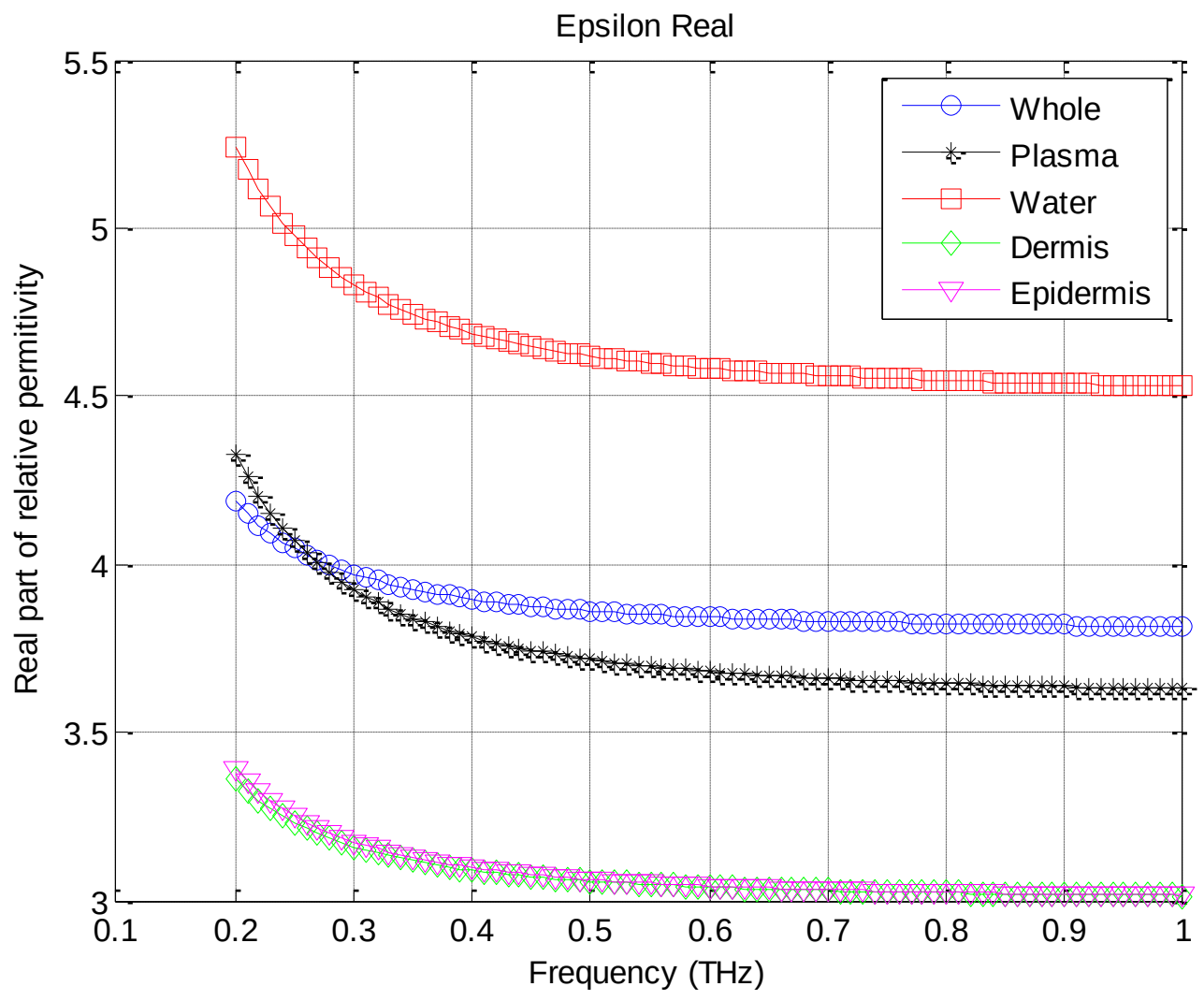

**Figure 5.2:** Real part of the relative permittivity.

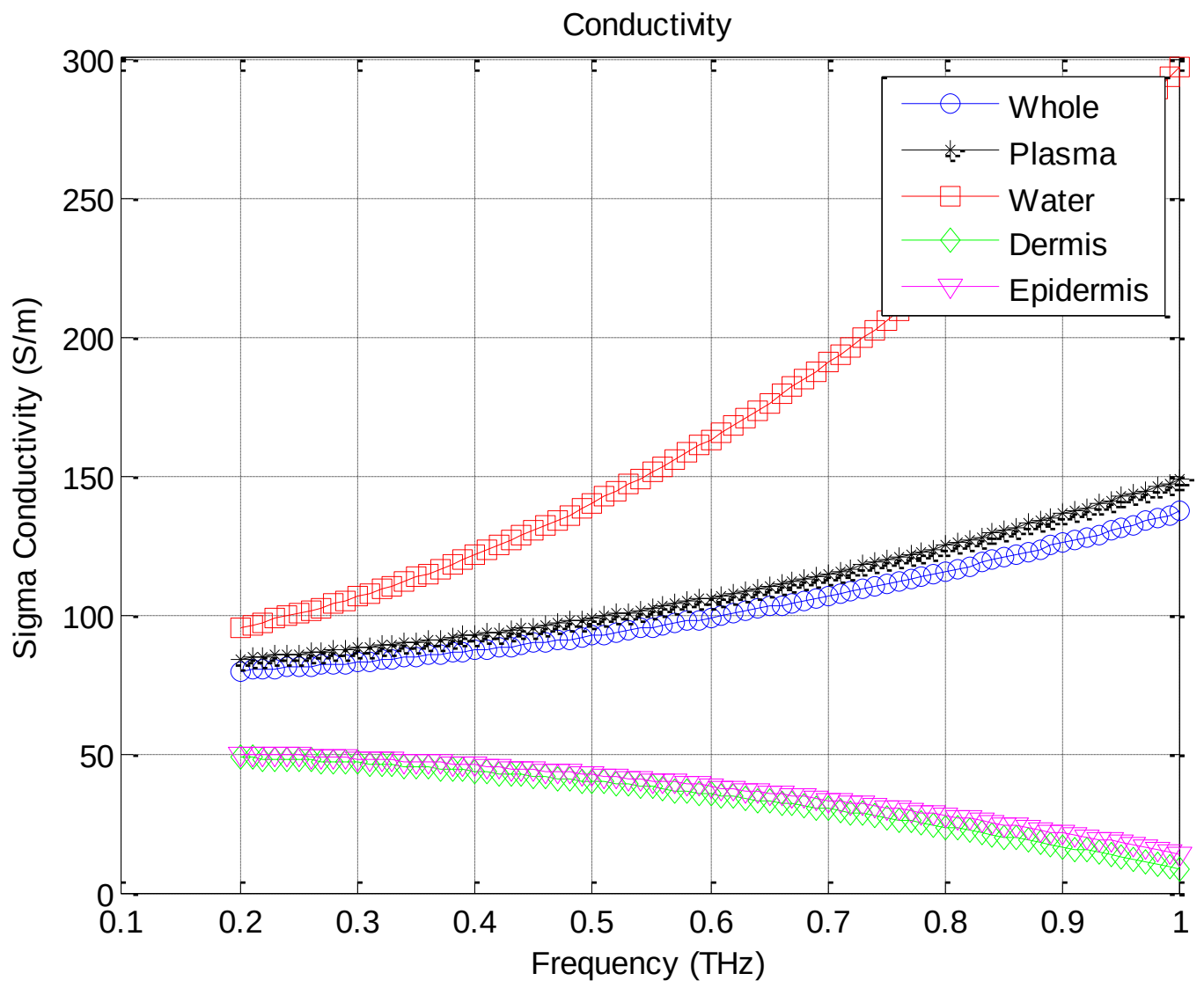

**Figure 5.3:** Conductivity of each constituent substance.

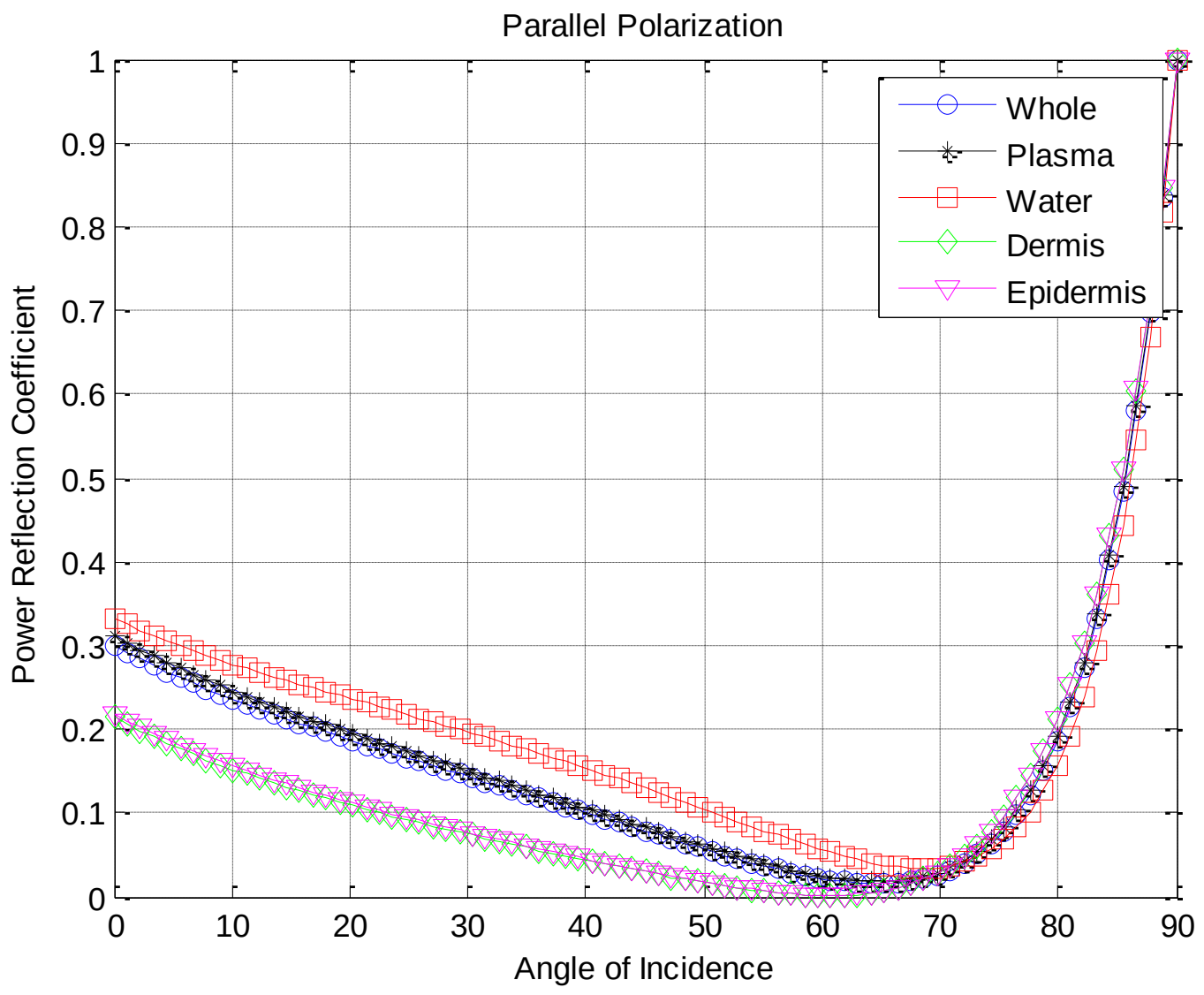

**Figure 5.4:** Parallel Polarization of each constituent substance.

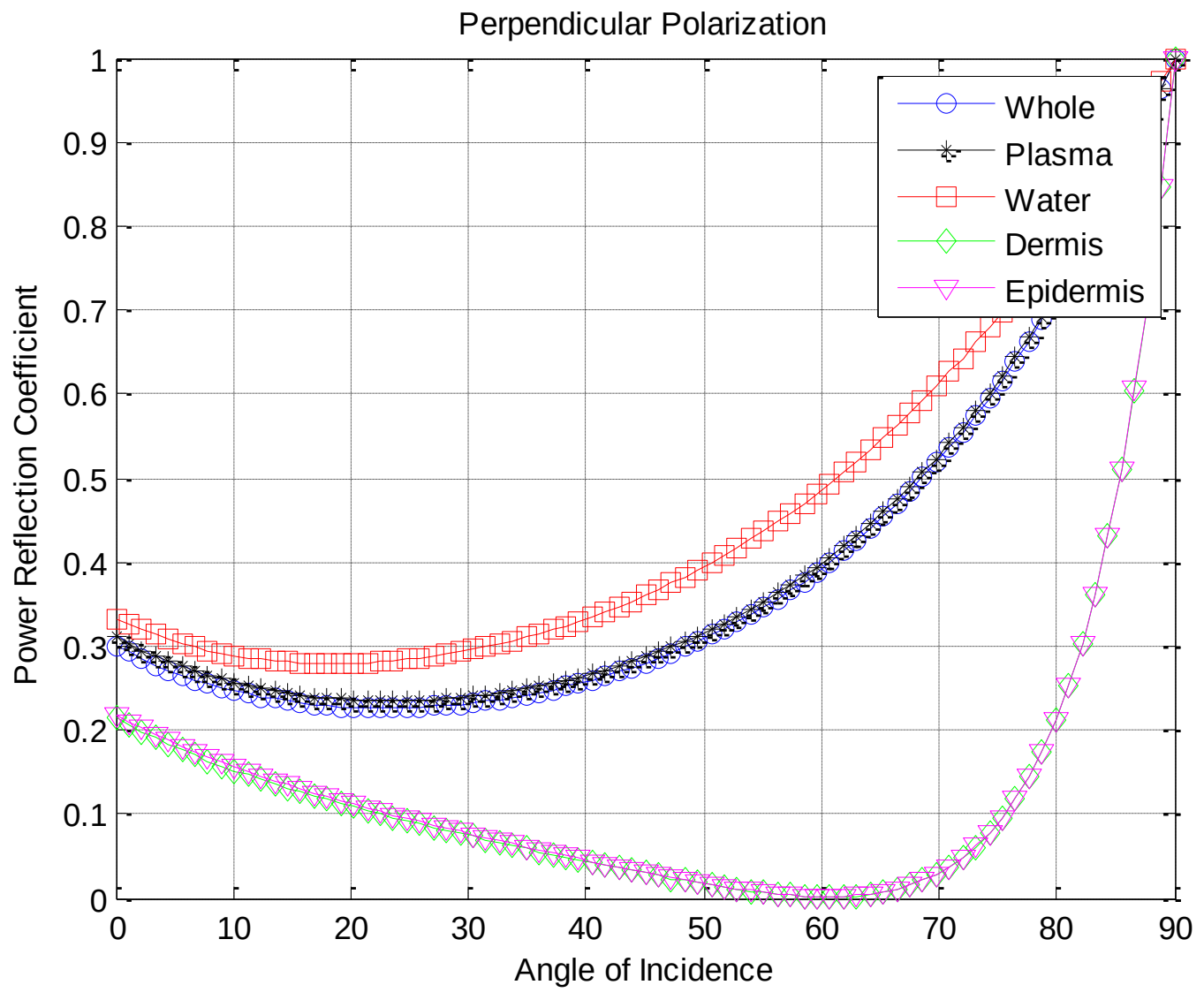

**Figure 5.5:** Perpendicular Polarization of each constituent substance.

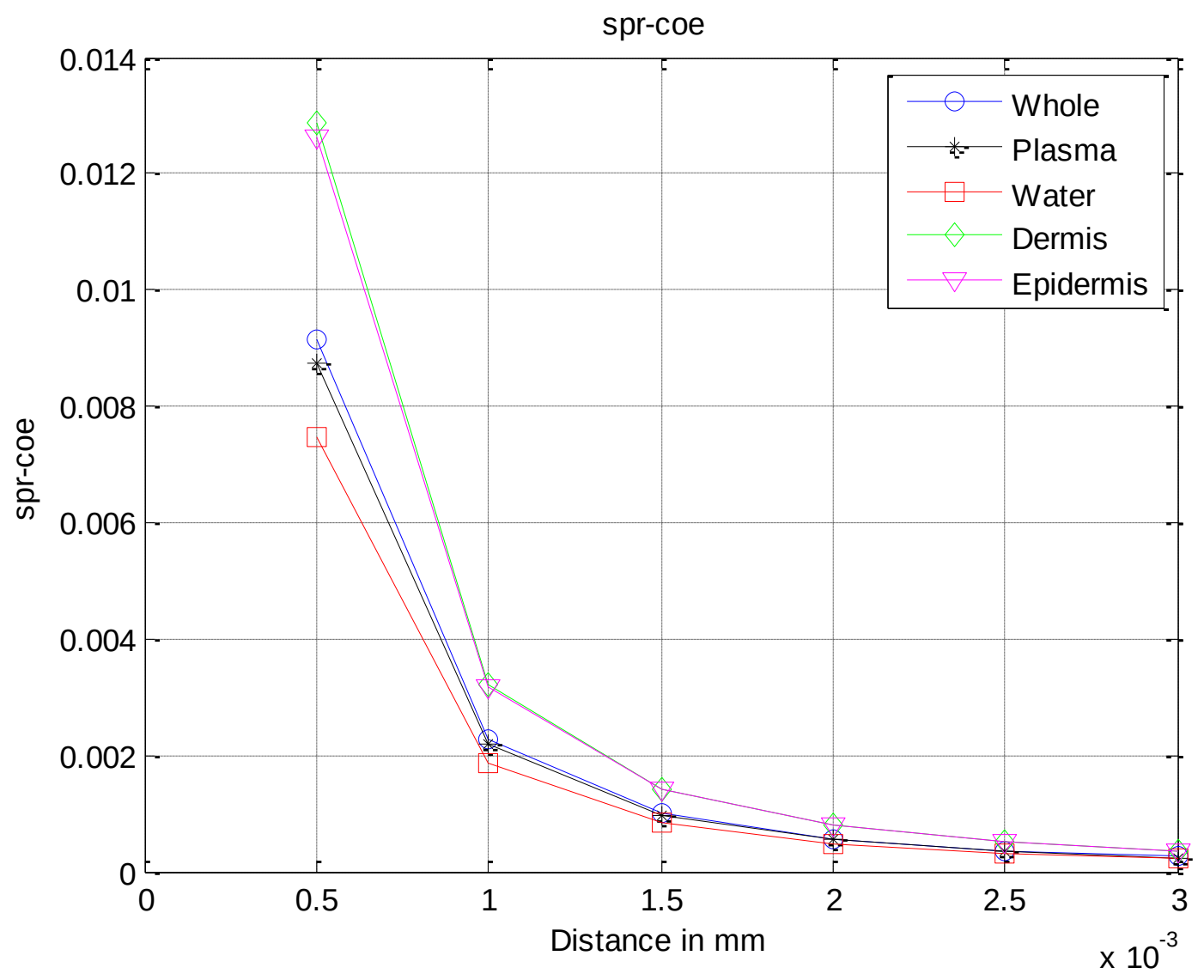

**Figure 5.6:** Spreading coefficient of each constituent substance.

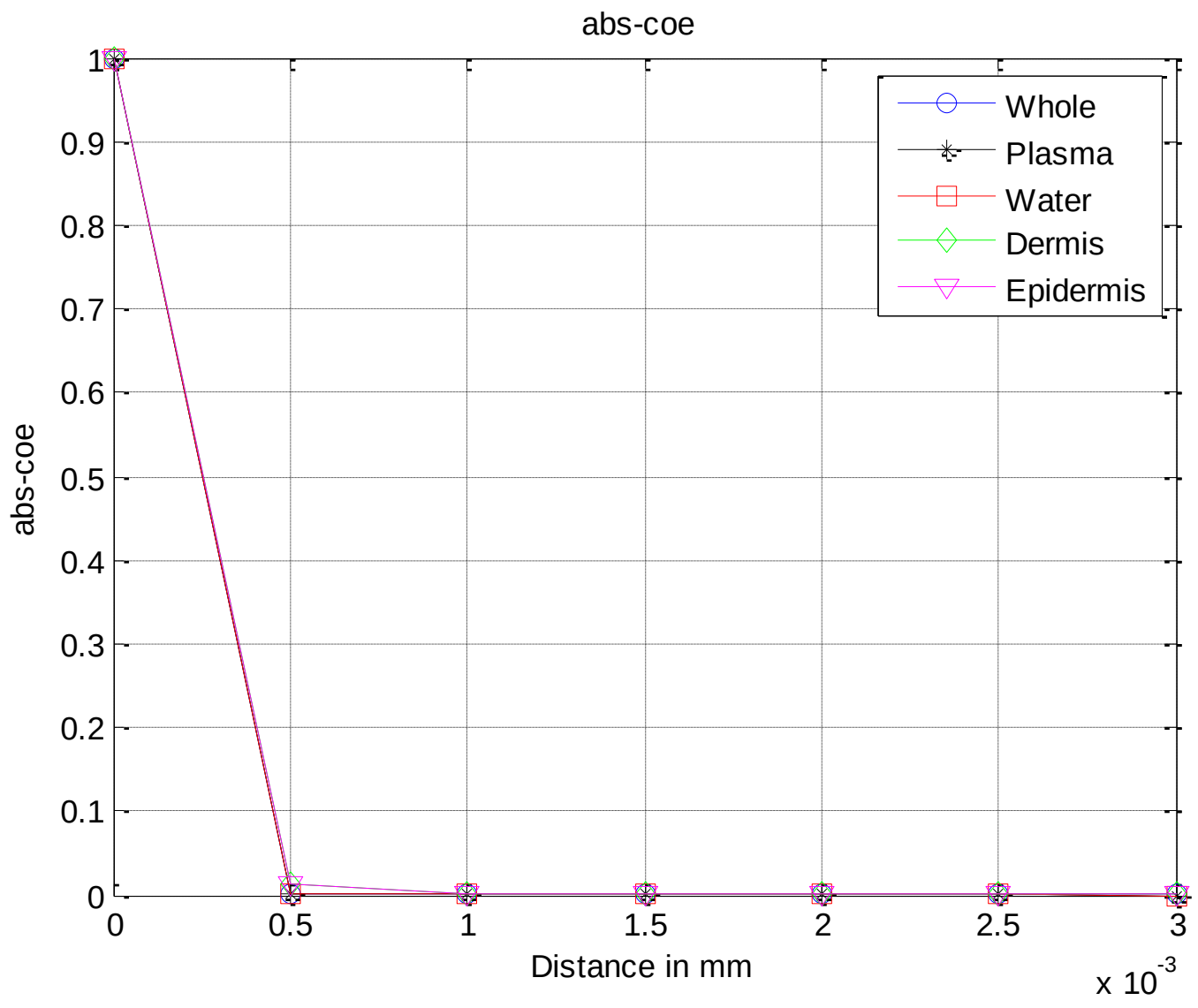

**Figure 5.7:** Absorption coefficient of each constituent substance.

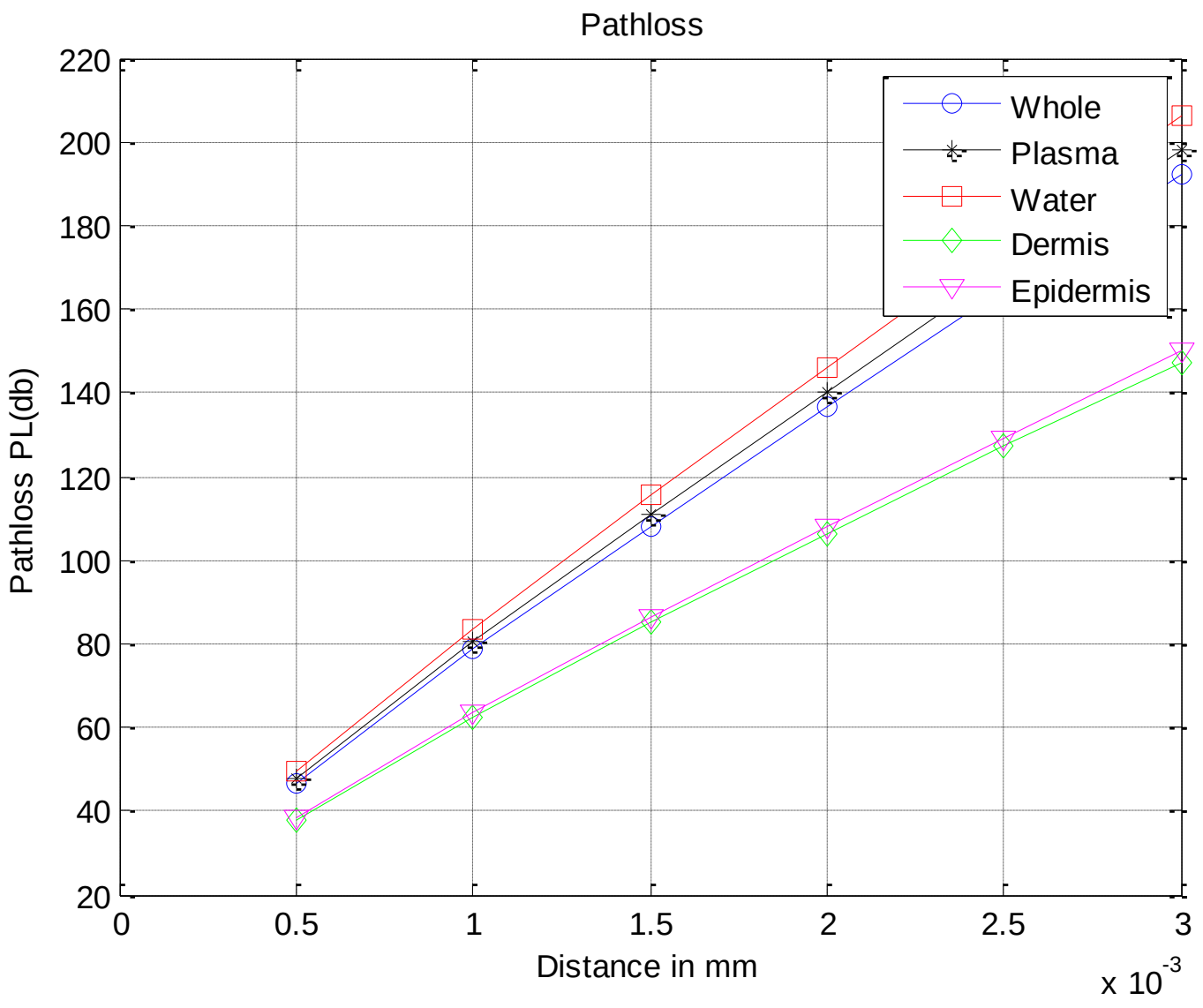

**Figure 5.8:** Path loss without considering the scattering loss of each constituent substance.

### 5.3.3 Intra-body Path Loss Due to Scattering by Human Tissue:

By comprehensively addressing the body physiology and the nanosensor perception, human body can be examined as a combination of various types of aggregated particles, such as cells, neurotransmitters, antibodies and molecules with distinguish geometrical analytic structures and constellations along with diverse electromagnetic characteristics. As previously defined, Scattering loss accounts for the attenuation due to the signal deflection from its directed trajectory to other routes. Therefore, it is indeed compulsory to emphasize the significant impact of Scattering due to those constituent particles and their non-uniformities characteristics on the signal transmission. Furthermore, the imperative of investigating the different dependent variables such as the size, shape, refractive index of each distinguished particle, the wavelength relevant to the incident beam, and the their implications on the wave propagation should not be neglected [46]. Both Rayleigh and Mie introduced theoretical models that expansively explain the scattering phenomenon affecting minor spherical obstacles. Rayleigh scattering model applied if the wavelength of the propagated EM wave is greater than the substance diameter. In contrast, Mie scattering model applied if the wavelength of the propagated EM wave is almost equivalent to the substance diameter [47]. However, if the constituent substance diameters are greater than the wavelength then reflection or geometrical scattering laws are applied [46]. The impact of scattering is examined by utilizing the geometry of Figure below, which illustrates a plane wave incident on a scattered particle located at the origin of a spherical coordinate system (r, θ, φ) [48].

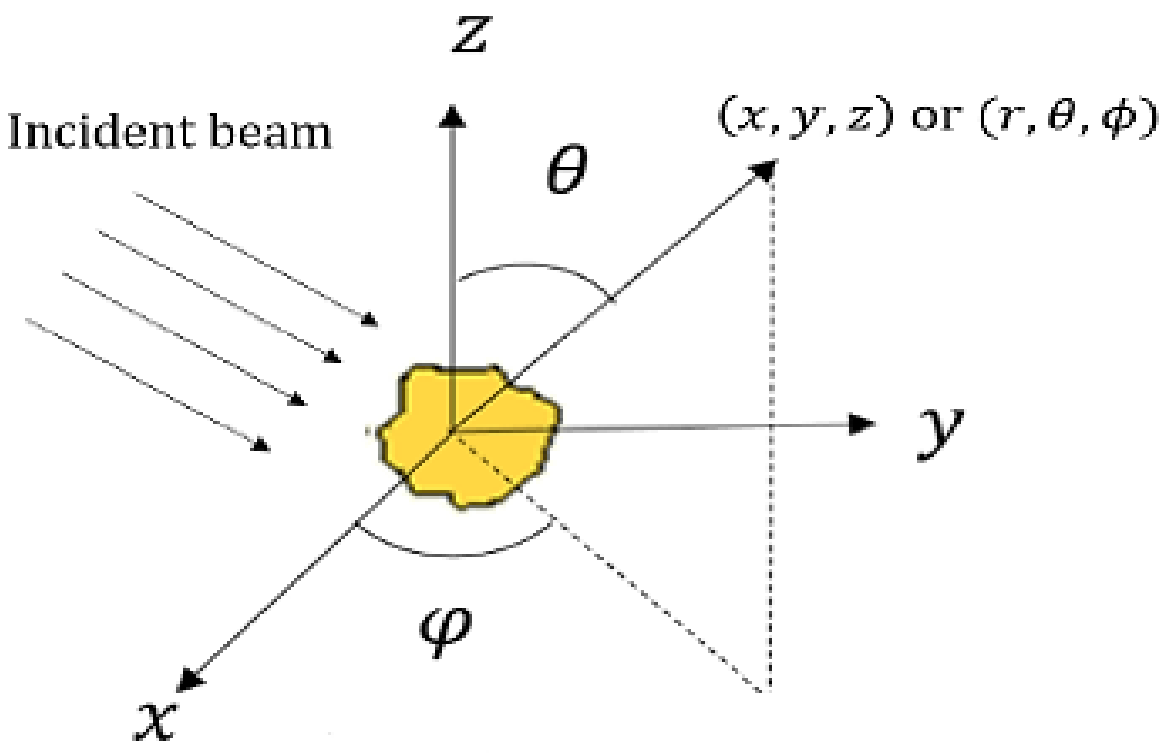

**Figure. 5.9:** Geometry of a scattering scenario showing the coordinates of both Cartesian and spherical coordinate systems.

The fundamental significant representative of a scattered wave is its intensity, $I_{sca}$, presented as [49]

$$I_{sca} = \frac{1}{(kr)^2} I_{inc}\, S(\theta, \Phi),$$

Where $k = 2\pi/\lambda$ is defined as the signal incident radiation factor,

$I_{inc}$ is the incident intensity, and $S(\theta, \Phi)$ is the representative equation of the scattered particle amplitude. Generally, $S(\theta, \Phi)$ is significantly dependent on various key parameters such as wavelength of the incident beam, the dimension, and its characterized form, in addition to the optical characteristics of the particle [49].

Besides the intensity equation, both the scattering cross-section and the scattering efficient performance are compulsory required to describe the scattering loss. The scattering cross-section, $\sigma_{\mathrm{sca}}$, is determined as the relation associating the power scattered due to the constituent substance and the incident power per unit area, and is expressed as

$$\sigma_{sca} = \frac{1}{(k)^2} \int_0^{2\pi} \int_0^{\pi} |\, S(\theta, \Phi)|^2 \sin\theta \, d\theta \, d\Phi,$$

Where the equation of the scattered particle amplitude, $S(\theta, \Phi)$, of a spherical object, that imitates the constituent particle, is described as

$$S(\theta, \Phi) = \frac{k^2}{4\pi} \int e^{-ik\xi \, \sin\theta(\xi \cos\Phi + \eta \sin\Phi)} \, (1 + \cos\theta)\, d\xi \, d\eta \,,$$

In which is defined as the magnitude of the planar aperture. Due to the fact that both a spherical object and an opaque disk have an analogous deflection configuration, the amplitude of a scattered spherical object is treated as if it is independent of the azimuthal angle $\Phi$ for the purpose of alleviating the computation complexity.

$$S(\theta) = \frac{k^2}{4\pi} \int e^{-ik\xi \, \sin\theta} \, (1 + \cos\theta)\, d\xi \, d\eta \,,$$

The integral assessment is demonstrated in [27].

Equivalently to absorption, $Q_{sca}$, an indicator that represents the scattering efficient performance, defined as the ratio relating the energy scattered by the constituent substance to the whole amount of energy concerted in the incident beam diverted by the geometrical cross-sectional structure of the substance and is expressed as

$$Q_{sca} = \sigma_{sca}/\sigma_g$$

Since these factors significantly dependent on the size of the substance, the importance of considering the scattering resulted due to both small substances and considerably large interior cells must be comprehensively addressed in order to come up with our optimum paradigm.

1) ***Scattering due to small substances:*** When a substance is much smaller compared to the wavelength, the wave produces a confined electric field that is almost uniform at any time interval. This electric field forms a dipole beam pattern surrounding the substance. The electric field alternations and the induced dipole alternations implying that the accelerated charges will emit radiations; and by referring to classical theory, the dipole will emit radiation in all directions. This scattering model is known as Rayleigh scattering [25].

The scattering efficiency of minor spherical absorbing substance is described as [48]

$$Q_{sca}^{small} = \frac{8}{3}\,\psi^4\ \mathbf{Re}\left(\frac{n^2-1}{n^2+2}\right)^2,$$

where $\psi = 2\pi r/\lambda_g$ is the dimensionless size factor of the substance. Subsequently, utilizing the same previously discussed approach, the obtained scattering coefficient for small particles is represented as

$$\mu_{sca}^{small} = \rho_v\ Q_{sca}^{small}\ \sigma_g,$$

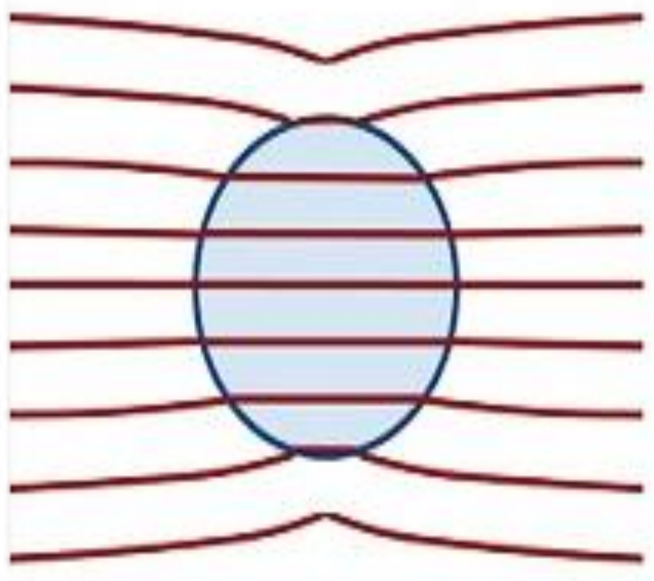

**Figure 5.10:** A uniform electric field in side particle.

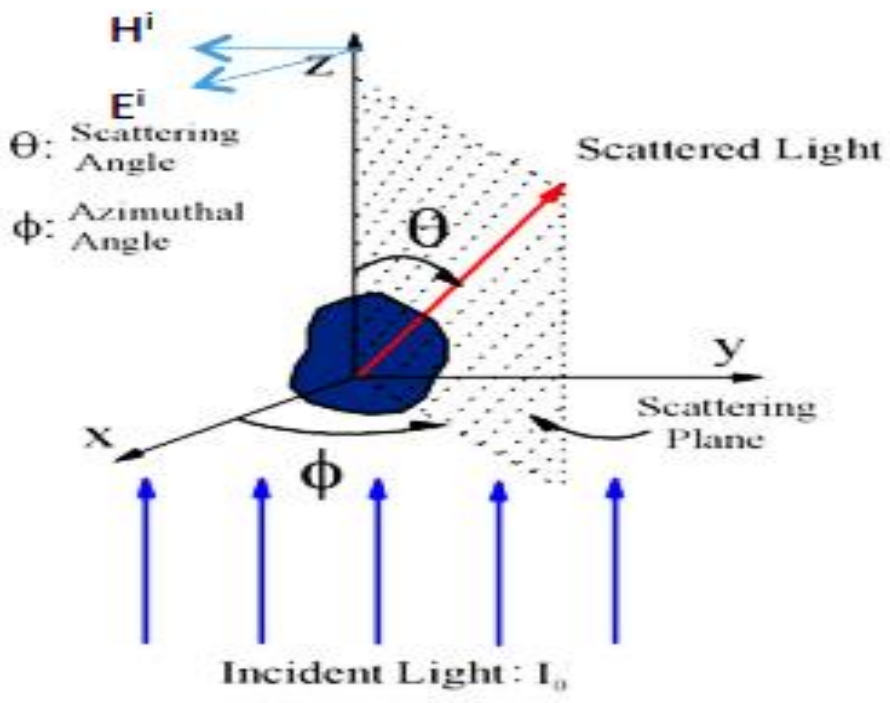

**Figure 5.11:** Coordinate geometry for the Rayleigh and Mie scattering.

2) ***<u>Scattering due to large interior Cells:</u>*** Scattering due to considerably large substance can be evaluated by utilizing van de Hulst approximation, which is known as well as the anomalous diffraction approximation [48]. Certainly, the whole amount of energy confined in the incident beam is extracted, the vanishing energy, is represented as the summation of scattering energy along with absorbing energy. Correspondingly, the evanescence efficiency is described as [48].

$$Q_{ext} = 2 - \frac{4}{p}\sin p + \frac{4}{p^2}(1 - \cos p),$$

Hence,

$$Q_{sca}^{large} = Q_{ext} - Q_{abs},$$

where $p = 4\pi r(n - 1/\lambda) = 2(n-1)\psi$ characterizes the phase offset of the EM signal crossing the central point of the substance. The fully comprehensive derivation of the equation above is presented in [23]. As an example, an adequate model of scattering due to several constituents inside the human blood can be demonstrated. Figure 5.4, the blood encompasses various integrated constituents.

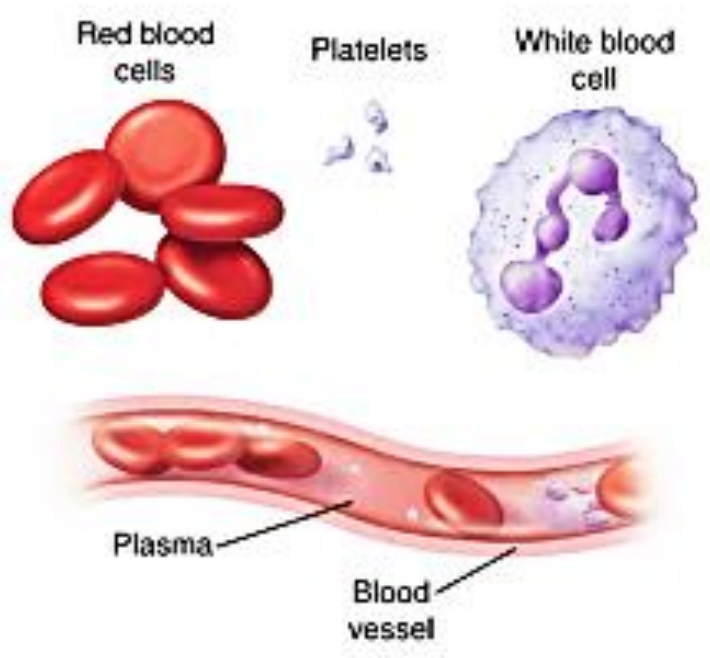

**Figure 5.12:** Blood Components

Blood plasma is the liquefied constituent of the blood and is primarily a combination of water (up to 95% by proportions) and delicate small substances of incorporated protein, glucose, etc. In addition, It consists of various kinds of blood cells which are considered as the most significant constituent substances of the blood, specifically, platelets (2 microns in diameter), red blood cells (7 microns), and white blood cell (up to 20 microns).

By substituting the equations, the scattering due to large substances can be represented as

$$\mu_{sca}^{large} = \rho_v \, Q_{sca}^{large} \sigma_g$$

By referring to light theory, a relatively large substance extracted from the beam is double the amount of the light diverted by its geometrical cross-sectional structure.

If a light interacts with a outsized substance, then beam is considered to entail a set of distinct rays. Certain rays when encountering the geometrical cross-sectional structure of a spherical object will be reflected at the substances interface whereas others deflected. The total energy incident on the substance interface is extracted from the beam due to scattering or absorption, By taking into account its efficiency parameter. There is, however, alternative cause of scattering due to the incident beam.

The intensity scattered inside the diffraction model is significantly subjected to various parameters, for instance the perimeter shape and the substance dimension as well as the light wavelength. The overall energy that occurs in the diffraction model is equivalent to the energy confined within the beam that is intercepted by the geometrical cross-sectional structure of the constituent substance. Consequently, the total efficiency factor related to the cross-sectional region obtained is approximately equal to 2 [24]. This result will be verified and broadly analyzed in the analytical outcomes segment.

Ultimately, diminution resulted from a scattered substance is represented by adding and in cooperating the scattering coefficient of large and small constituent substances.

As a result,

$$L_{sca} = e^{-(\mu_{sca}^{small} + \mu_{sca}^{large})d},$$

where d is defined as the distance of a propagated signal.

## 5.4 Numerical results:

This part will discuss the analytical models for different losses including spreading, absorption and scattering. For this purpose, the three main body constituents will be analyzed. Those constituents are water, blood, skin and fat.

### A. Molecular Absorption

The figure below demonstrates different molecular absorption coefficients referred as $\mu_{abs}$ that is provided in the above equation for various human tissues at the Terahertz range of frequencies. It is clear from the figure that the molecular absorption has more impact on the blood than the rest of tissues. This note is predicted since the blood is composed of a higher amount of water compared to the rest. The reason behind having higher absorption in the Terahertz band is due to the rotation move of water in this band.

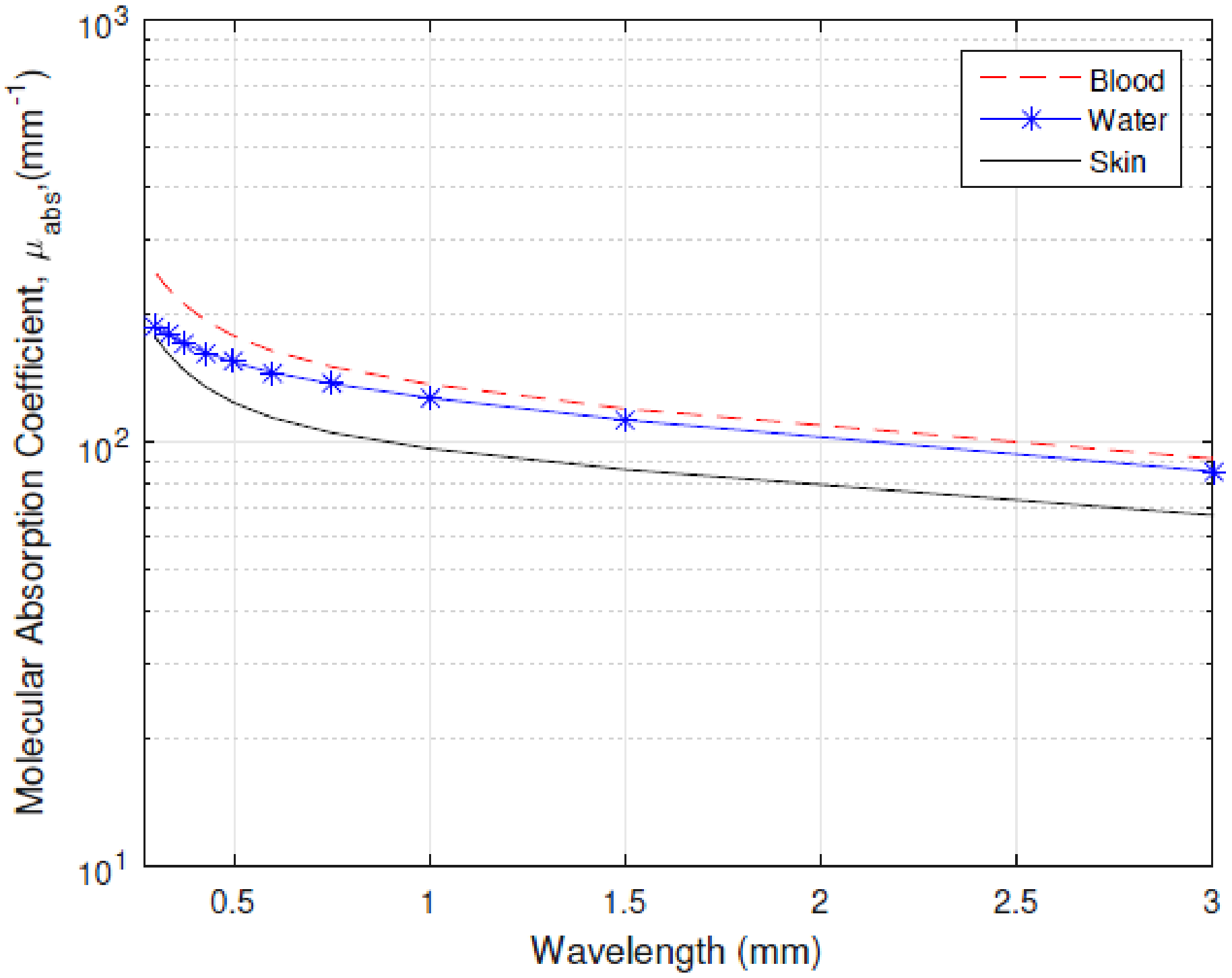

**Figure 5.13:** Molecular absorption coefficient, $\mu_{abs}$, for different human tissues vs. wavelength at THz ( $\lambda$ = 300 µm to 3 mm).

On the other hand, the figure below illustrates the reduction in the molecular absorption for the blood tissue by the amount of order of magnitude at the optical frequencies range between 400 and 750 THz, compared to its value at the low band of Terahertz. Another note can be observed is that, as the frequency approaches the optical band, the propagation wave wavelength gets similar to the cell size. Hence, it is concluded that molecular absorption in the blood is higher than in water by four orders of magnitude. Due to that, it causes the cells to vibrate which results in heat generation.

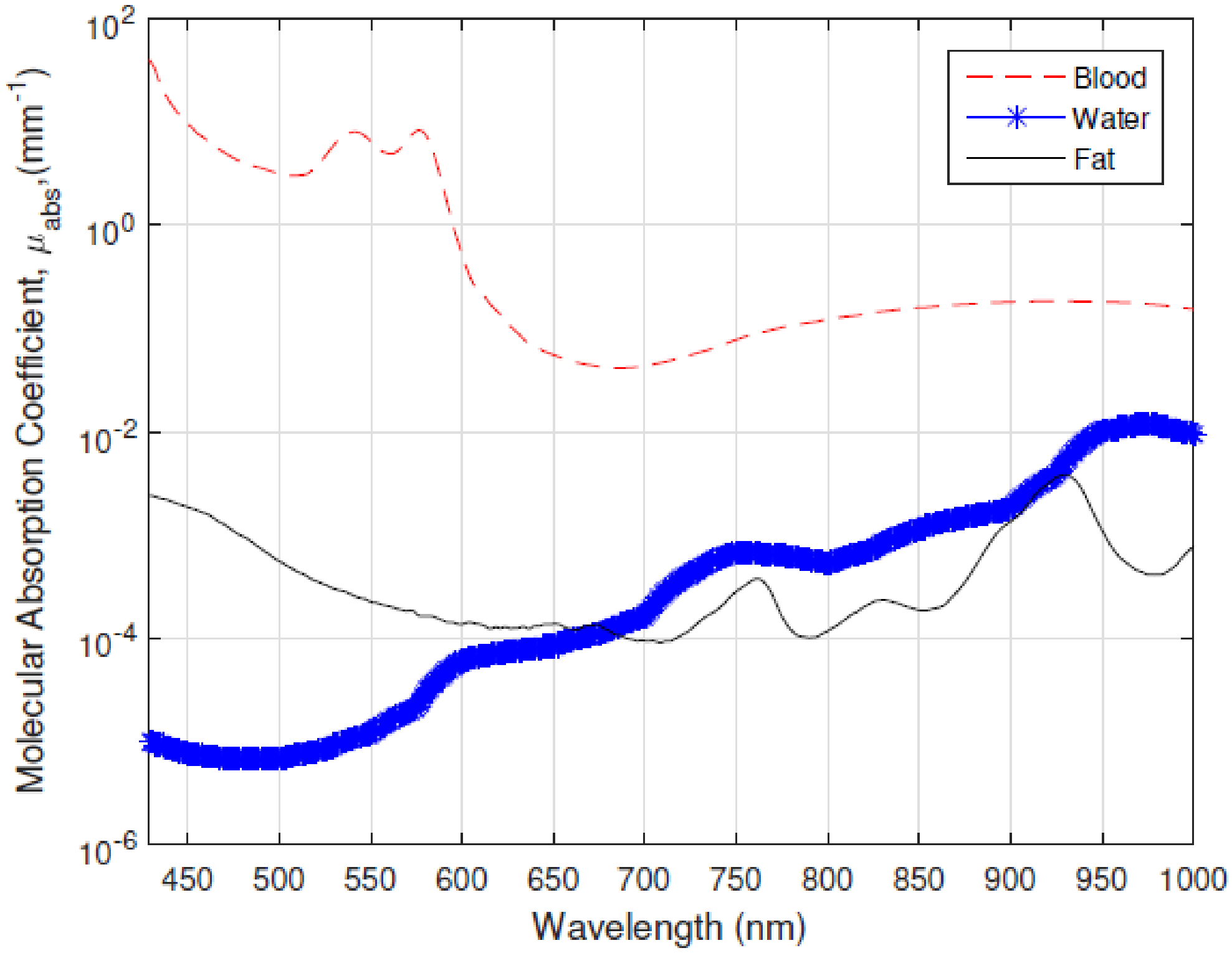

**Figure 5.14:** Molecular absorption coefficient, $\mu$ abs, for different human tissues vs. wavelength at optical window ( $\lambda$ = 450 nm to 1000 nm).

## B. Scattering

In order to examine the impact of scattering, both (23) and (26) are used taking into account the radii of different particles of the body as illustrated in table VII. It is noticeable that scatterers size at the Terahertz band is way smaller than the THz propagation wavelength. The results are demonstrated in the figure below and represent the scattering coefficient, μsca which is very small compared to the absorption one.

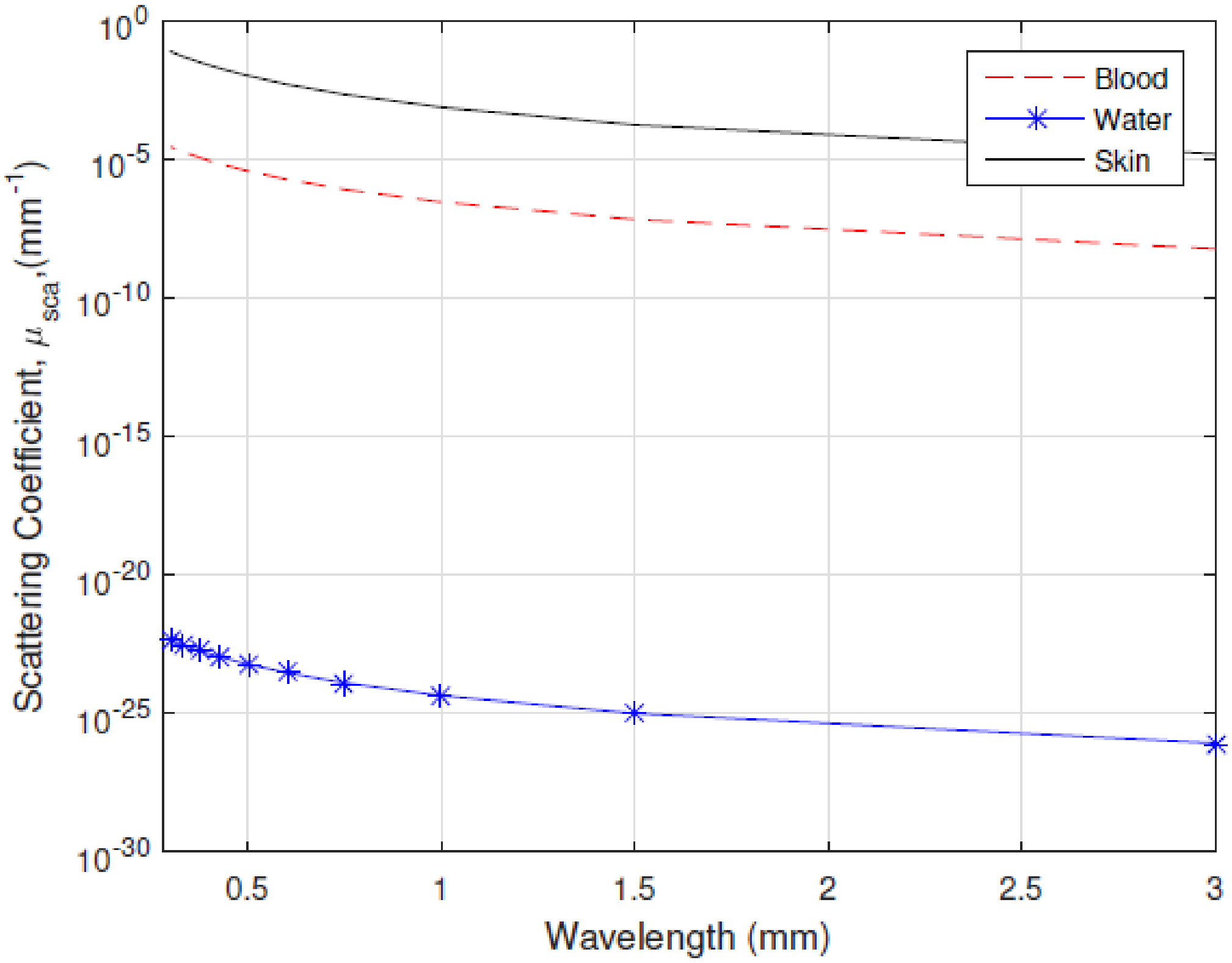

**Figure 5.15:** Scattering coefficient, $\mu$ sca, for different human tissues vs. wavelength at THz ( $\lambda$ = 300 µm to 3 mm).

This outcome is reasonable since the scattering is major only for the wavelengths that are much smaller than the dimensions of the scatterer. This is considered as an advantage of deploying the Terahertz band for the purpose of intra-body communication since the signal will not be suffered from scattering. Therefore, the only type of losses will present and contribute to the total path loss at the Terahertz, are the spreading and absorption. In figure below, the in-vivo scattering effects have been examined for optical frequencies.

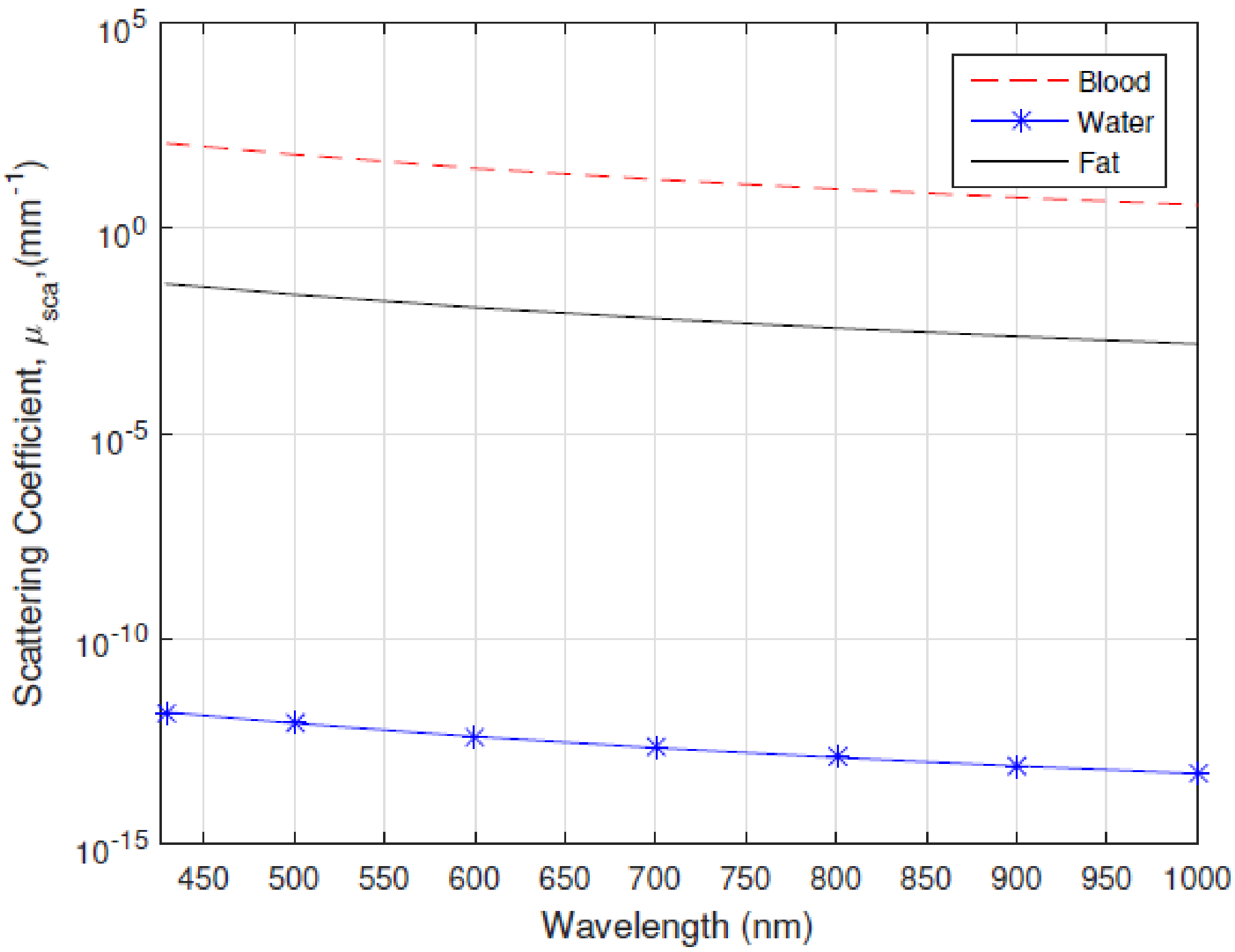

**Figure 5.16:** Scattering coefficient, $\mu$ sca, for different human tissues vs. wavelength at optical window ( $\lambda$ = 450 nm to 1000 nm).

It is concluded from the two above figures that scattering is more obvious at optical frequencies. For further investigations, the scattering efficiency can be calculated as a function of particles radius. The figure below shows the obtained results.

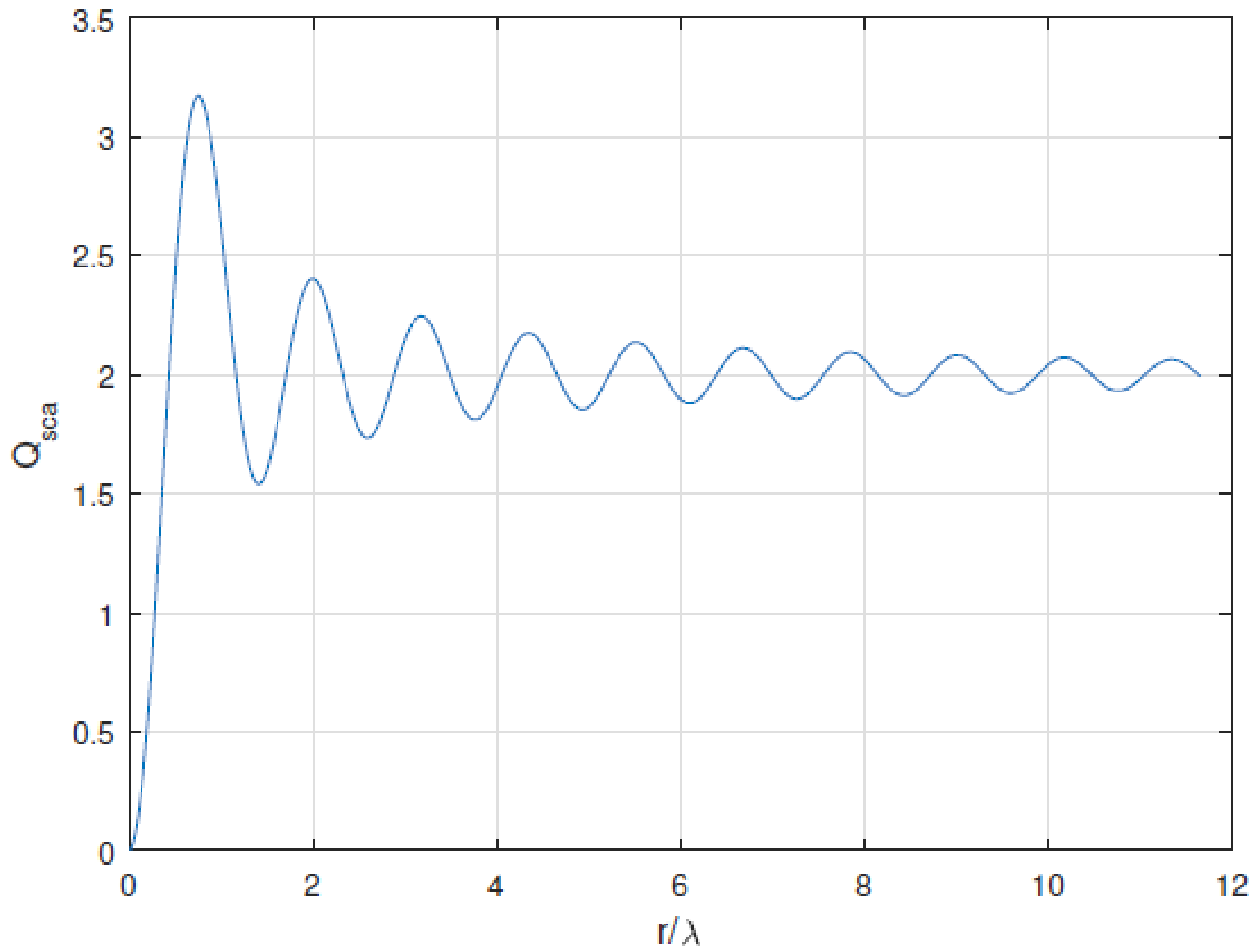

**Figure 5.17:** Scattering efficiency of a blood particle at $\lambda$ = 450 nm.

As it can be seen from the figure, the scattering amplitude at the optical frequencies looks like sinusoid. Those results could help predicting the scattering impact for different types and sizes of blood cell. It is noticed that as the particle size exceed the 1, the scattering efficiency approaches 2. Hence, blood cells are taken into consideration; their large size will yield to scattering and then absorption within the same cell. Despite we are operating in the near or far field region of the antenna. According to Fraunhofer distance [14], if the value of $\frac{2\pi}{\lambda}r \gg 1$, then it is in the far field of the antenna. Otherwise, it is in the near field. At the optical frequencies, at a wavelength of 600 nm, a 1 µm transmission distance will ensure operating in the far field. In the case of THz frequencies, the transmission distance should exceed 100 µm when assuming the far field operation.

## C. Path Loss

This section will verify the validity of the theoretical model with the help of COMSOL Multiphysics. A medium that is homogenous and characterized by identical parameters values used in the theoretical model has been considered by thinking of a cell made up of number of different layers. The source of EM wave radiation will be the point dipole antenna. A prefect matched layer (PML) surrounds the entire medium. The PML is employed to simulate the infinite environment, also its thickness being $\frac{\lambda}{2}$. The overall path loss encountered between a pair of nano-devices functioning at THz frequencies given in formula is shown in Figure. below. This path loss is experienced when operating at short distance communication link. It can be noticed that depicted model matches the FEM simulation which confirms the accuracy of developed model. It is worth mentioning that latest improvements in THz technologies contributed in the emerging of novel THz transmitters and receivers which, in turns, supported the connection among nanodevices and laid the ground to potential iWNSNs bio-convenient applications. On the second hand, Figure below represents the path loss has been suffered from at the optical frequency window where we can directly note that numerical outcomes are in accordance with the simulated results. The suggested scenarios are substantial for the IBC analysis. According to the application, the user has the freedom to choose the operating frequency and the distance will be traveled by the electromagnetic wave.

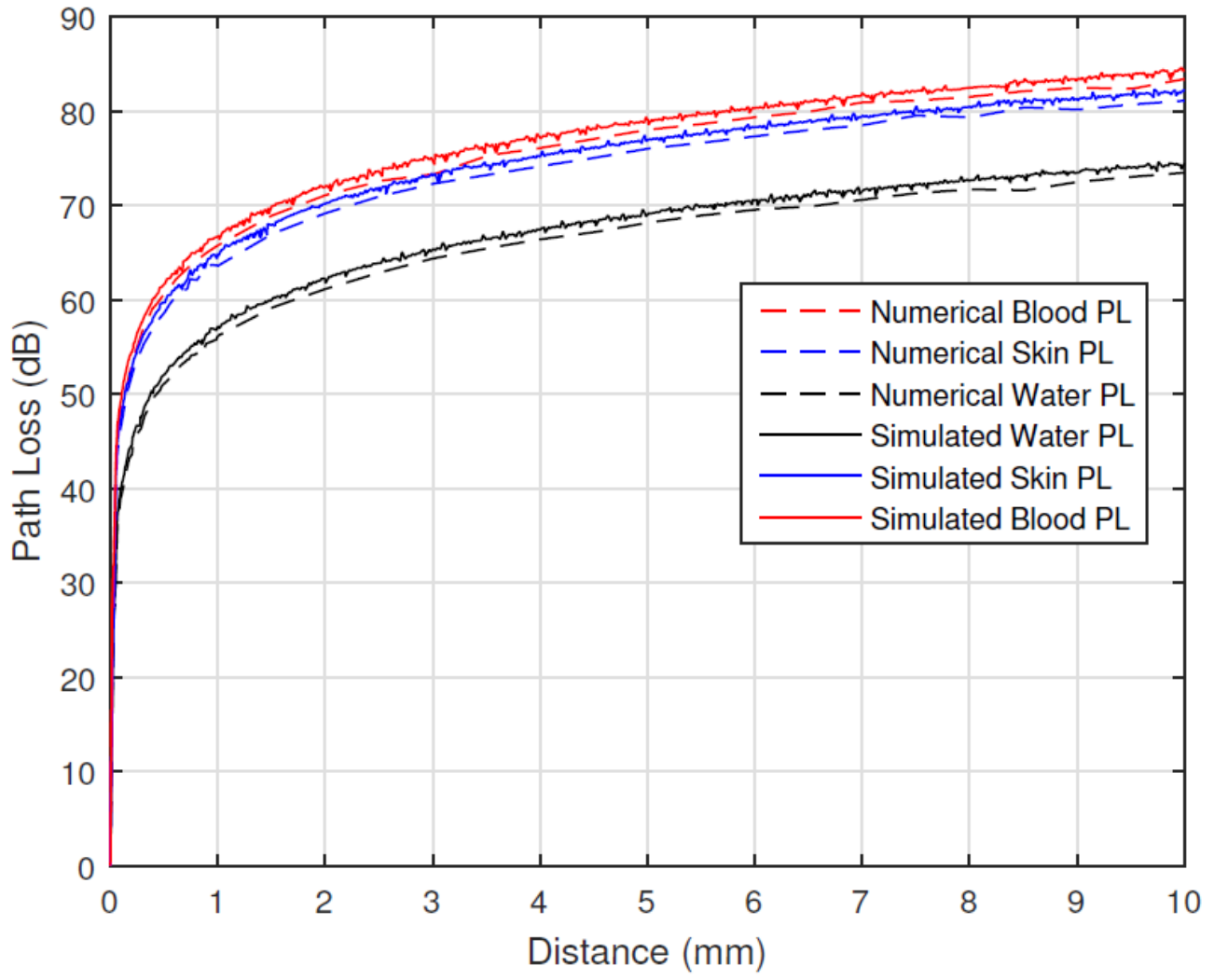

**Figure 5.18:** the overall path loss factor, $L_{tot}$, at a wavelength= 300 nm when considering short range link (0.01 mm ~ 10 mm).

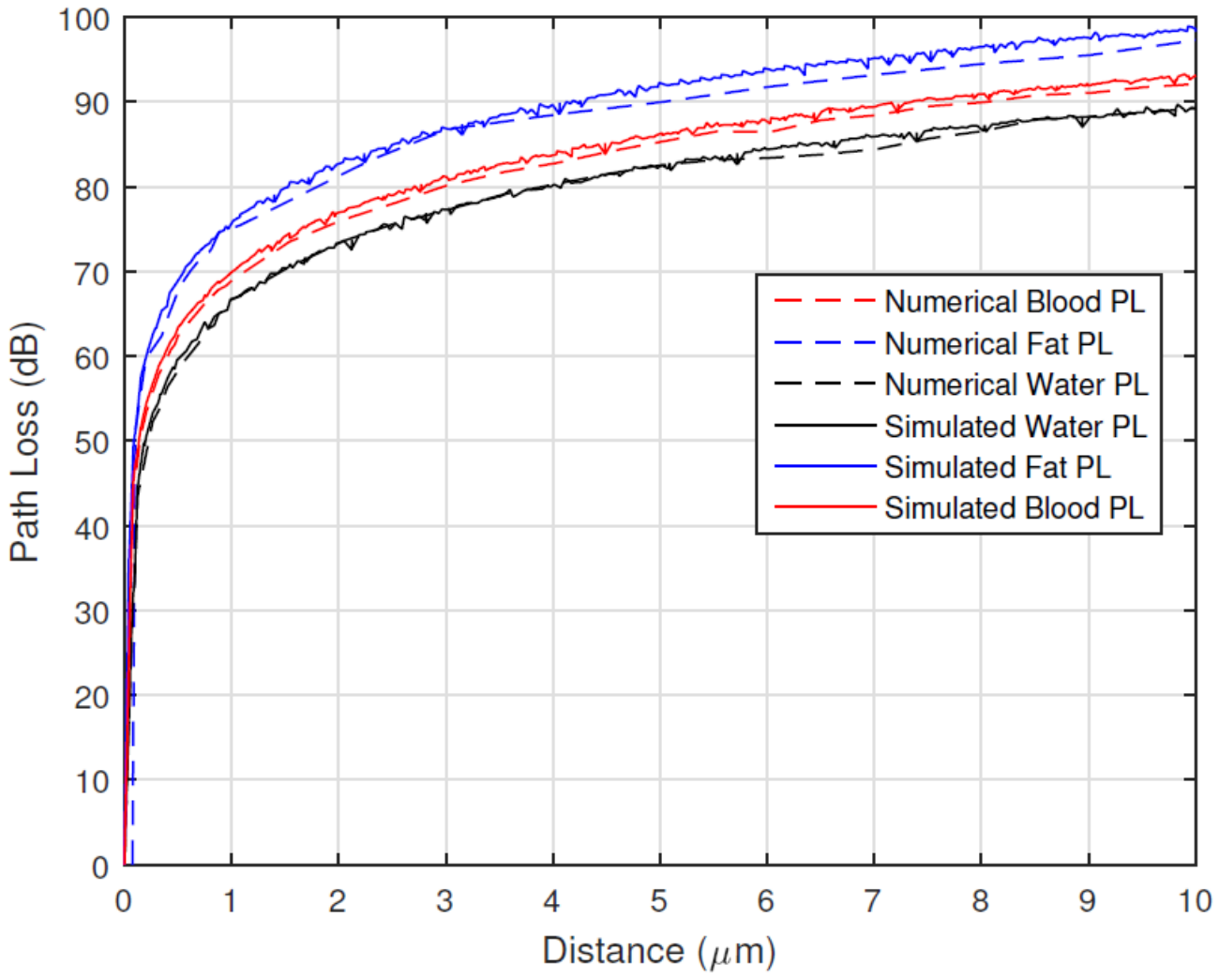

**Figure 5.19:** the overall path loss factor, $L_{tot}$, at a wavelength= 600 nm when considering short range link (0.01 mm ~ 10 mm).

## 5.5 Link Budget Analysis:

If one is estimating the many losses encountered as a signal, in any communication systems design, is propagated to the receiver from the transmitter, it is important to have link budget analysis as the starting point. To predict the required transmitter power and receiver sensitivity, there is a need to go, through a comprehensive model, predict all of the losses within the communication link. The link budget estimations in RF systems have been extensively studied but the link budget analysis, between nanodevices within the human body, lacks in the literature under consideration. The link budget analysis can only be performed when the losses and gains from the nanodevices through the human body to a communication system are taken into account. The following is the equation for the link budget;

$$\mathbf{P_R\ (dB) = P_T + G_T - Losses + G_R} \quad (28)$$

Where;

$\mathbf{P_R}$ – Is the received power and $\mathbf{P_T}$ – Is the transmitted power

$\mathbf{G_R}$ – Is the gain of receiving antenna and $\mathbf{G_T}$ – Is the gain of the transmitting antenna

Recall what was mentioned in section 3, wireless systems are necessary to achieve particular minimum conveyance quality. For this reason, the most obvious and self-evident method to calculate the needed transmit power is basically the link budget. In fact , it organizes all equations which relates the transmitted power at $R_x$ to the incoming power at $T_x$, all in the form of the SNR (Signal to Noise Ratio).

1) *THz Band:* as deduced from Fig.8, the path loss experienced by a 1 Terahertz wave when it travels over a distance of 1 mm in the blood is 65.8 dB. Obtained from Table below, a magnitude of 1mW (equivalent to -30 dBW) was selected to be the transmitted power.
   The amount of power detected at receiver side in is

$P_R(dB) = -30 - 65.8 = -95.8$

Converted to decibel scale, this is about 263 pW. Suppose an SNR equal to 10 dB,

$SNR = P_R - N$

$10 = -95.8 - N$

Therefore, it is required, as minimum threshold value, a receiver sensitivity of -105.8 dBW (26.3 pW). Relying on Table below, there exist a number of different receivers able to correctly detect the received signal. This re-affirms the viability of Intrabody Communication based on THz band.

2) *Optical Window:* extracted from Figure below, a propagating 499 THz signal (wavelength of 600 nm) over 10 μm distance through blood, will undergo a path loss of 88.6 dB. This specific penetration distance was selected to guarantee that a feasible optical link is set up. Referring to Table 4, we will assign a value of 100mW (-10 dBW) to the transmitted power. Different existing technologies imposed a trade-off between attainable transmit powers. Hence, the incoming power presented in equation above is

$P_R(dB) = -10 - 88.6 = -98.6$ ,

or in linear scale 138 pW. The SNR is assumed to take a value of 10 dB.

$SNR = P_R - N$

$10 = -98.6 - N$

Thus, it is required, as minimum threshold value, a receiver sensitivity of -108.6 dBW (13.8 pW). From Table below, photodiodes optical detectors made up of silicon, will have a high capability of recognizing received signal. Again, this emphasizes that Intrabody communication means are practically feasible.

It must be mentioned that optical generators demonstrated in Table below are with high directionality achieving gains can reach up to 15 dB according to the width of the beam ($\Delta\theta$) given in equation of section 2. This causes the intrabody penetration distance to be increased by few extra millimeters. However, to perform a fair comparison with the Terahertz spectral band, the gained outcomes assume both techniques are omnidirectional.

## *Chapter6:*

## *EM Channel Characterization accounting for the Reflection at In-Vivo Terahertz Nano-Communications in WBANs:*

### 6.1 Background on the theoretical law of Reflection:

Reflection is a concept that we meet in our daily lives, particularly when we stand in front of a mirror. This is the process by which light travels and hits an object whether opaque or transparent, some light rays bounce back (in the case of the opaque object) while in a transparent object, partly the light rays bounce back while some pass through the object. The process by which light bounces off an object is called reflection.

To well understand the concept of wave reflection it is important to consider how the waves bounce off a reflecting surface. The process could be well explained by understanding the law of reflection. As regards to the law of reflection, the wave moving toward an object or a surface is known as the incident ray while that which bounces off the surface is the reflected ray. The law under consideration states that; if a perfect line is drawn in between the incident and reflected beams, the incident angle is equal to the reflected angle [48]. The concept is depicted in the diagram below;

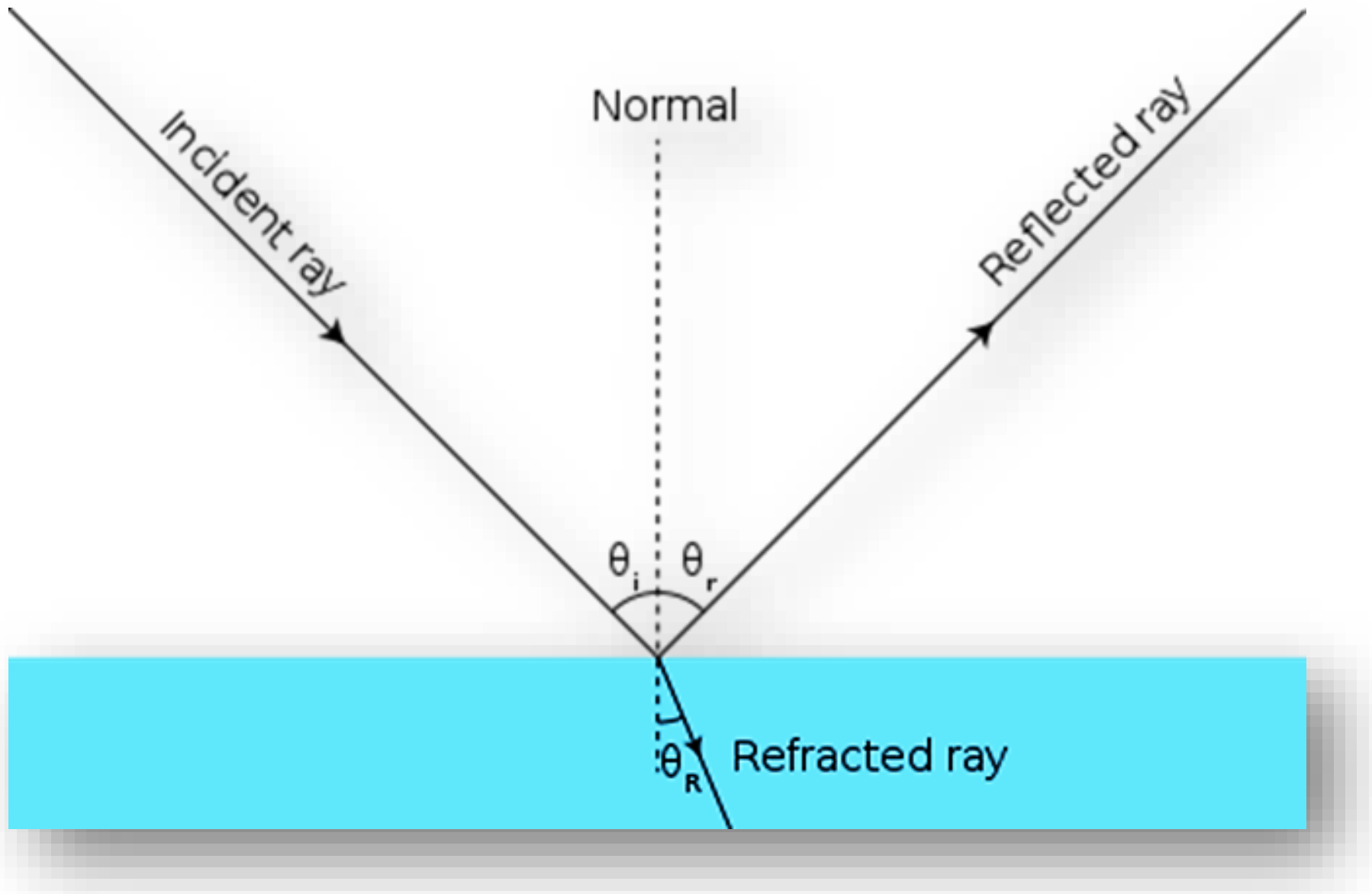

**Figure 6.1:** Representation on the law of reflection

It is also important to note that reflection of a wave does not always occur as showed above. The diagram is a depiction of a reflection on a perfectly smooth surface or a surface where all the wave rays reflect off a surface at the same angle which is known as a specular reflection. The above does not happen on a regular surface where there are a lot of imperfections which when hit by wave rays, the rays are reflected off the surface at different angles, where the reflection is referred to as diffuse [30].

## 6.2 Reflection of waves in intra-body communication (IBC):

The reflection of waves in the human body, though follows the general concepts and laws regarding wave reflection as discussed in the introduction, is a little bit different from the behaviors of waves in air or atmosphere (outside or on the human body)[31]. Also, the propagation of waves in the human body differs in different tissues and organs. That is, in soft tissues, the propagation of the various types of waves differs from that in bones. The fact that the body contains different inhomogeneous medium such as skin, blood, and fat of different masses implies that a propagated wave in the human body would be absorbed or would pass through various mechanical characteristics [31]. As the above occurs, some energies of the propagated wave are reflected but behave differently for different tissues and organs. Therefore, the reason for the reflection of a wave in the human body could be attributed to the different inhomogeneous medium inside the human body. The discussion above implies that the reflection of the wave in the human body can be well-understood and thoroughly comprehended when we have prior knowledge of the typical wave permeability for different biological materials [32]. These materials could include air, lung, fat, water, blood, muscle, and skull bone among others. It is possible to calculate the reflection coefficients for reflections of a wave in different parts of the human body, using the following formula which represents Fresnel equations for calculating reflection of TM and TM waves:

$$R(\mathrm{TM}) = \frac{n_{21}^2\ \cos(\theta) - \sqrt{n_{21}^2 - sin^2(\theta)}}{n_{21}^2\ \cos(\theta) + \sqrt{n_{21}^2 - sin^2(\theta)}} \quad : \ R(\mathrm{TE}) = \frac{\cos(\theta) - \sqrt{n_{21}^2 - sin^2(\theta)}}{\cos(\theta) + \sqrt{n_{21}^2 - sin^2(\theta)}}$$

It is also important to note that, in the context of THZ waves, the reflection of a wave in the human body critically depends on the actual physical dimensions regarding the wavelength of the wave under consideration [33]. Also, the reflections are not on flat and infinite surfaces hence are a combination of interactions between diffuse and specular reflections [33].

## 6.3 Transverse Electric (TE) and Transverse Magnetic (TM) Waves:

The concepts of TE and TM waves are as a result of waves being propagated in a homogeneous waveguide like coaxial cables, fiber optical cables, and hollow metal pipes among others[30]. To well understand the concept of TE and TM waves, we consider the following diagram of a waveguide, where the direction of the wave is depicted by (z).

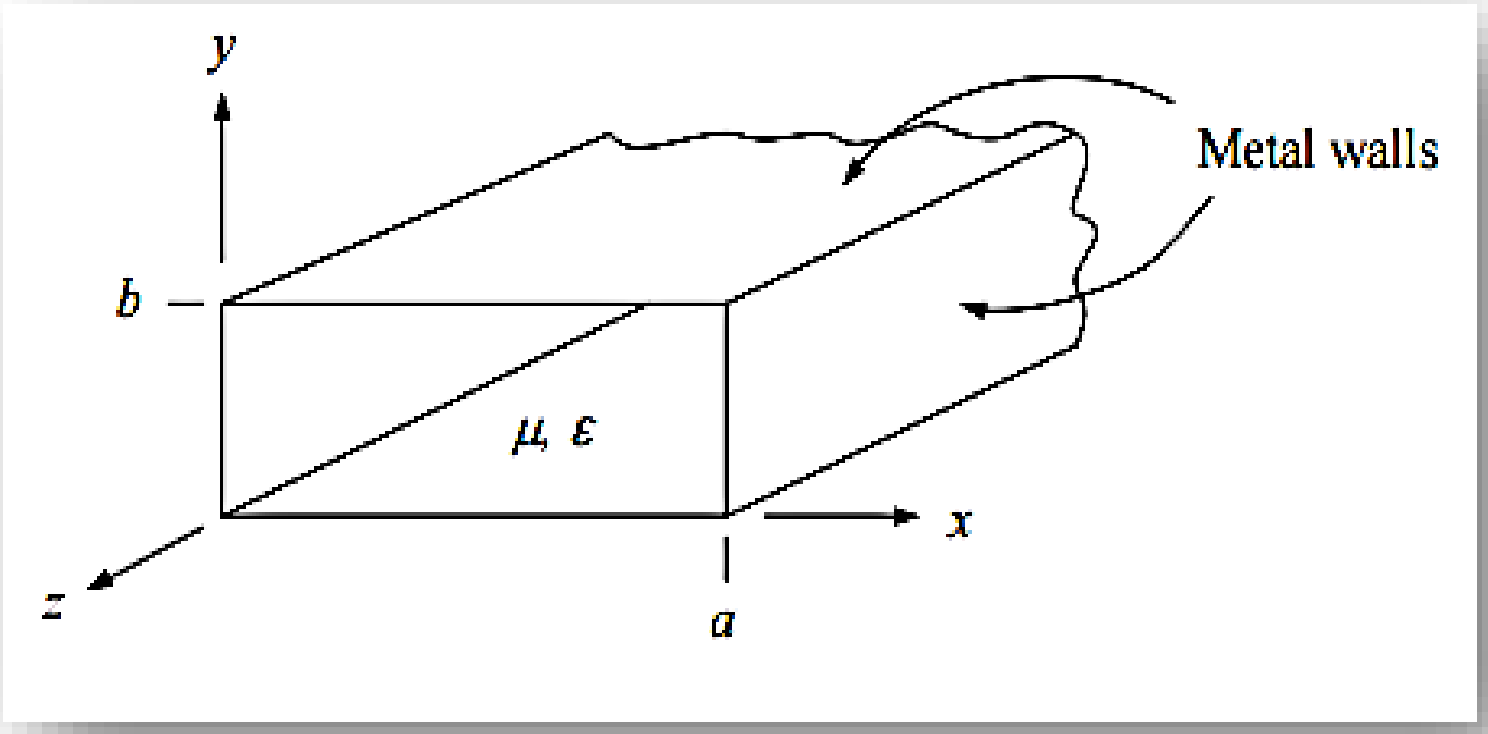

**Figure 6.2:** A Homogeneous Waveguide.

In the diagram, (y and x) represent the axial components of the transverse elements of the two (TE and TM) waves, (E and H). Using the Maxwell curl equations, the two elements can be used to produce the simple algebraic equations of the TE and TM waves by taking into consideration the electric (E) and Magnetic (H) values of the waves. It is also important to note that, from the Maxwell curl equations, the transverse components of (E) and (H) can be calculated thus allowing us to differentiate TE waves from TM waves. Simply, the TE and TM are a type of electromagnetic waves which are right-angled reflected in a homogeneous waveguide[30]. The difference for the two is that the TE wave has axial components Ez equal to zero and Hz not equal to zero. On the contrary, the TM wave has axial components Ez not equal to zero and Hz equal to zero. The diagram below can also be used to understand the difference between the two waves where the red and blue fields represent the TE and TM waves respectively [28].

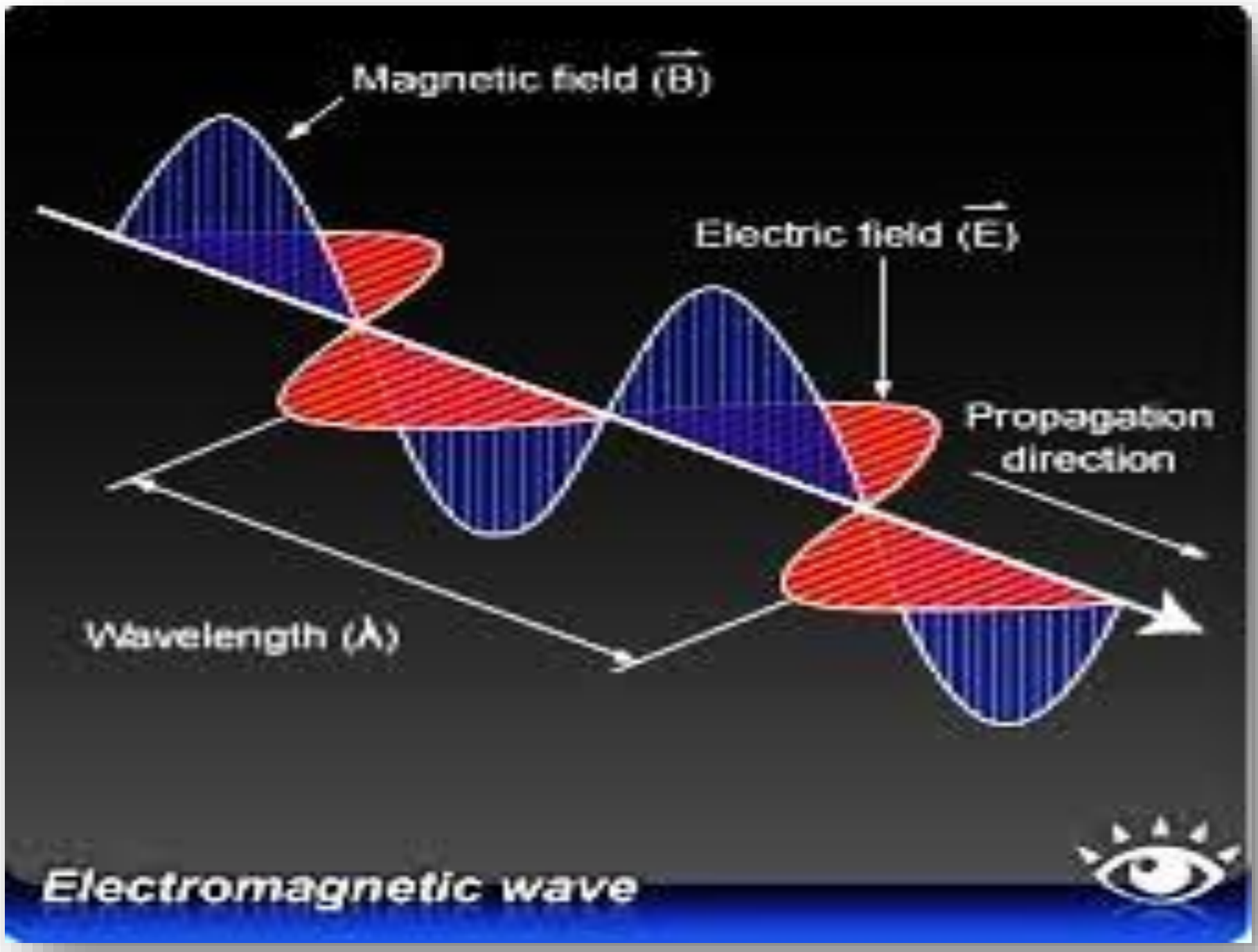

**Figure 6.3:** TE and TM Waves illustration.

As regards to reflection, the TE and TM waves' relationship can be explained on the basis of the normal law of reflection where both have equal angles of incidence and reflection[34]. The difference occurs in polarization where the TE wave result in perpendicular polarization while the TM wave brings about parallel polarization as seen in the diagram below[35].

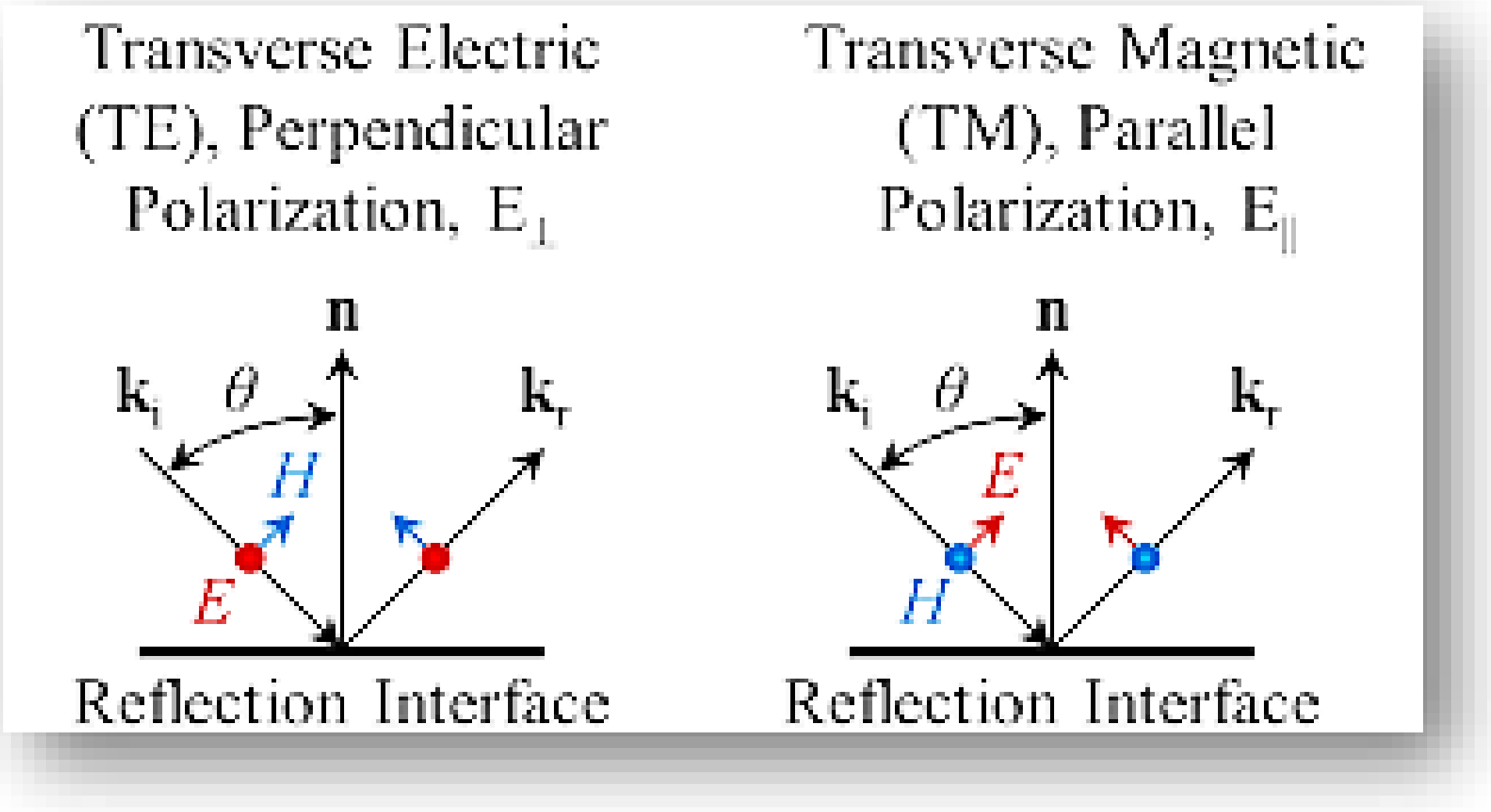

**Figure 6.4:** TE and TM Reflection Relationship.

In the diagram, ki is the incident ray while kr the reflected ray, just like in any other reflection. E and H are transverse components of the TE and TM waves.

## 6.4 The Brewster's Angle:

The concept of polarization discussed in the TE and TM waves' reflection behavior brings about the concept of Brewster's angle [36]. It is also known as the polarization angle and is that angle of incidence at which polarized wave is transmitted perfectly trough a transparent surface which is dielectric whereby there is no reflection. The angle is illustrated in the figure below:

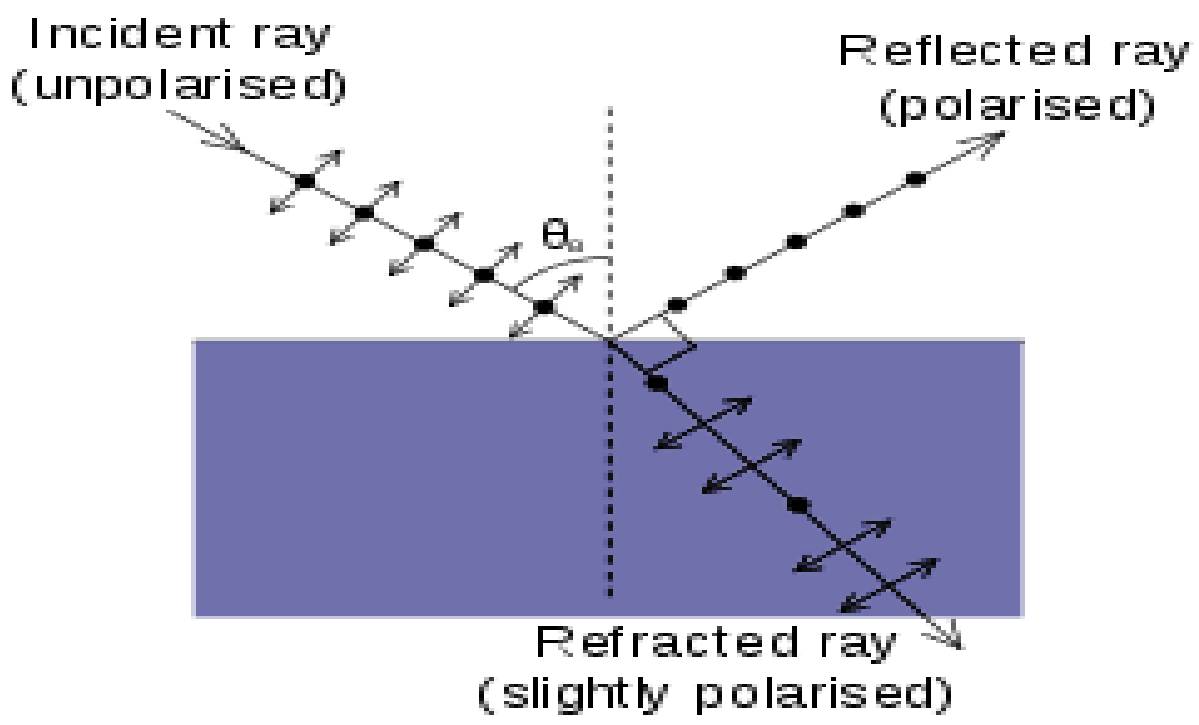

**Figure 6.5:** Brewster's Angle.

The Brewster's angle is between the reflected (polarized) and refracted (slightly polarized) rays.

Given the explanation above, the effect of Brewster's angle is to reduce reflections, like from sun rays, which reflect on horizontal surfaces. In the context of waves, Brewster's angle principle can be applied to remove reflections from transparent objects including water and air [35].

## 6.5 Total Internal Reflection :

To understand the concept of total internal reflection, it is important first to define the critical angle. The critical angle is that angle of incidence which is the minimum for total internal reflection to occur [36]. Therefore, total internal reflection will occur only and only if the angle of incidence (for the incident ray) is above the critical angle. Consequently, when a propagated wave hits a medium, at an angle larger than the critical angle, the total internal reflection phenomenon occurs. Hence, we are inclined to conclude that the reasons behind the occurrence of the phenomenon are as a result of a greater angle of incidence than the critical angle and that the wave in a medium of a higher refractive index reaches to an obstacle of a medium with a lower refractive index [36].Thus, total internal reflection cannot occur when wave's travels from air to other medium. Finally, as mentioned above that the phenomenon of total internal reflection result in reflecting the entire wave, and preventing the wave from propagating to the desired medium. Our main objectives in this project are to provide a good communication between the wearable and the implanted Nano-sensors in order to provide the doctor by all the needed patient health information. So in order to achieve our objective we need to avoid the phenomena of total internal reflection that might occurs when the wave is propagating from the wearable device into the implanted Nano-sensors inside the human body.

## 6.6 Electrical Properties of Human Tissues in the THz Band:

The biological tissues electrical features in the operating Terahertz band must be taken into consideration in order to be able to evolve a multiple layer propagation model. Our operation scale is named as the mesoscopic length scale, which is in between the two famous scales of molecular and macroscopic scales. In the Terahertz frequency spectrum, the disordered nature and the microstructure of the biological substance and supracellular orientation as well in similar matters, usually adopting fractal structure mode, trigger various polarization techniques which contain several relaxation times and asymmetric responses in the real time domain. Specifically, the Debye Relaxation Model that we founded to describe the dielectric response of tissues with high water content in the frequency domain. This model characterizes the molecules rearrangement that may include rotational and translational diffusion, hydrogen bond configuration, and structural reorganization. .

## 6.7 Numerical Results Interpretation:

Regarding pure materials, a number of Debye procedures are feasible where the complex quantity of permittivity is being as follows:

$$\epsilon = \epsilon_\infty + \sum_{j=1}^{n} \frac{\Delta\epsilon}{1 + jw\tau_j},$$

Where $\epsilon_\infty$ indicates the high frequency limit permittivity, $\Delta\varepsilon = \varepsilon_{j+1}, \varepsilon_j$ are average values, taking place at different time of permittivity, $\tau_j$ represents the relaxation time associated with the $j^{th}$ Debye type relaxation process, $\omega$ is the angular freguency which is equal to $2\pi f$.

To ensure that the complex permittivity for polar liquids is provided with ideal approximation, particularly when dealing with frequencies up to 1 THz, the Double Debye formulas are used.

$$\epsilon = \epsilon_\infty + \frac{\epsilon_1 - \epsilon_2}{1 + jw\tau_1} + \frac{\epsilon_2 - \epsilon_\infty}{1 + jw\tau_2}.$$

The above equation is already rational. The real and imaginary sections were detached. Relying on the data values gathered in Table below, $\varepsilon'' and\ \varepsilon'$ were derived. These two numerical values are then used to obtain the refractive index $n$ according to the formula below:

$$n = \sqrt{\frac{\sqrt{\epsilon'^2 + \epsilon''^2} + \epsilon'}{2}}$$

**Table 6:** Permittivity values and Relaxation time

| Model | $\epsilon_\infty$ | $\epsilon_1$ | $\epsilon_2$ | $\tau_1$(ps) | $\tau_2$(ps) |
|---|---|---|---|---|---|
| Water [13] | 3.3 | 78.8 | 4.5 | 8.4 | 0.1 |
| Whole Blood [16] | 2.1 | 130 | 3.8 | 14.4 | 0.1 |
| Skin [13] | 3.0 | 60.0 | 3.6 | 10.6 | 0.2 |

The varied response of TE and TM waves reflection coefficients as function of incident angle was acquired with the aid of equation above. Obtained results were graphed in Figures below by taking into account the blood, skin and fat material parameters associated with 1 THz. The total internal reflection takes place at the critical angle that happens only at the blood-fat interface despite the polarization mode. However, when TM polarization type is the case, the signal experiences total transmission (refraction condition) occurs at an angel referred as the Brewster angle and exists only for TM polarization mode. Figure below provides a core reason for employing specifically the THz frequency band in intrabody communication. It indicates clearly that, regardless of polarization mode, waves are almost completely transferred across the three-layer biological tissue model with neglected reflection effect as long as they are propagating within the THz band.

### 6.6.1 TE and TM Waves Reflection From implanted sensor to wearable sensor:

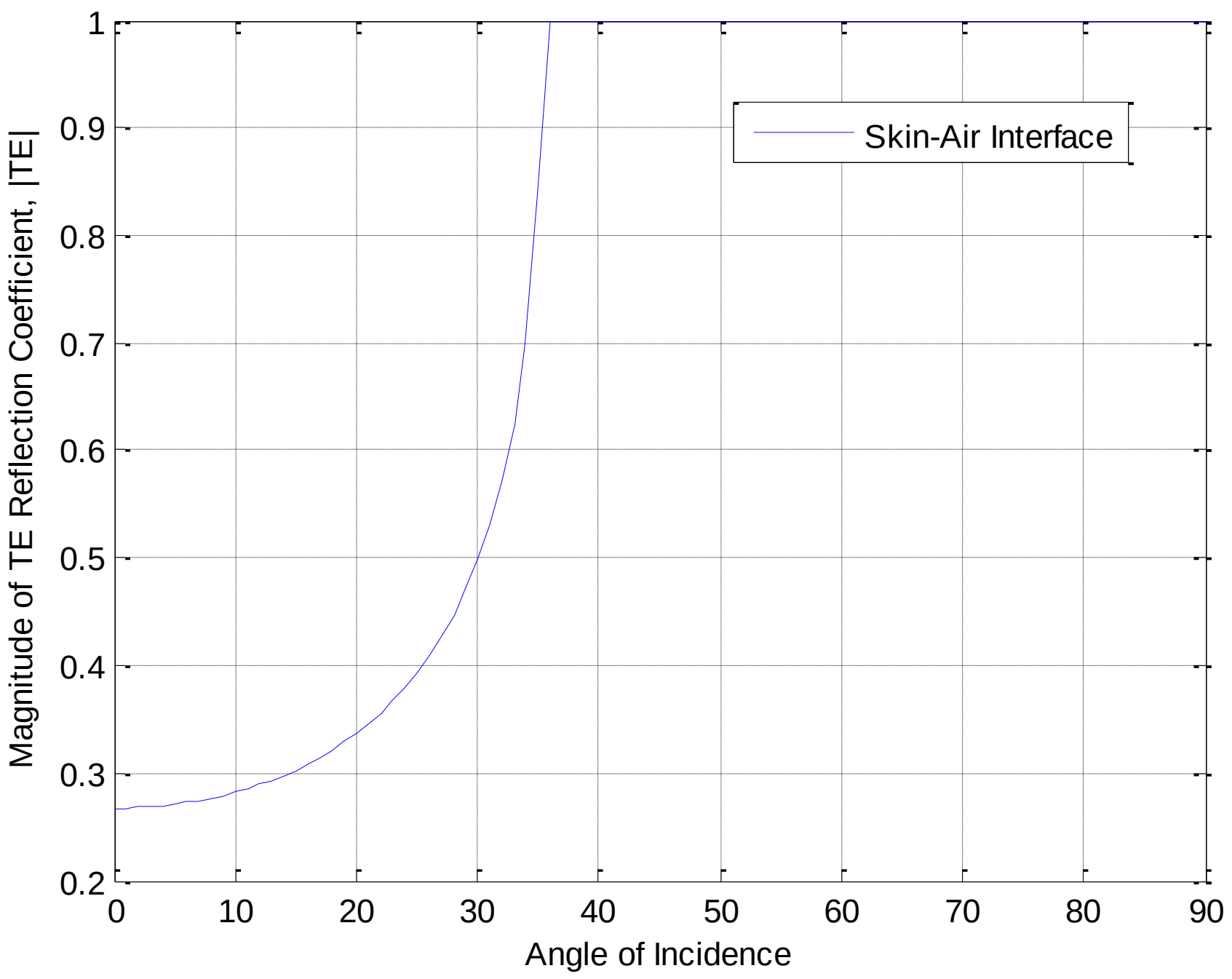

**Figure 6.6:** Representation of TE Reflection at Skin to Air Interface.

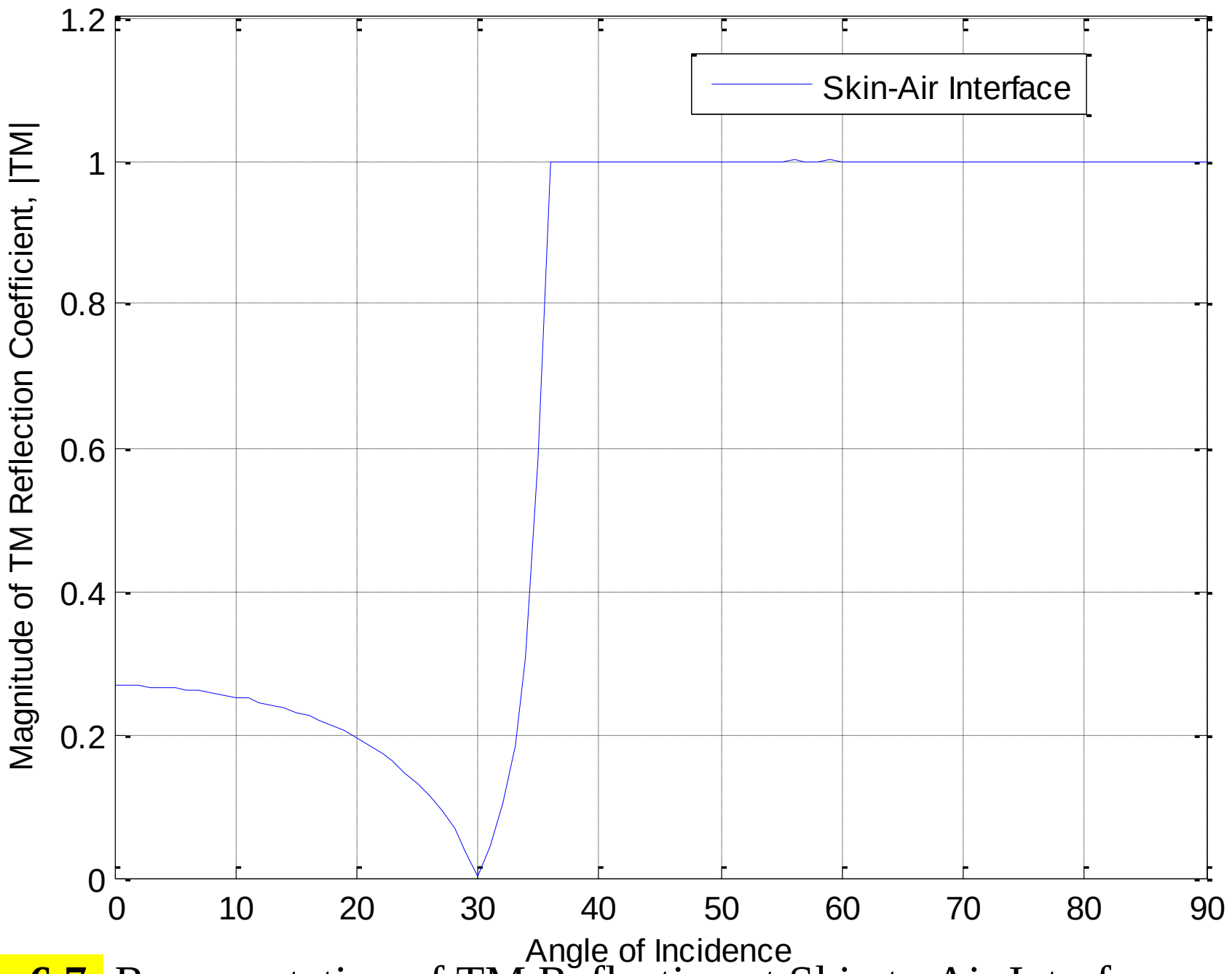

**Figure 6.7:** Representation of TM Reflection at Skin to Air Interface.

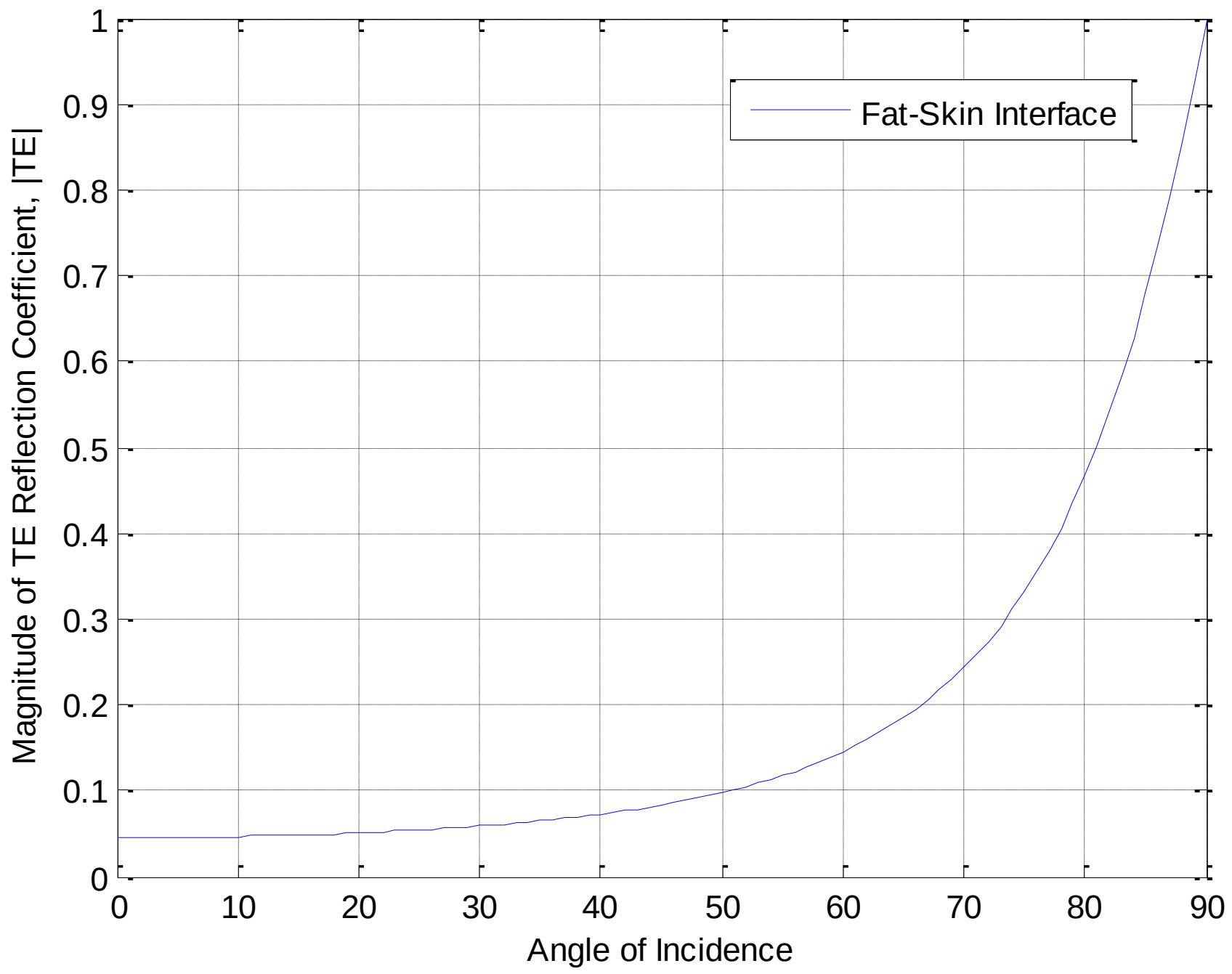

**Figure 6.8:** Representation of TE Reflection at Fat to Skin Interface.

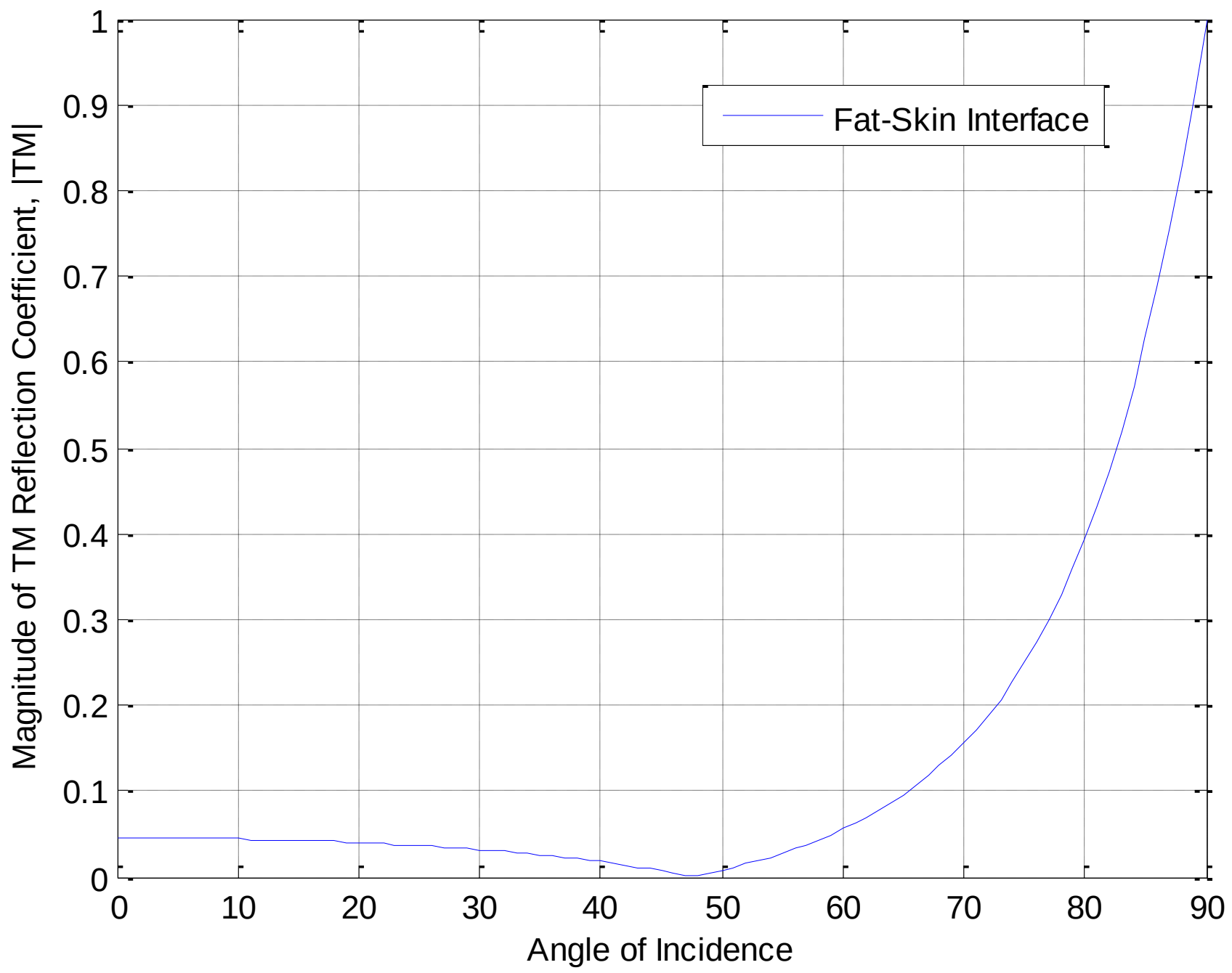

**Figure 6.9:** Representation of TM Reflection at Fat to Skin Interface.

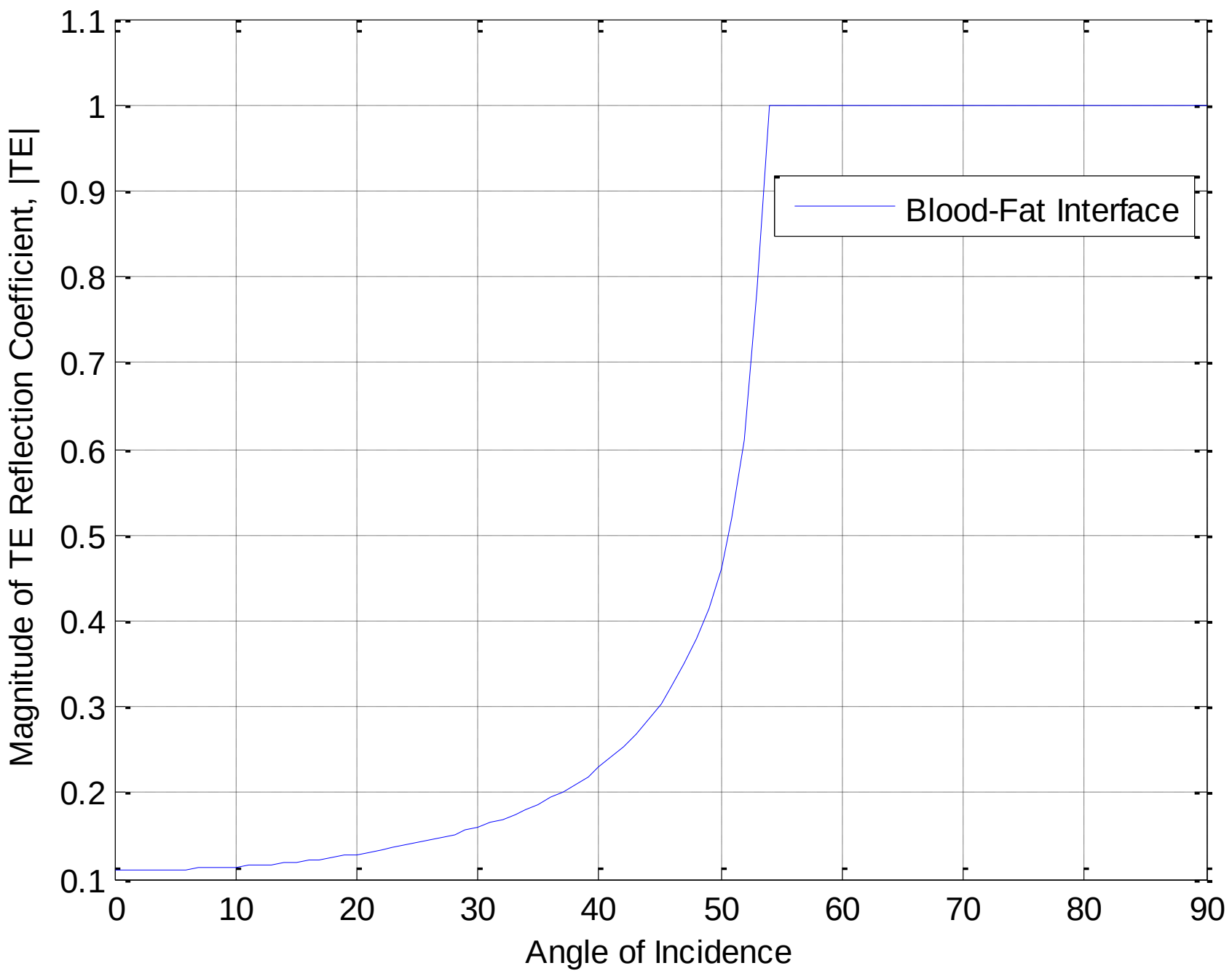

**Figure 6.10:** Representation of TE Reflection at Blood to Fat Interface.

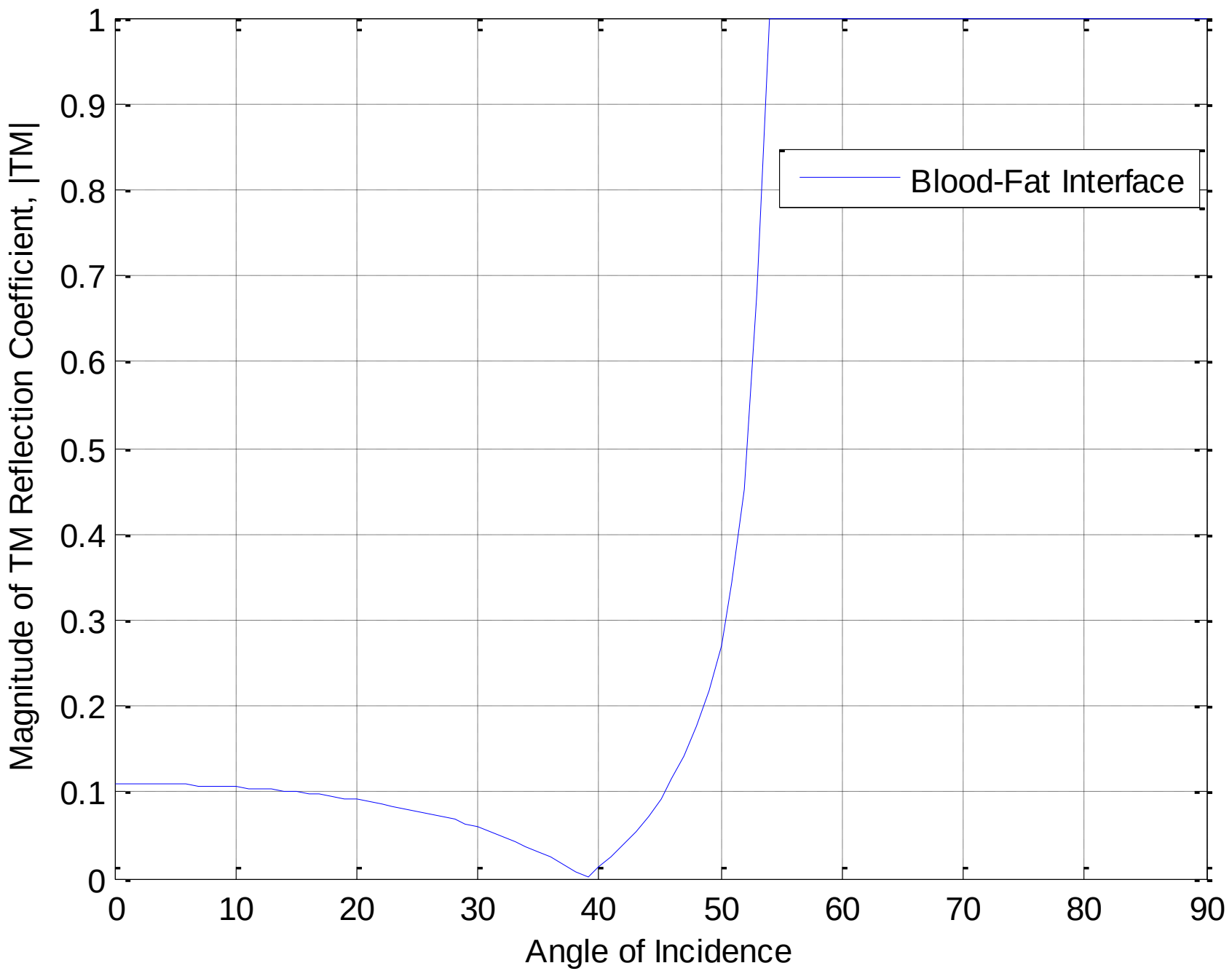

**Figure 6.11:** Representation of TM Reflection at Blood to Fat Interface.

### 6.6.2 TE and TM Waves Reflection From wearable sensor to implanted sensor:

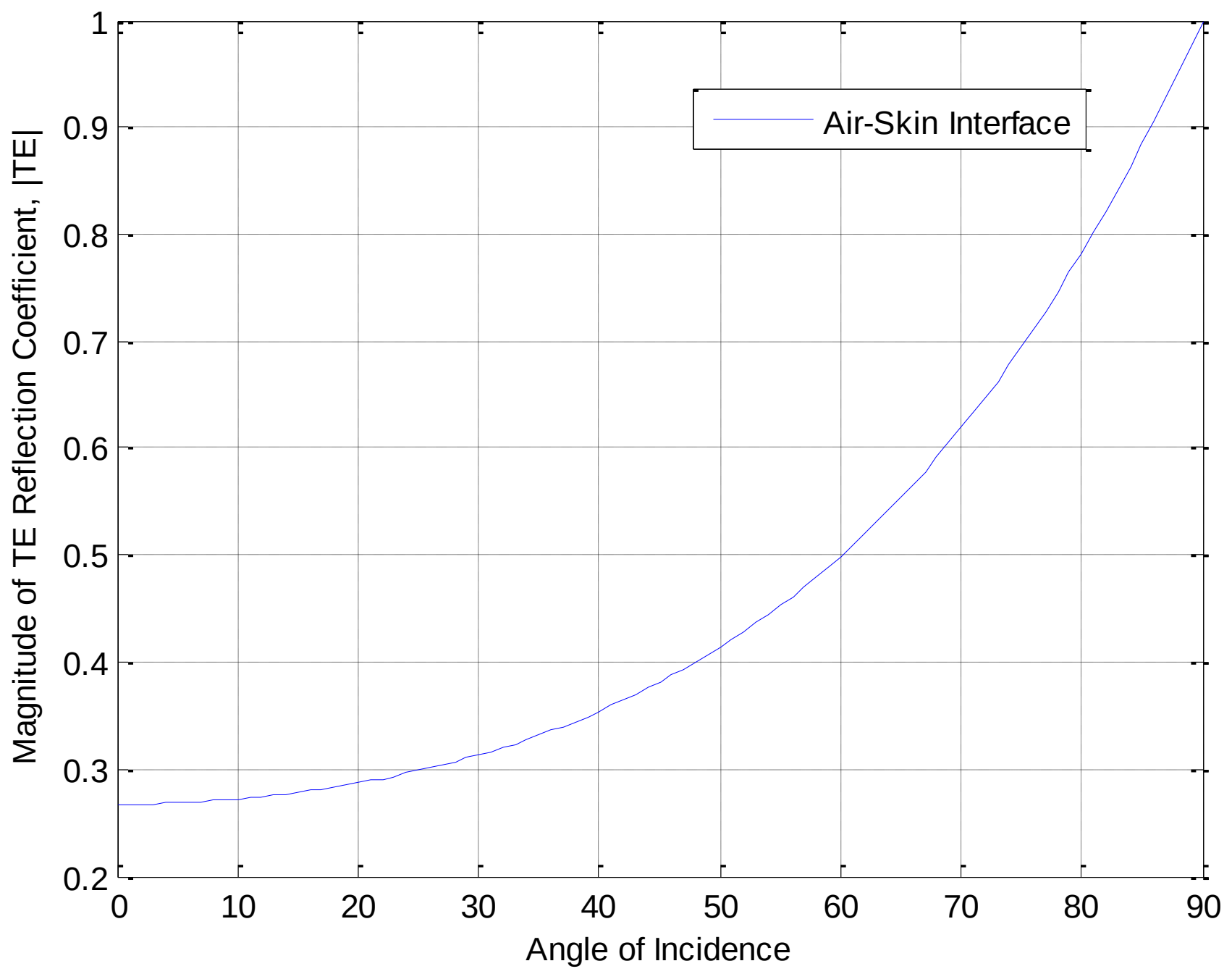

**Figure 6.12:** Representation of TE Reflection at Air to Skin Interface.

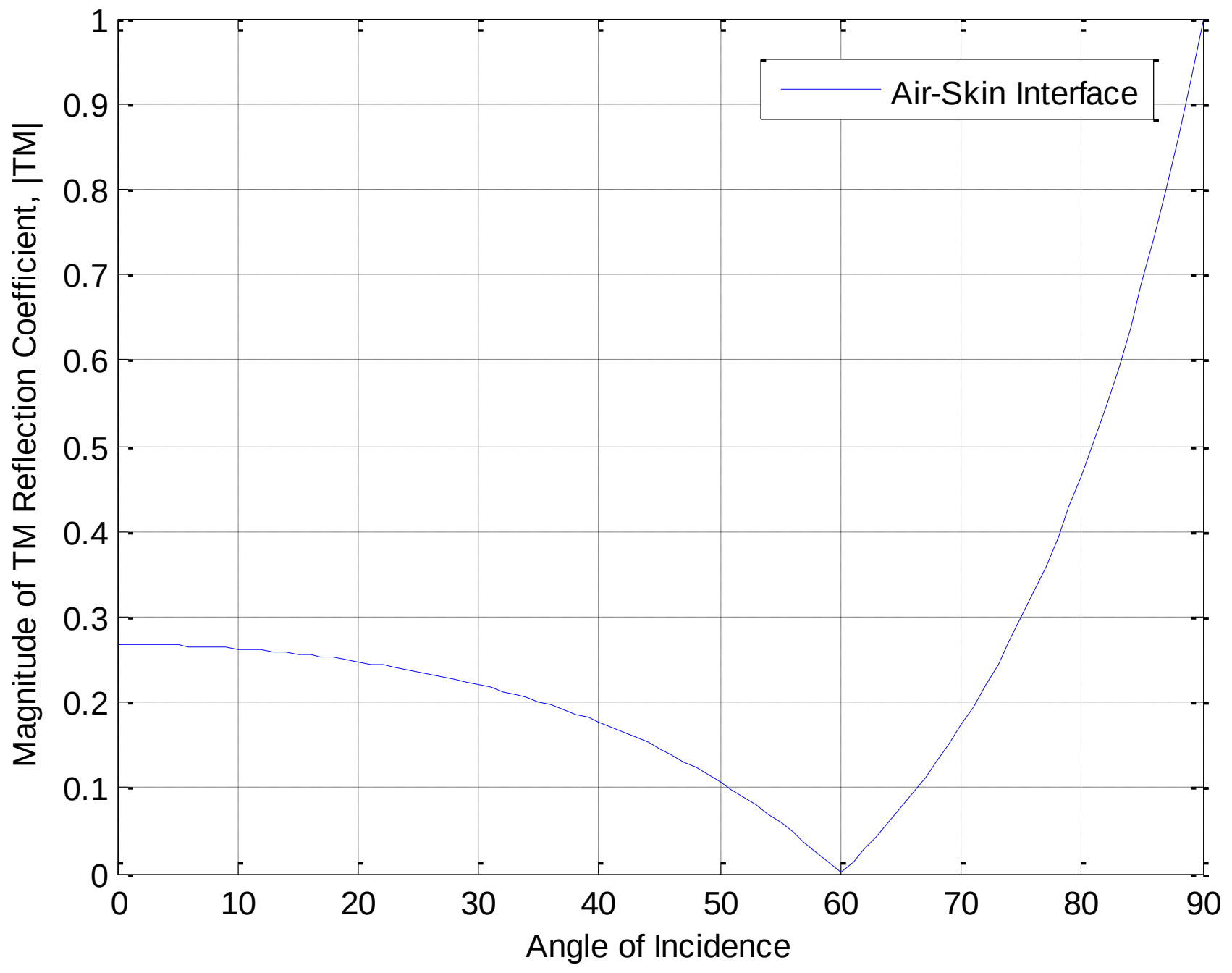

**Figure 6.13:** Representation of TM Reflection at Air to Skin Interface.

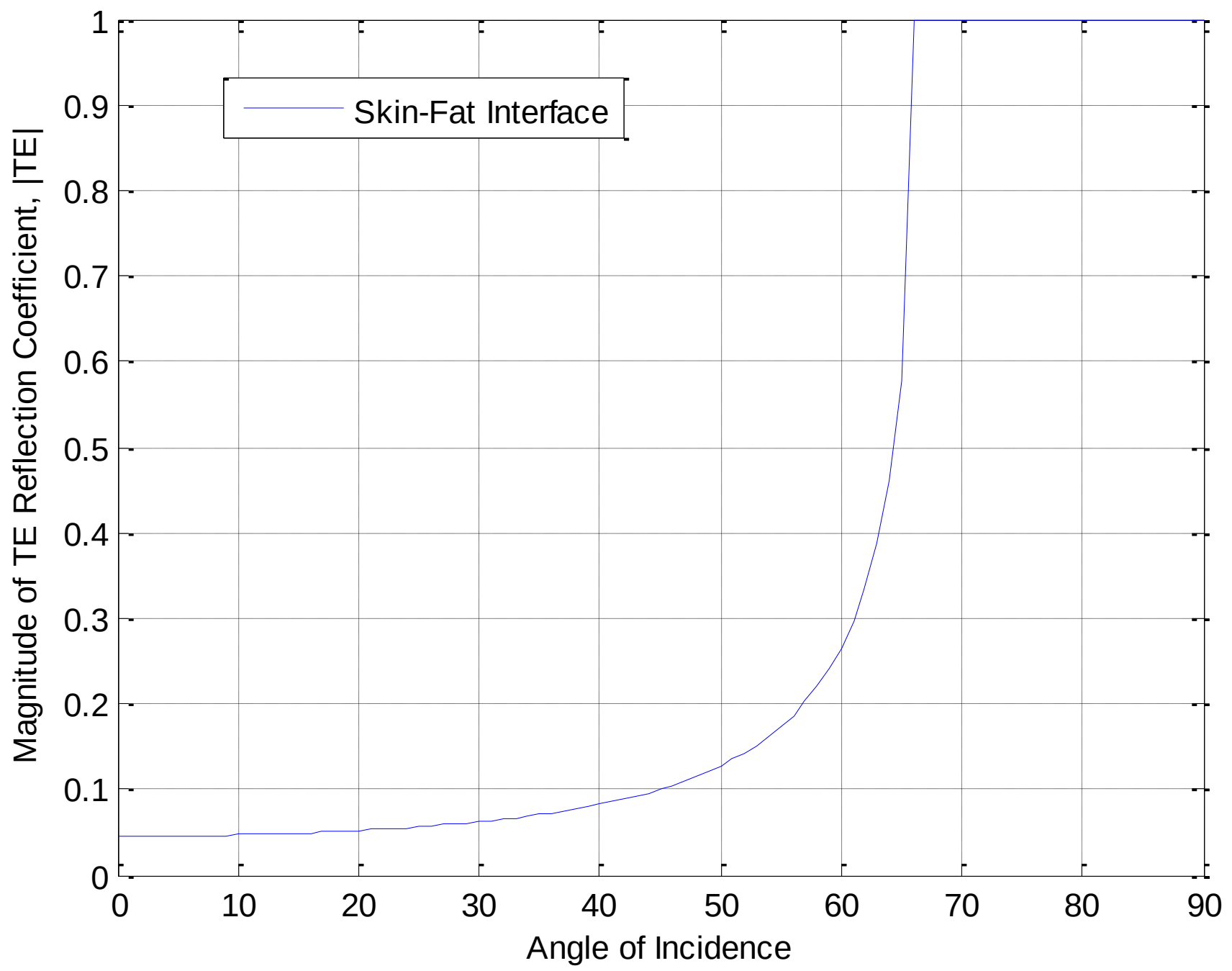

**Figure 6.14:** Representation of TE Reflection at Skin to Fat Interface.

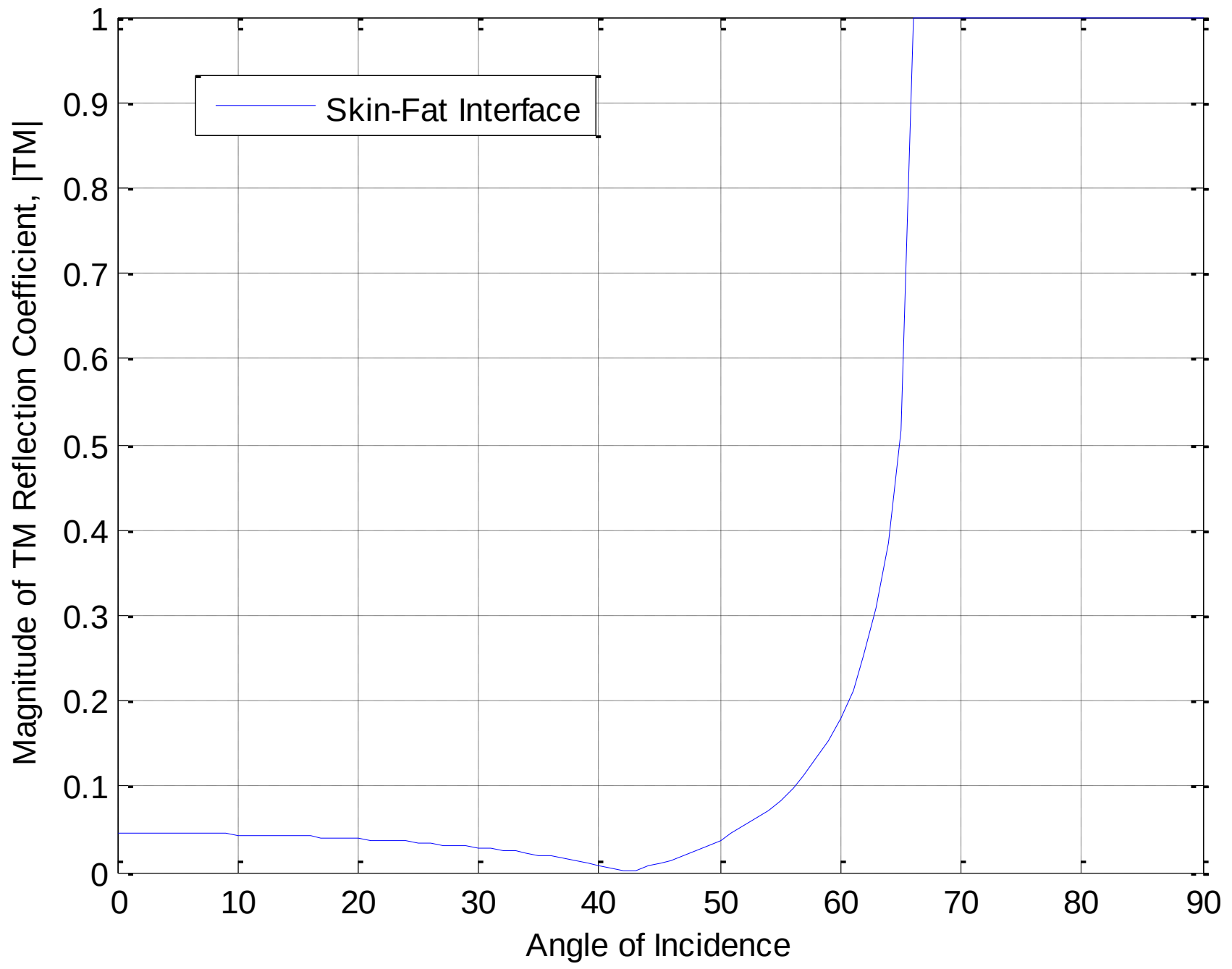

**Figure 6.15:** Representation of TM Reflection at Skin to Fat Interface.

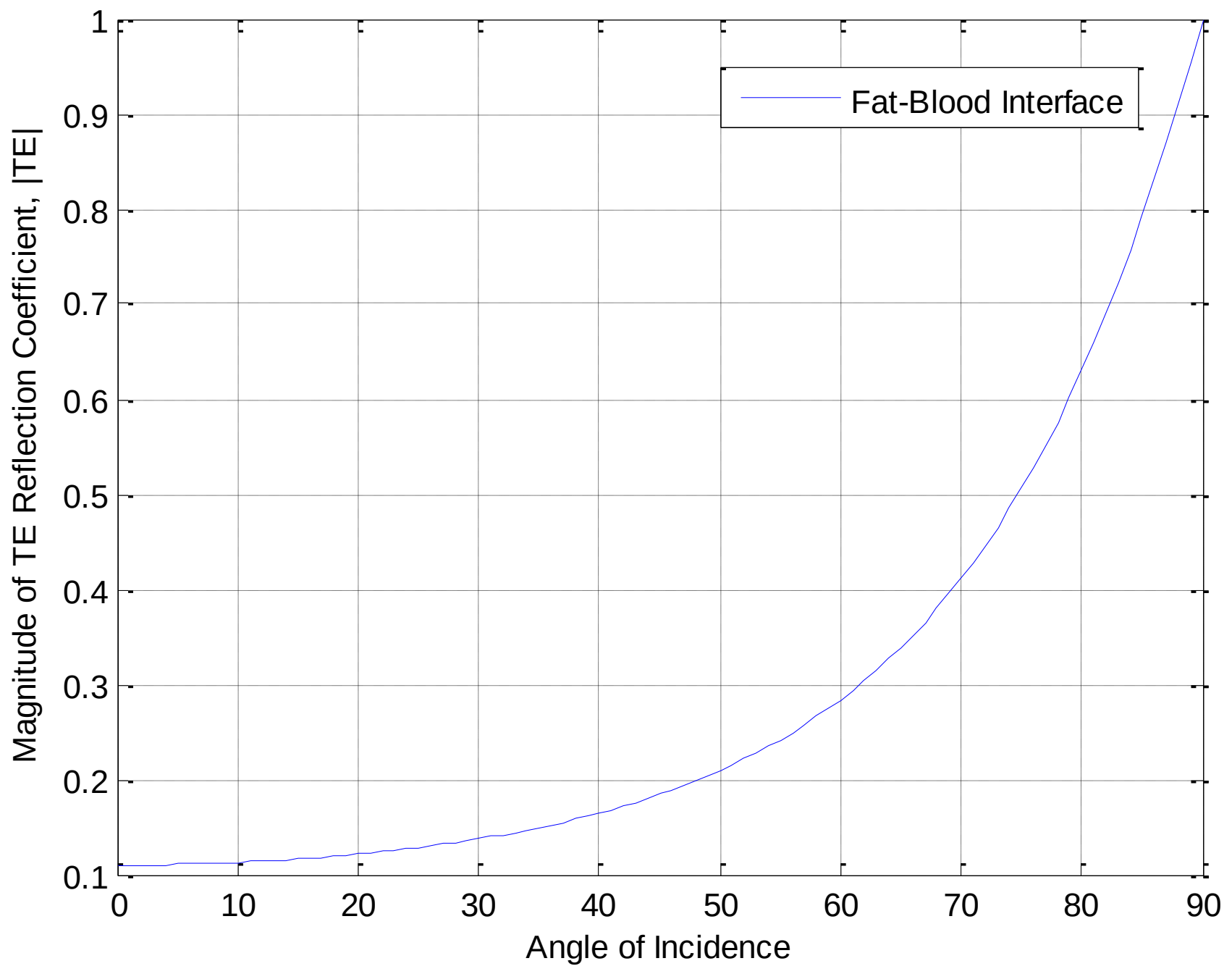

**Figure 6.16:** Representation of TE Reflection at Fat to Blood Interface.

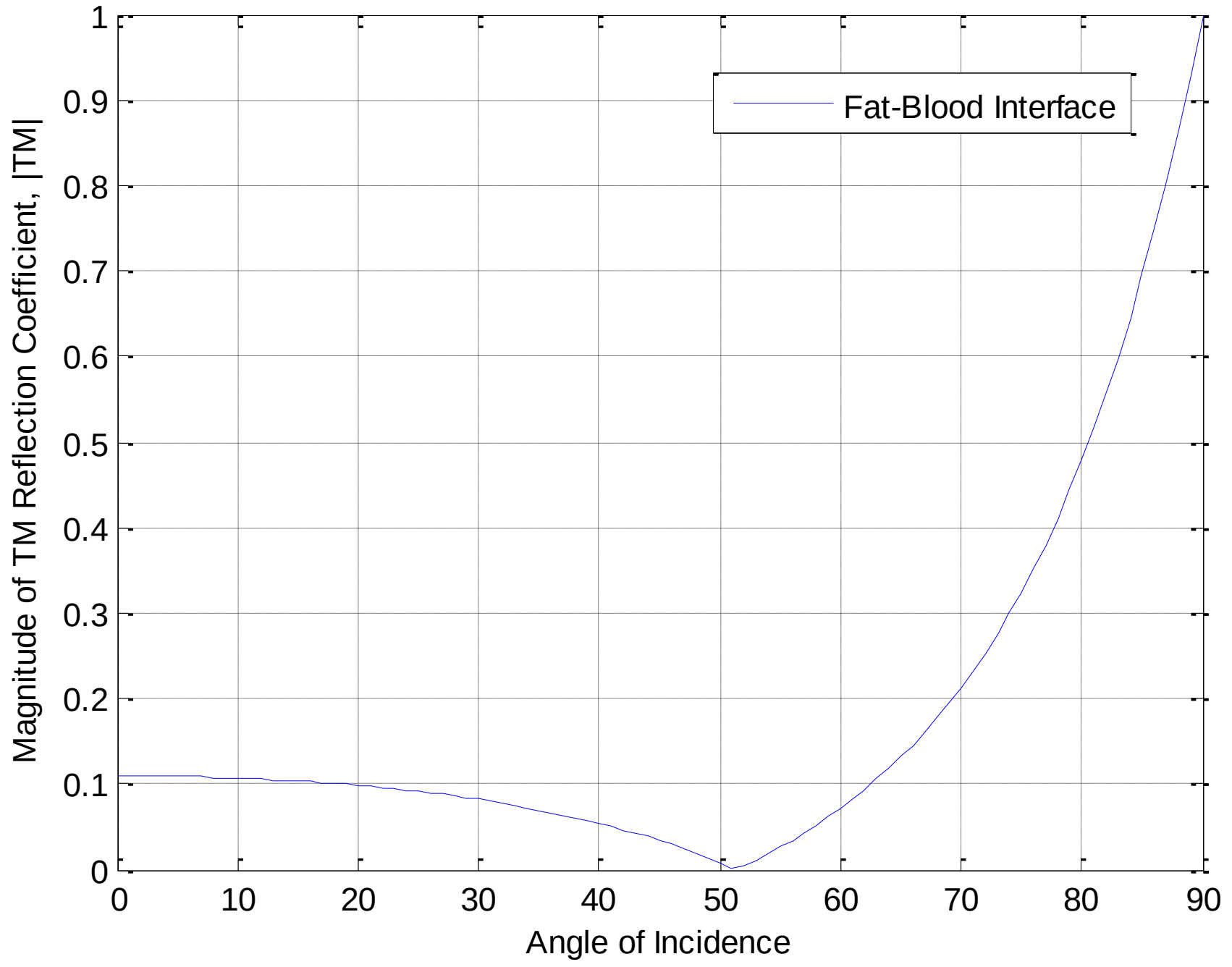

**Figure 6.17:** Representation of TM Reflection at Fat to Blood Interface.

# *Chapter7:*

# *Cross-Layer Bio-Electromagnetic Terahertz Wave Propagation for In Vivo Wireless Nanosensor Network:*

## 7.1 Overview:

Recently there have been many significant improvements in various fields such as: embedded systems, sensing technologies, wireless communications and nanotechnologies. Such advancements resulted in intelligent systems that are capable of observing human health conditions in continuous manner. Especially the Intra Body Communication that is a promising technology for a plenty of future applications, for instance, *in-vivo* health monitoring, drug delivery programs and the minimally invasive operations using nanorobots. However, in all these mentioned applications, the study of successful wireless communication built between the intrabody nano-sensor and the wearable nodes becomes necessary. In the time that there exist a wide variety of wireless modern technologies to facilitate the IBCs, a huge progress is witnessed in the area of nanoelectronics and nanophotonics is motivating the advancement of antennas and detectors that works in the terahertz range (0.1-10THz). Moreover, the rapid evolution of ultrafast lasers motivated the foundation of recent Terahertz time domain spectroscopy. It is used to inspect and describe different biomaterials since majority of biological molecular vibrations and skeleton rotations belong to the same THz spectrum radiation band. This spectrum part is rarely used, yet, it is expected to considerably contribute to the potential upcoming health care techniques. Because of the non-ionizing radiation characteristics, it is classified as fully safe for our biological tissues aside from its minor scattering losses. Up to date, the initial researches in regard the characterization of the THz frequency channel concentrated on analyzing EM wave propagation impacts through individual tissue type. None the less, all the above mentioned applications demand an effective path of signals that ensures valid communication link between the nanosensor and the external control device. This reveals the serious need to closely explore the effect of bio-EM wave penetration through multi-tissue layers instead of individual layer essentially that the current literature is in need of comprehensive intrabody communication model that even consider the impact of cross-layer on the overall reflection magnitude. To effectively overcome this lack, the present contributions suggest an approach of analyzing signal penetration based upon planar sequential tissues to frame the multi-layer effect via equivalent impedance that interpret the dispersion in electrical properties of layers and their thicknesses.

In this chapter, we will develop an analytical model relying on the multi-layer propagation across different biological tissues. We will investigate two opposite scenarios, the wave propagation from the wearable device to the nano-sensor and vice versa. Results being proposed at the end of this chapter are significant to boost the intrabody communication links quality that, in turn, influences the state of the art of bionanosensors technologies. In addition, we will outline in this chapter the cascaded model of body tissues. Also, the features of EM waves, in particular THz band, will be tested. Lastly, we are going to include some numerical results that illustrate the wave reflection and transmission attributes associated with the operating THz band. These are all in order to highlight the need for addressing cross-layer reflectivity across the below proposed model.

## 7.2 Cascaded Multilayer Intra-body Model:

To simulate the intra-body medium, we first think of the order of the layers being as follows: blood, fat, skin then air, from the innermost layer until the outermost one respectively. The initial assumption is that the nanosensor is already operating in the terahertz frequency band and situated in the blood layer. This nanosensor at the same time is transmitting data to an on-body sensor that, in turn, contacts with an outward medical infrastructure. The layered media shown in the figure below analyze and link different field measurements, for example the electric field and reflection coefficient at the left side of every interface. The configuration of adjacent biological layers is presented in the following figure. The term $n_i$ stands for the refractive index corresponding to the $i^{th}$ layer necessary to compute the refraction angle $\theta_i$ across every interface. This can be done easily employing the Snell's law.

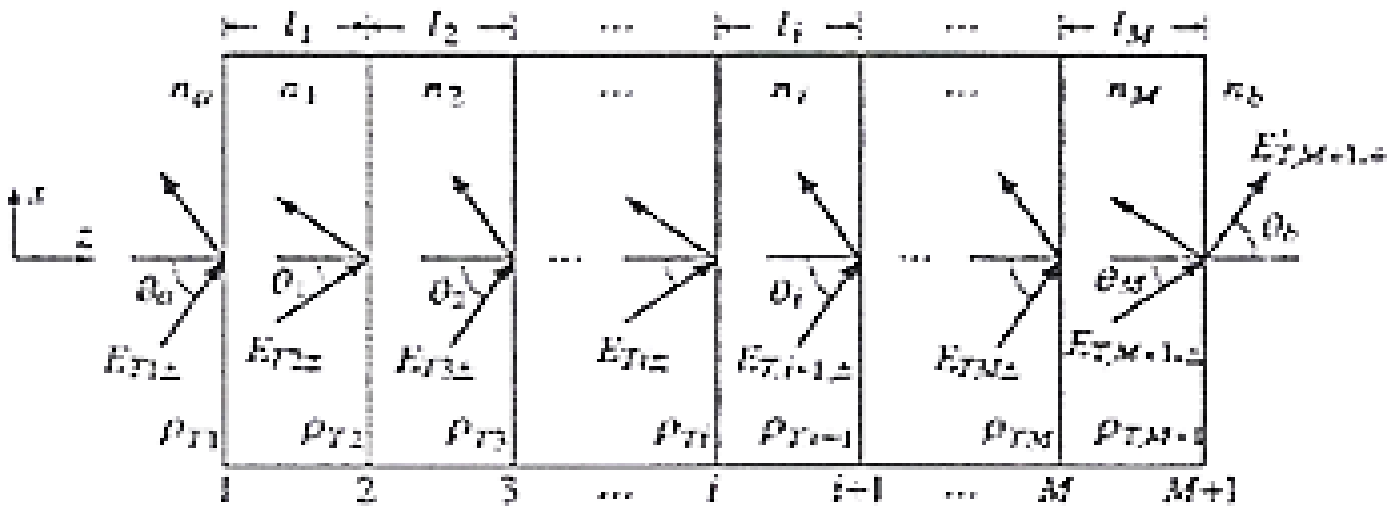

**Figure 7.1:** multiple layer dielectric model for the transverse

We will start by defining the phase thickness $\delta_i$ as follows:

$$\delta_i = \frac{2\pi}{\lambda} l_i n_i \cos\theta_i, \quad i = 1, 2, ...M + 1,$$

In which $k_0$=$2\pi/\lambda$, λ denotes the wavelength, $l_i$ represents the layer $i$ thickness. Moreover, the equation describing the transverse reflection coefficients that takes place at M+1 interfaces is :

$$\rho_i = \frac{n_{i-1} - n_i}{n_{i-1} + n_i}, \quad i = 1, 2, ..., M + 1,$$

Where we establish $n_0 = n_a$ also $n_{M+1} = n_b$.

To find out the recursive formulas of electric field for a layer, we have to deal first with the interface $i + 1$ and apply the propagation matrix to the fields being at the left side of it, so we enable them to propagate to the right of interface $i$. Next, we apply another transfer matrix to enable propagation towards left side of interface.

$$\begin{bmatrix} E_{i+} \\ E_{i-} \end{bmatrix} = \frac{1}{\tau_i} \begin{bmatrix} 1 & \rho_i \\ \rho_i & 1 \end{bmatrix} \begin{bmatrix} e^{j\delta_i} & 0 \\ 0 & e^{-j\delta_i} \end{bmatrix} \begin{bmatrix} E_{i+1,+} \\ E_{i+1,-} \end{bmatrix},$$

The parameter $\tau$ denotes for the transmission coefficient. Performing matrix multiplication would yield the following:

$$\begin{bmatrix} E_{i+} \\ E_{i-} \end{bmatrix} = \frac{1}{\tau_i} \begin{bmatrix} e^{j\delta_i} & \rho_i e^{-j\delta_i} \\ \rho_i e^{j\delta_i} & e^{-j\delta_i} \end{bmatrix} \begin{bmatrix} E_{i+1,+} \\ E_{i+1,-} \end{bmatrix},$$

For the $i = M, M-1, \ldots, 1$ the iteration will be initialized to be the left of $(M+1)^{th}$ interface. This can be achieved by executing an extra matching to the right of that interface.

$$\begin{bmatrix} E_{M+1,+} \\ E_{M+1,-} \end{bmatrix} = \frac{1}{\tau_{M+1}} \begin{bmatrix} 1 & \rho_{M+1} \\ \rho_{M+1} & 1 \end{bmatrix} \begin{bmatrix} E'_{M+1,+} \\ 0 \end{bmatrix}.$$

Referring to the equation above it follows that the reflection responses, $\Gamma_i = E_{i-}/E_{i+}$, will fulfill similar recursions. The aim behind proceeding these recursive phases is to extract the composite reflection response $\Gamma_i$ into air medium $n_a$.

$$\Gamma_i = \frac{\rho_i + \Gamma_{i+1} e^{-2j\delta_i}}{1 + \rho_i \Gamma_{i+1} e^{-2j\delta_i}}, \quad i = M, M-1, \ldots, 1,$$

The initial assumption is that $\Gamma_{M+1} = \rho M + 1$. Likewise, the overall transverse electrical (TE) and transverse magnetic (TM) recursions at each interface are demonstrated as

$$\begin{bmatrix} E_i \\ H_i \end{bmatrix} = \begin{bmatrix} \cos\delta_i & j\eta_i \sin\delta_i \\ j\eta_i^{-1} \sin\delta_i & \cos\delta_i \end{bmatrix} \begin{bmatrix} E_{i+1} \\ H_{i+1} \end{bmatrix}, \quad i = M, M-1, \ldots 1.$$

The phenomenon of cross-layer intrabody wave propagation could be well understood by simply looking for an alternative effective media to substitute with the actual cascaded layers. These media should have equivalent impedance that considers disparity among layers. This method is identical to that one utilized in transmission line theory (TL). Essentially, the below recursion is being satisfied by the wave impedances$Z = E_i/H_i$. The initial condition is that $Z_{M+1} = \eta_b$

$$Z_i = \eta_i \frac{Z_i + j\eta_i \tan\delta_i}{\eta_i + jZ_{i+1} \tan\delta_i}, \quad i = M, M-1, \ldots, 1.$$

The importance of the equation above is that it enable us to figure out the corresponding impedance at each of the skin-fat and the fat-blood interfaces, which are, in turns, crucial for calculating the separate cross-layer reflection parameters, ($|\Gamma^2|$). This step allows us to anticipate the portion of the transmitted power across each interface included in the previously described human-tissue model. It is good to mention that such structure is not only restricted to nanobiosensing applications, it can extend to cover communication link from the wearable device to the nanosensors. This topology would open new horizon to come up with applications such as drug delivery and early cancer detection.

## 7.3 Numerical Results:

In this part, we need to consider the practical values of the different electrical parameters of intrabody characteristics (all summed up in above table). These parameters will enable us to evaluate numerically the analytical model depending on the successive tissue layers.

A. ***Scenario 1: The Propagation of EM Wave from Inside the Human Body towards Outside.***

To be feasible to operate at THz frequency band, the case in which the bionanosensor being established within the internal human tissues transmits information data to the outer wearable device must be considered. Figure 7.2 demonstrates the changes in the amount of reflected wave power corresponding to the THz band different frequencies (0.1~1 THz) and at the blood-fat interface. The fat-skin-air will be regarded in terms of their equivalent impedances. Referring back to Figure 7.2, it is obvious that reflection percentage at blood-fat interface is restricted to 30.08% of the total incident power. Experienced at the last interface, the skin-air wave transmittance is figured out and results are illustrated in Figure 7.3. The results conclude that leastwise 51% of the power reaches successfully the external on-body device. It is good to note that each of Figure 7.2. and Figure 7.3 follow a periodic manner. The reason behind this is that effective thickness of fat and skin layers is altered in response to frequency variations.

Figure 7.4 and Figure 7.5 shows the equivalent impedance resistance and reactance, respectively, when we operate from inside towards outside. When testing the results of the resistance in Figure 7.4 along with their associated reactive values in Figure 7.6, it can be deduced that the parallel medium has combatively smaller reactance than its equivalent resistance. This implicates that cascaded layers equivalent impedance is approximately purely resistive, paving the way to dielectric medium in which THz signals are able to propagate without being exposed to severe losses. For instance, operating at 1 THz and bearing in mind that $z = R(1 + j\frac{X}{R})$, the percentage of reactance to the resistance is just 0.2%. This denotes that the phase shift between the magnetic and electric fields of the THz wave is zero.

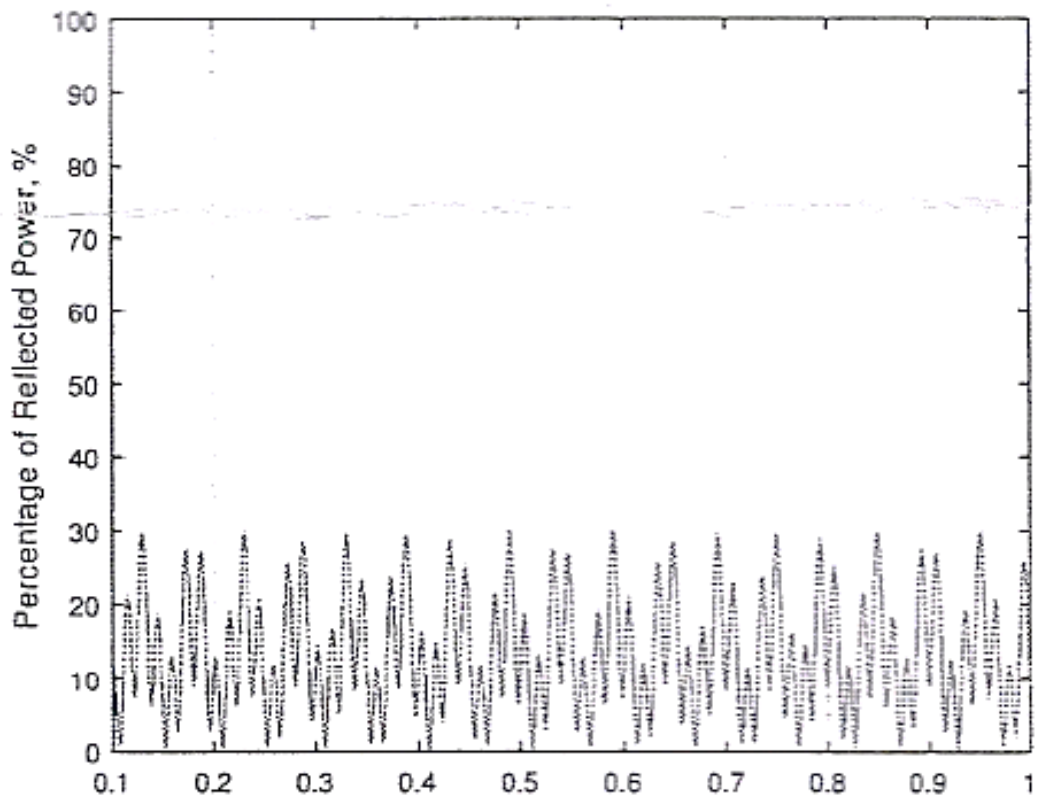

**Figure 7.2:** The portion (in percentage) of THz power incident at the interface being reflected when propagating from inner to external medium.

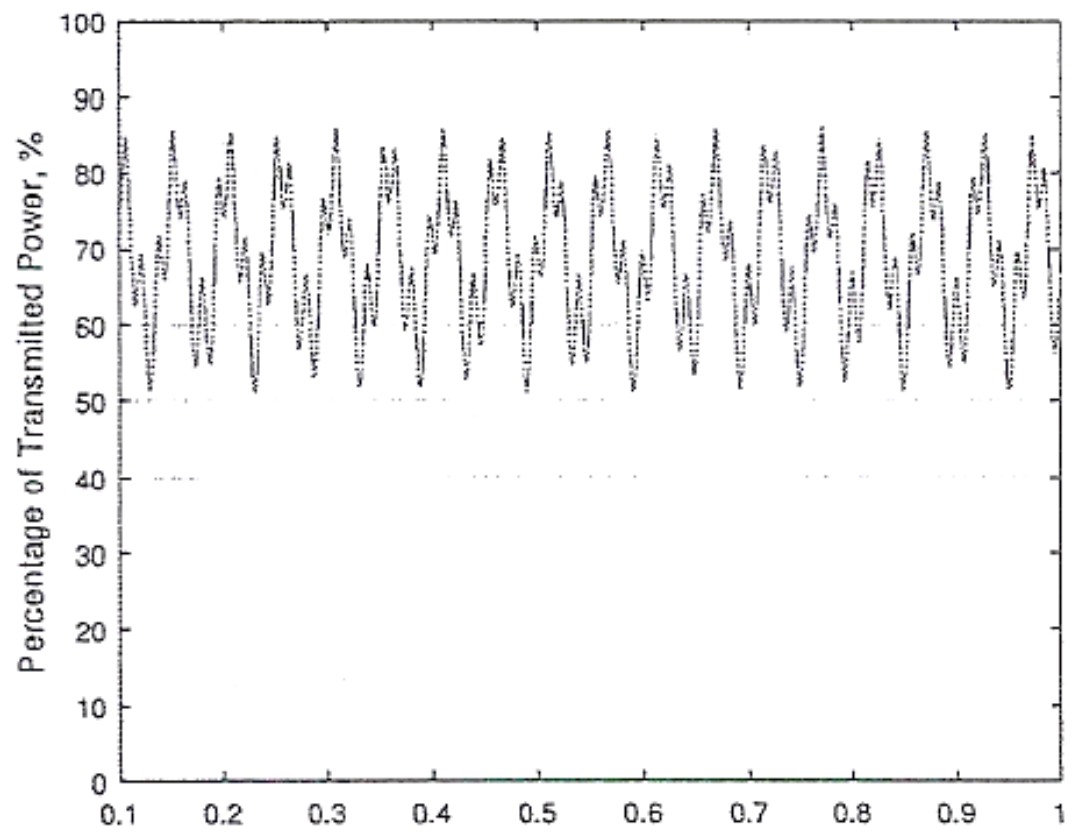

**Figure 7.3:** The portion (in percentage) of THz power incident at the interface being successfully transmitted when propagating from inner to external medium.

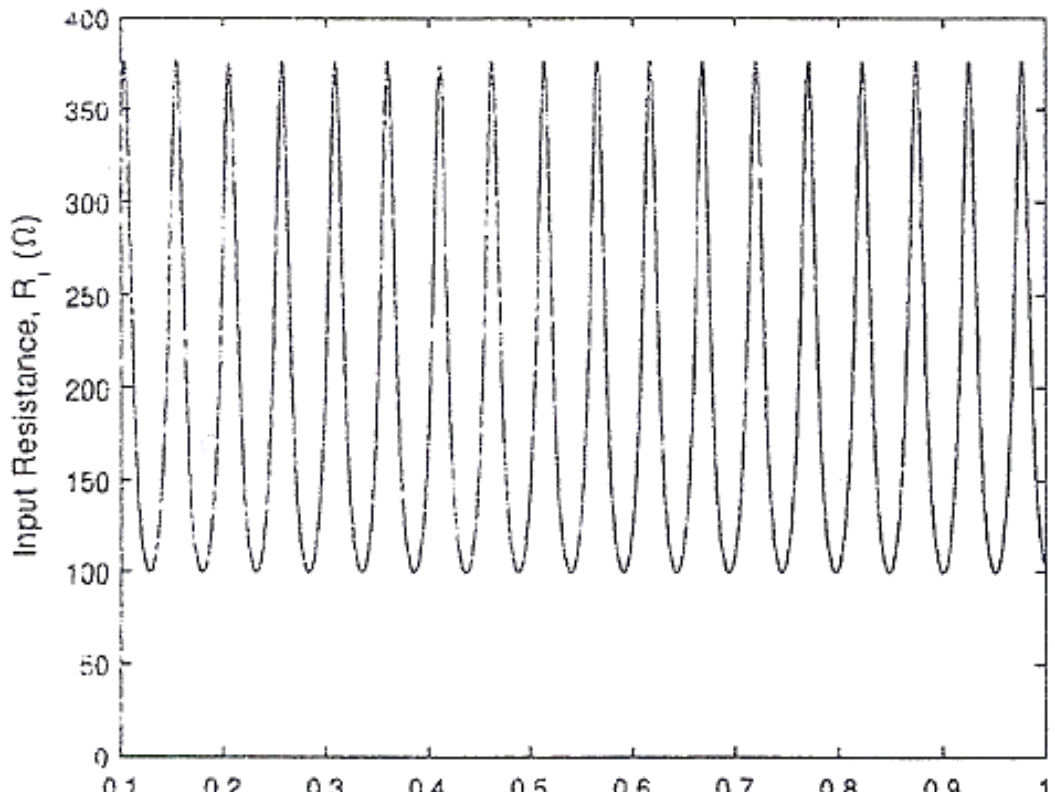

**Figure 7.4:** Equivalent resistance input values as a function of Terahertz-range frequency when propagating from inner to external medium (fat-skin air-interface).

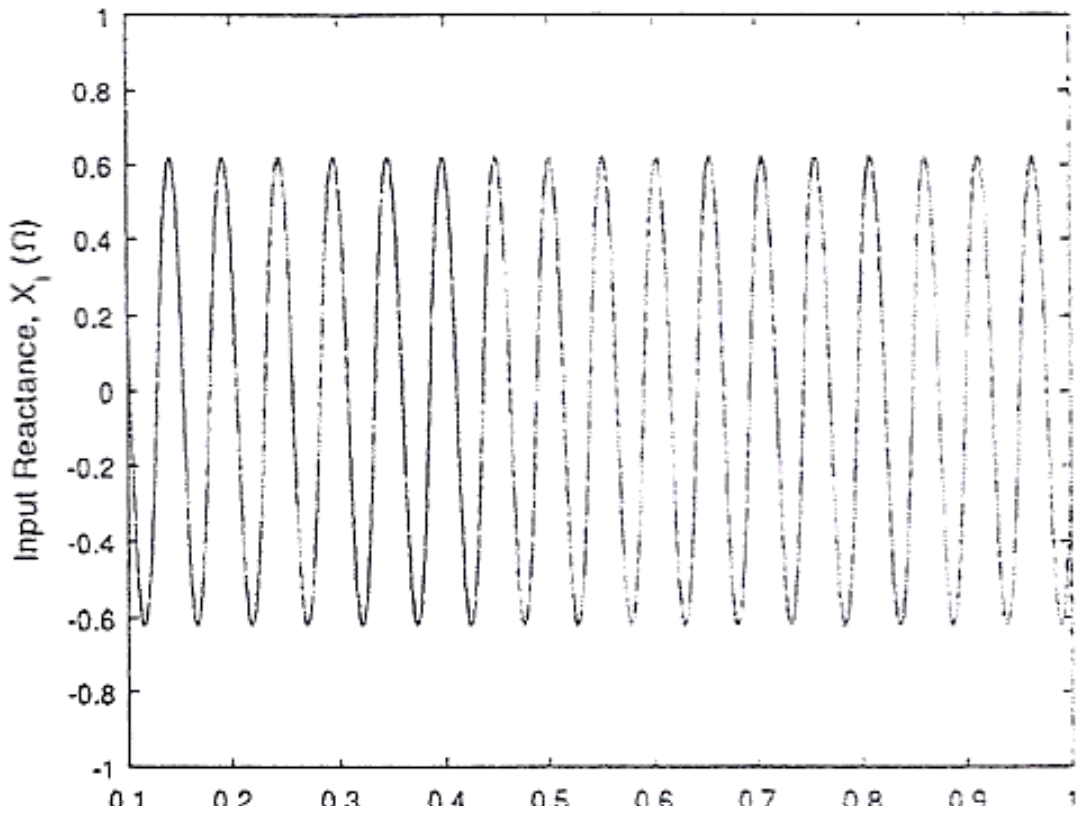

**Figure 7.5:** Equivalent reactance input values as a function of Terahertz-range frequency when propagating from inner to external medium (fat-skin-air interface).

### *B. Scenario 2: The Propagation of EM Wave from Outside the Human Body towards Inside.*

Taking into consideration the opposite scenario, where now the wearable device transfers information data to the blood embedded bionanosensor. Metrics were examined and analysis in the preceding case are exactly the same of those will be applied in here. Fig. 6 sketches the wave reflectivity fluctuations at the air-skin interface. An effective medium with parallel impedance will be used to represent the skin-fat-blood interface. It can be shown that the reflection percentage at the skin-air interface may amount to 30.19%. The opposite parameter, transmittance, is computed at the final fat-blood interface at which a minimum of 63% of energy is sent to the embedded nanosensor. The achieved results point out the necessity to consider the dissipated power due to cross-layer wave penetration encountered in the communication channel linking between the inner bionanosensor and the outer device. In spite of the multiple layer reflection effect, the Terahertz frequency range emerged as an effective tool to realize the Intrabody Communication. Furthermore, Figure 8 and Figure 9 clarify, respectively, the input equivalent resistance and reactance for skin-fat-blood. The final results re-emphasized our prior returns suggesting that, at THz band, the effective medium is about to be perfectly resistive with minimal insignificant conductivity.

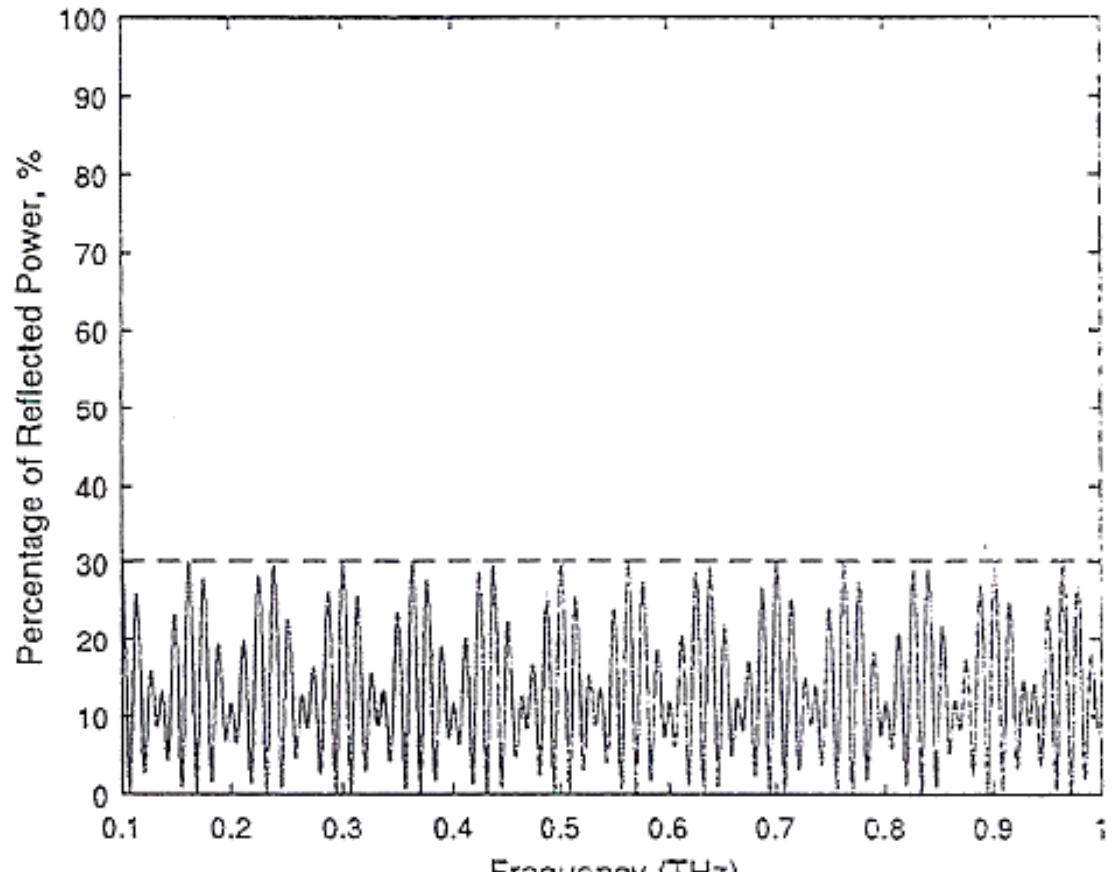

**Figure 7.6:** The portion (in percentage) of THz power incident at the interface being reflected when propagating from external to internal medium.

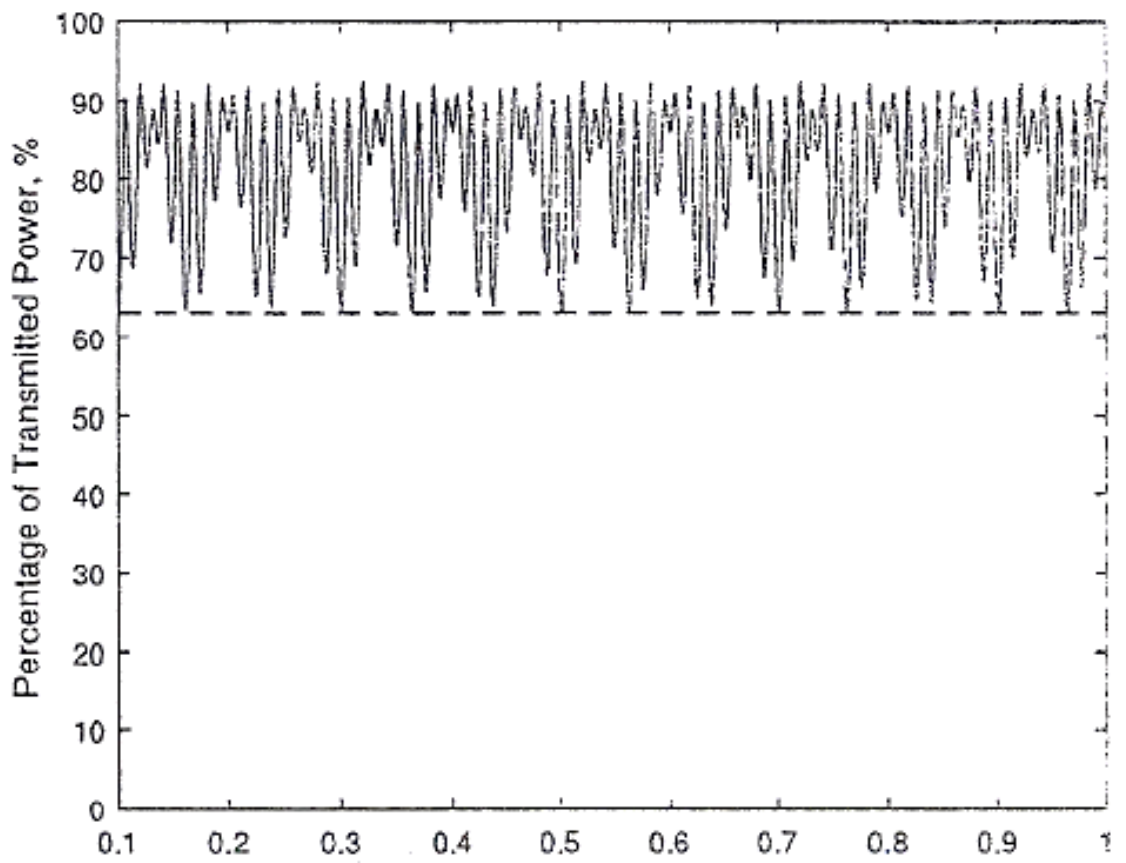

**Figure 7.7:** The portion (in percentage) of THz power incident at the interface being successfully transmitted when propagating from external to internal medium.

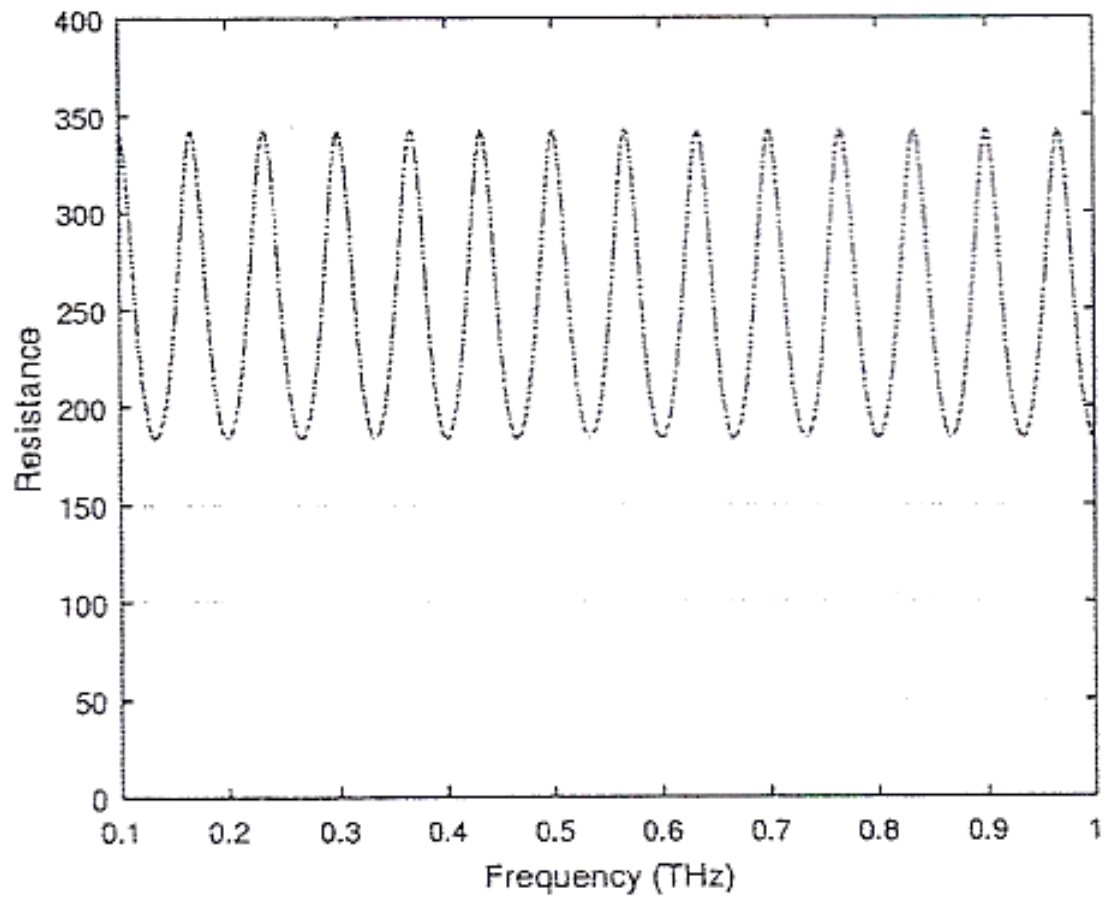

**Figure 7.8:** Equivalent resistance input values as a function of Terahertz-range frequency when propagating from external to internal medium.

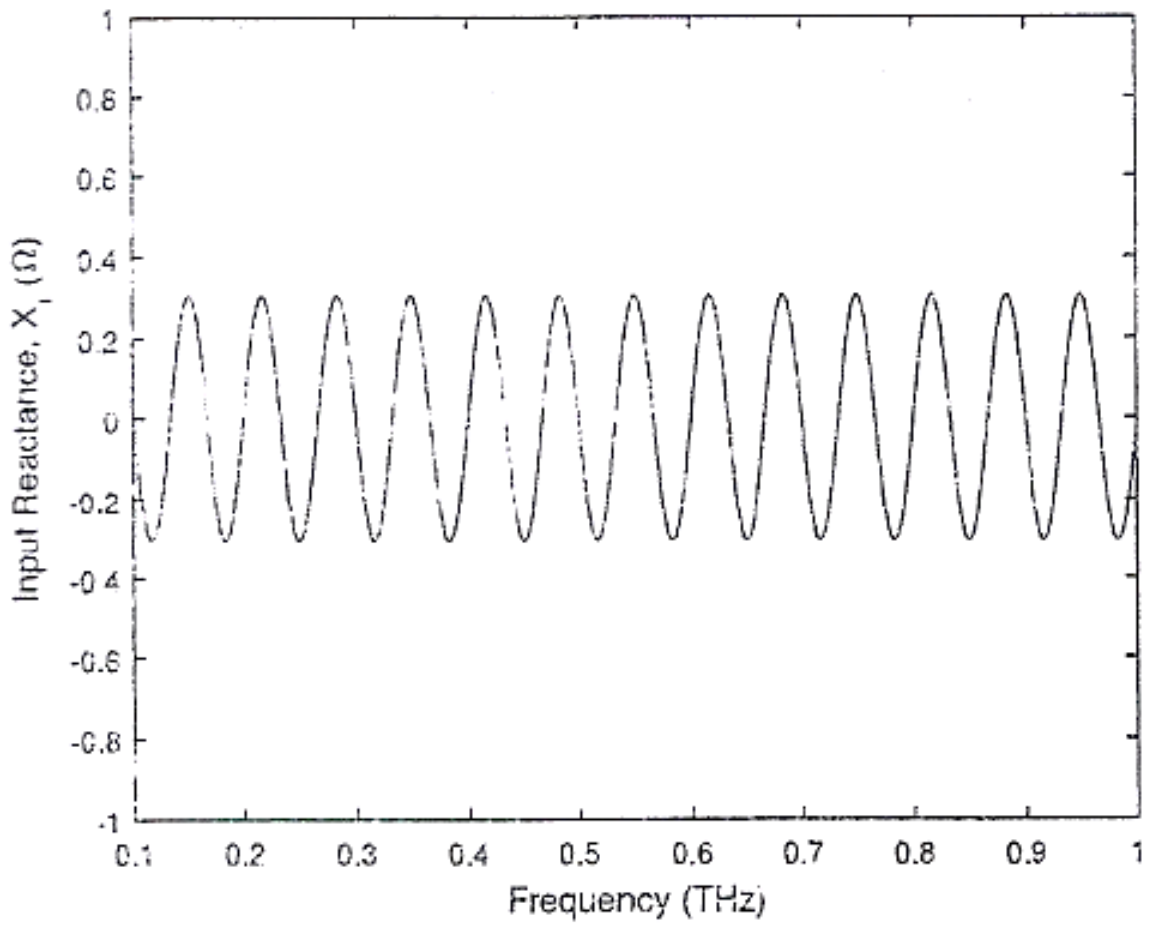

**Figure 7.9:** Equivalent reactance input values as a function of Terahertz-range frequency when propagating from external to internal medium.

# *Conclusion:*

In this report, a general overview of the in vivo communication and networking was demonstrated. The overview emphasizes on the state of art of this particular type of communication, the modeling and characterization of communication channel, and the principle of MIMO. It also includes in-depth discussion on potential investigation areas covering the development of parametric models and the comprehensive study of MIMO technology. Frequency domain, power levels and SAR requirements are few aspects that should be wisely considered in any forthcoming research allied with this field.

We introduced in this report the Nanotechnology field as well. This field can be classified as relatively new and modern, and although the complete scope of contributions of these technological advances in the field of medicine stays unexplored, latest advances propose that nanotechnology will have a pioneering impact on disease prevention, disease identification and cure. Recent improvements in modern health care approaches were highlighted and discussed. Different applications used in the diagnosis and treatment of diseases were considered. These comprise but not restricted to drug delivery, gene delivery, immune system support and health monitoring. In addition, the bionanosensors technology was deliberated concentrating on their potential ability to sense any biochemical and biophysical wave related to the existence of diseases at the molecular or even cellular levels. Several features of nanoparticles were also revealed. Also, the possible anticipated challenges and the prospective future directions convoyed with this emerging technology were shown. It is worth to mention that nanotechnology field doesn’t stand by itself as a one evolving scientific discipline but rather a combined collaboration between number of sciences such as physics, chemistry, Biology as well as material science, all as contributions to come up with a joint expertise essential to develop these nanotechnologies.

The growth of nano-networks facilitated the exploration of different approaches to realize successful communication between nanodevices. Although molecular nano-communications grabbed major attention in the previous years, it still has few limitations. This prompted the study of nano-communications from the electromagnetic viewpoint. This project examines path loss at THz for nano-scale WBANs based on a basic biological model. Numerical results indicated that path loss at the millimeter level is not substantially larg. This shows the probability of achieving EM nano-communication among devices. The analytical results offered in this report were supplementary proved by comparison with future experimental measurements.
In this project, we have presented the deep analysis of the cross-layer bio-electromagnetic Terahertz wave propagation across a cascaded human tissue model. Two different scenarios were investigated in which the nanosensor communicates either with a wearable device in order to sense signals and track the human significant parameters. The second scenario demonstrates the wearable device as a device that interacts with a nanosensor for the purpose of drug delivery or cancer early detection. It was discovered that the inhomogeneity and the discrepancy nature of human biological tissues will cause a periodic type of spectral responsivity for signal reflections that can occupy up to 30% of an incident amount of power. This affirms the need to consider for the cross-layer reflection to yield a precise characterization of the communication link between the nanosensors and the on-body device all operating in Terahertz band.

The scope of future Intrabody communication for In vivo wireless nanosensor body networks has been shown to be viable and practical when operating at the Terahertz spectral band. The mathematical framework based on cascaded tissue layers was developed and employed to inspect the behavior of the reflection coefficient at the three main interfaces of blood-fat, fat-skin and skin-air. These were fully examined at both polarization modes (TE & TM). Many crucial findings in this report will extremely develop a perfect intrabody communication model.

Moreover, in this report, we established a channel model aimed to predict the THz affect and the deploy of optical bio-electromagnetic waves. We presented novel model since it takes into consideration the overall joint effect of three key propagation phenomena encountered in the communication field of Human body. These three main factors are molecular absorption, scattering and spreading. We did incorporate the nanoantenna directivity in order to precisely quantify the effect of spreading. The study of molecular absorption effect illustrated that the molecules of blood are more absorbent compared to any other composite. Nevertheless, absorption due to blood cells is single order of magnitude lower in optical window; this is approximately between 400 and 750 THz. Furthermore, scattering factor has been exactly figured out by considering the scattered nanoparticles size with respect to wavelength in each of Terahertz spectral band and the window of optical frequency. It has been also concluded that when the particle size parameter ψ becomes greater than 1, the efficiency of scattering approaches value 2, which successfully agrees with principles of classical optics. The resultant total effects of these three main propagation phenomena in the field of intrabody communication will certainly motivate the design and utilization of iWNSNs. This project also offered detailed analysis of the Terahertz and photonic devices existing in the literature. Such study is vital for preparing link budget analysis between different nanodevices functioning in the human body. The findings delivered in this report point out those nano-communication based devices have the potential to play important role in future healthcare technologies by improving the quality of the human life.

# *References:*

[1] R. M. Shubair, "Robust adaptive beamforming using LMS algorithm with SMI initialization," in *2005 IEEE Antennas and Propagation Society International Symposium*, vol. 4A, Jul. 2005, pp. 2–5 vol. 4A.

[2] R. M. Shubair and W. Jessmi, "Performance analysis of SMI adaptive beamforming arrays for smart antenna systems," in *2005 IEEE Antennas and Propagation Society International Symposium*, vol. 1B, 2005, pp. 311–314 vol. 1B.

[3] F. A. Belhoul, R. M. Shubair, and M. E. Ai-Mualla, "Modelling and performance analysis of DOA estimation in adaptive signal processing arrays," in *10th IEEE International Conference on Electronics, Circuits and Systems, 2003. ICECS 2003. Proceedings of the 2003*, vol. 1, Dec. 2003, pp. 340–343 Vol.1. 19

[4] R. M. Shubair and A. Al-Merri, "Robust algorithms for direction finding and adaptive beamforming: performance and optimization," in *The 2004 47th Midwest Symposium on Circuits and Systems, 2004. MWSCAS '04*, vol. 2, Jul. 2004, pp. II–589–II–592 vol.2.

[5] E. Al-Ardi, R. Shubair, and M. Al-Mualla, "Direction of arrival estimation in a multipath environment: An overview and a new contribution," in *ACES*, vol. 21, 2006.

[6] G. Nwalozie, V. Okorogu, S. Maduadichie, and A. Adenola, "A simple comparative evaluation of adaptive beam forming algorithms," *International Journal of Engineering and Innovative Technology (IJEIT)*, vol. 2, no. 7, 2013.

[7] M. A. Al-Nuaimi, R. M. Shubair, and K. O. Al-Midfa, "Direction of arrival estimation in wireless mobile communications using minimum variance distortionless response," in *Second International Conference on Innovations in Information Technology (IIT'05)*, 2005, pp. 1–5.

[8] M. Bakhar and D. P. Hunagund, "Eigen structure based direction of arrival estimation algorithms for smart antenna systems," *IJCSNS International Journal of Computer Science and Network Security*, vol. 9, no. 11, pp. 96–100, 2009.

[9] M. AlHajri, A. Goian, M. Darweesh, R. AlMemari, R. Shubair, L.Weruaga, and A. AlTunaiji, "Accurate and robust localization techniques for wireless sensor networks," June 2018, arXiv:1806.05765 [eess.SP].

[10] J. Samhan, R. Shubair, and M. Al-Qutayri, "Design and implementation of an adaptive smart antenna system," in *Innovations in Information Technology, 2006*, 2006, pp. 1–4.

[11] M. AlHajri, A. Goian, M. Darweesh, R. AlMemari, R. Shubair, L.Weruaga, and A. Kulaib, “Hybrid rss-doa technique for enhanced wsn localization in a correlated environment,” in *Information and Communication Technology Research (ICTRC), 2015 International Conference on*, 2015, pp. 238–241.
[12] M. S. Khan, A. D. Capobianco, S. M. Asif, D. E. Anagnostou, R. M. Shubair, and B. D. Braaten, “A Compact CSRR-Enabled UWB Diversity Antenna,” *IEEE Antennas and Wireless Propagation Letters*, vol. 16, pp. 808–812, 2017.
[13] R. M. Shubair and Y. L. Chow, “A closed-form solution of vertical dipole antennas above a dielectric half-space,” *IEEE Transactions on Antennas and Propagation*, vol. 41, no. 12, pp. 1737–1741, Dec. 1993.
[14] A. Omar and R. Shubair, “UWB coplanar waveguide-fed-coplanar strips spiral antenna,” in *2016 10th European Conference on Antennas and Propagation (EuCAP),* Apr. 2016, pp. 1–2.
[15] R. M. Shubair and H. Elayan, “In vivo wireless body communications: State-of-the-art and future directions,” in *Antennas & Propagation Conference (LAPC), 2015 Loughborough*. IEEE, 2015, pp. 1–5.
[16] H. Elayan, R. M. Shubair, J. M. Jornet, and P. Johari, “Terahertz channel model and link budget analysis for intrabody nanoscale communication,” *IEEE transactions on nanobioscience*, vol. 16, no. 6, pp. 491–503, 2017.
[17] H. Elayan, R. M. Shubair, and A. Kiourti, “Wireless sensors for medical applications: Current status and future challenges,” in *Antennas and Propagation (EUCAP), 2017 11th European Conference on*. IEEE, 2017, pp. 2478–2482.
[18] H. Elayan and R. M. Shubair, “On channel characterization in human body communication for medical monitoring systems,” in *Antenna Technology and Applied Electromagnetics (ANTEM), 2016 17th International Symposium on*. IEEE, 2016, pp. 1–2.
[19] H. Elayan, R. M. Shubair, A. Alomainy, and K. Yang, “In-vivo terahertz em channel characterization for nano-communications in wbans,” in *Antennas and Propagation (APSURSI), 2016 IEEE International Symposium on*. IEEE, 2016, pp. 979–980.
[20] H. Elayan, R. M. Shubair, and J. M. Jornet, “Bio-electromagnetic thz propagation modeling for in-vivo wireless nanosensor networks,” in *Antennas and Propagation (EUCAP), 2017 11th European Conference on*. IEEE, 2017, pp. 426–430. 21

[21] H. Elayan, C. Stefanini, R. M. Shubair, and J. M. Jornet, “End-to-end noise model for intra-body terahertz nanoscale communication,” *IEEE Transactions on NanoBioscience*, 2018.

[22] H. Elayan, P. Johari, R. M. Shubair, and J. M. Jornet, “Photothermal modeling and analysis of intrabody terahertz nanoscale communication,” *IEEE transactions on nanobioscience*, vol. 16, no. 8, pp. 755–763, 2017.

[23] H. Elayan, R. M. Shubair, J. M. Jornet, and R. Mittra, “Multi-layer intrabody terahertz wave propagation model for nanobiosensing applications,” *Nano Communication Networks*, vol. 14, pp. 9–15, 2017.

[24] H. Elayan, R. M. Shubair, and N. Almoosa, “In vivo communication in wireless body area networks,” in *Information Innovation Technology in Smart Cities*. Springer, 2018, pp. 273–287.

[25] M. O. AlNabooda, R. M. Shubair, N. R. Rishani, and G. Aldabbagh, “Terahertz spectroscopy and imaging for the detection and identification of illicit drugs,” in *Sensors Networks Smart and Emerging Technologies (SENSET), 2017*, 2017, pp. 1–4.

[26] K. YANG, “Characterisation of the in-vivo terahertz communication channel within the human body tissues for future nano-communication networks.” Ph.D. dissertation, Queen Mary University of London, 2016.

[27] I. F. Akyildiz, F. Brunetti, and C. Bla´zquez, “Nanonetworks: A new communication paradigm,” *Computer Networks*, vol. 52, no. 12, pp. 2260–2279, 2008.

[28] I. F. Akyildiz and J. M. Jornet, “The internet of nano-things,” *IEEE Wireless Communications*, vol. 17, no. 6, pp. 58–63, 2010.

[29] R. A. Freitas, “Nanotechnology, nanomedicine and nanosurgery,” *International Journal of Surgery*, vol. 3, no. 4, pp. 243–246, 2005.

[30] C. M. Pieterse and M. Dicke, “Plant interactions with microbes and insects: from molecular mechanisms to ecology,” *Trends in plant science*, vol. 12, no. 12, pp. 564–569, 2007.

[31] R. E. Smalley, M. S. Dresselhaus, G. Dresselhaus, and P. Avouris, *Carbon nanotubes: synthesis, structure, properties, and applications*. Springer Science & Business Media, 2003, vol. 80.

[32] Y. Liu, T. P. Ketterl, G. E. Arrobo, and R. D. Gitlin, “Modeling the wireless in vivo path loss,” in *RF and Wireless Technologies for Biomedical and Healthcare Applications (IMWS-Bio), 2014 IEEE MTT-S International Microwave Workshop Series on*. IEEE, 2014, pp. 1–3.

[33] C. He, Y. Liu, T. P. Ketterl, G. E. Arrobo, and R. D. Gitlin, “Performance evaluation for mimo in vivo wban systems,” in *RF and Wireless Technologies for Biomedical and*

*Healthcare Applications (IMWS-Bio), 2014 IEEE MTT-S International Microwave Workshop Series on*. IEEE, 2014, pp. 1–3.

[34] A. K. Skrivervik, "Implantable antennas: The challenge of efficiency," in *Antennas and Propagation (EuCAP), 2013 7th European Conference on*. Ieee, 2013, pp. 3627–3631.

[35] H. G. Schantz, "Near field phase behavior," in *IEEE Antennas and Propagation Society International Symposium*, vol. 3. IEEE; 1999, 2005, p. 134.

[36] K. Ito, N. Haga, M. Takahashi, and K. Saito, "Evaluations of body-centric wireless communication channels in a range from 3 mhz to 3 ghz," *Proceedings of the IEEE*, vol. 100, no. 7, pp. 2356–2363, 2012.

[37] J. M. Jornet and I. F. Akyildiz, "Graphene-based plasmonic nano-antenna for terahertz band communication in nanonetworks," *IEEE Journal on selected areas in communications*, vol. 31, no. 12, pp. 685–694, 2013.

[38] L. P. Gin´e and I. F. Akyildiz, "Molecular communication options for long range nanonetworks," *Computer Networks*, vol. 53, no. 16, pp. 2753–2766, 2009.

[39] ——, "Channel modeling and capacity analysis for electromagnetic wireless nanonetworks in the terahertz band," *IEEE Transactions on Wireless Communications*, vol. 10, no. 10, pp. 3211–3221, 2011.

[40] I. A. Ibraheem, N. Krumbholz, D. Mittleman, and M. Koch, "Low-dispersive dielectric mirrors for future wireless terahertz communication systems," *IEEE microwave and wireless components letters*, vol. 18, no. 1, pp. 67–69, 2008.

[41] R. M. Goody and Y. L. Yung, *Atmospheric radiation: theoretical basis*. Oxford University Press, 1995.

[42] T. W. Crowe, D. W. Porterfield, J. L. Hesler, W. L. Bishop, D. S. Kurtz, and K. Hui, "Terahertz sources and detectors," in *Defense and Security*. International Society for Optics and Photonics, 2005, pp. 271–280.

[43] S. Logothetidis, "Nanotechnology in medicine: the medicine of tomorrow and nanomedicine," Hippokratia, vol. 10, no. 1, pp. 7–21, 2006.

[44] I. F. Akyildiz, J. M. Jornet, and M. Pierobon, "Nanonetworks: A new frontier in communications," Commun. ACM, vol. 54, no. 11, pp. 84–89, Nov. 2011. [Online]. Available:http://doi.acm.org/10.1145/2018396.2018417

[45] J. Wilkinson, "Nanotechnology applications in medicine." Medical device technology, vol. 14, no. 5, pp. 29–31, 2003.

[46] O. V. Salata, "Applications of nanoparticles in biology and medicine," Journal of nanobiotechnology, vol. 2, no. 1, p. 1, 2004.

[47] E. S. Foundation and E. M. R. Councils, Nanomedicine: An ESF-European Medical Research Councils (EMRC) Forward Look Report, ser. Forward Look report. ESF, 2005. [Online]. Available: https://books.google.com/books?id=5joFOAAACAAJ

[48] R. A. Freitas, "Current status of nanomedicine and medical nanorobotics," Journal of Computational and Theoretical Nanoscience, vol. 2, no. 1, pp. 1–25, 2005.

[49] S. Sarkar and S. Misra, "From micro to nano: The evolution of wireless sensor-based health care," IEEE Pulse, vol. 7, no. 1, pp. 21–25, Jan 2016.

[50] P. Martins, D. Rosa, A. R Fernandes, and P. V. Baptista, "Nanoparticle drug delivery systems: recent patents and applications in nanomedicine," Recent Patents on Nanomedicine, vol. 3, no. 2, pp. 105–118, 2013.

[51] B. S. Zolnik, A. Gonzalez-Fernandez, N. Sadrieh, and M. A. Dobrovolskaia, "Minireview: nanoparticles and the immune system," Endocrinology, vol. 151, no. 2, pp. 458–465, 2010.

[52] S. Parveen, R. Misra, and S. K. Sahoo, "Nanoparticles: a boon to drug delivery, therapeutics, diagnostics and imaging," Nanomedicine: Nanotechnology, Biology and Medicine, vol. 8, no. 2, pp. 147–166, 2012.

[53] S. Sahoo, S. Parveen, and J. Panda, "The present and future of nanotechnology in human health care," Nanomedicine: Nanotechnology, Biology and Medicine, vol. 3, no. 1, pp. 20–31, 2007.

[54] D. R. Dreyer, S. Park, C. W. Bielawski, and R. S. Ruoff, "The chemistry of graphene oxide," Chemical Society Reviews, vol. 39, no. 1, pp. 228– 240, 2010.

[55] N. M. Iverson, P. W. Barone, M. Shandell, L. J. Trudel, S. Sen, F. Sen, V. Ivanov, E. Atolia, E. Farias, T. P. McNicholas et al., "In vivo biosensing via tissue-localizable near-infrared-fluorescent single-walled carbon nanotubes," Nature nanotechnology, vol. 8, no. 11, pp. 873–880, 2013.

[56] H. Shao, J. Chung, L. Balaj, A. Charest, D. D. Bigner, B. S. Carter, F. H. Hochberg, X. O. Breakefield, R. Weissleder, and H. Lee, "Protein typing of circulating microvesicles allows real-time monitoring of glioblastoma therapy," Nature medicine, vol. 18, no. 12, pp. 1835–1840, 2012.

[57] J. D. Driskell, C. A. Jones, S. M. Tompkins, and R. A. Tripp, "One-step assay for detecting influenza virus using dynamic light scattering and gold nanoparticles," Analyst, vol. 136, no. 15, pp. 3083–3090, 2011.

[58] D. R. Cooper, D. Bekah, and J. L. Nadeau, "Gold nanoparticles and their alternatives for radiation therapy enhancement," Frontiers in chemistry, vol. 2, p. 86, 2014.

[59] A. Verma, N. Kumar, R. Malviya, and P. K. Sharma, "Emerging trends in noninvasive insulin delivery," Journal of pharmaceutics, vol. 2014, 2014.

# *Appendix:*

## Algorithm:

- Initialize frequency [f] range in Hz
- Initialize theta [θ] range in radian
- Initialize the following variables

  *Perm, Free_sp, Epsilon_2* [$\varepsilon_2$]

  *Epsilon_S* [$\varepsilon_s$]*,Epsilon_inf* [$\varepsilon_i$]

  *Tau1* [$\tau_1$], *Tau2* [$\tau_2$]
- Find Frequency in Radian [ω =2*π*f]
- Find values of A,B,C,D, E and F

  $A = (\varepsilon_2 - \varepsilon_s) * (\omega * \tau_1) / (1+(\omega * \tau_1)^2)$ & $B = (\varepsilon_2 - \varepsilon_i) * (\omega * \tau_2) / (1+(\omega * \tau_2)^2)$

  $C = \varepsilon_i + (\varepsilon_s - \varepsilon_2) / (1+(\omega * \tau_1)^2)$ & $D=(\varepsilon_2 - \varepsilon_i) / (1+(\omega * \tau_2)^2)$

  $E = \cos(\theta)$ & $F = \sin(\theta)$
- Find Epsilon [ε] values:

  $\varepsilon_{img} = A + B$ & $\varepsilon_{real} = C + D$

  $\varepsilon_{complex} = \varepsilon_{real} - j\ \varepsilon_{img}$ & $\varepsilon_{mag}$= absolute( $\varepsilon_{complex}$)
- Find values of Y1, Z1, Y2 and Z2

  $Y1 = abs( ( -B*E) + \sqrt{(B - F^2)} )$ & $Z1 = abs( ( -B*E) + \sqrt{(B - F^2)} )$

  $Y2 = abs( E - \sqrt{(B - F^2)} )$ & $Z2 = abs( E + \sqrt{(B - F^2)} )$

- Find Skin Conductivity [σ] & Skin Penetration Depth [$S_{pd}$]

  $\sigma = \varepsilon_{img} * \omega * Free_sp$ & $S_{pd} = \frac{1}{\sqrt{\left(\frac{(w*perm*\sigma)}{2}\right)}}$

- Find Value of Parallel(P1) & Perpendicular (P2) Power Reflection Coefficient (P1)

  $P1 = \left(\frac{Y1}{Z1}\right)^2$ & $P2 = \left(\frac{Y2}{Z2}\right)^2$
- Plot Skin Conductivity [σ] vs Frequency[f]
- Plot Skin Penetration Depth vs Frequency [f]
- Plot Power Reflection Coefficients [P1 & P2] vs Theta [θ]

## Flow Chart:

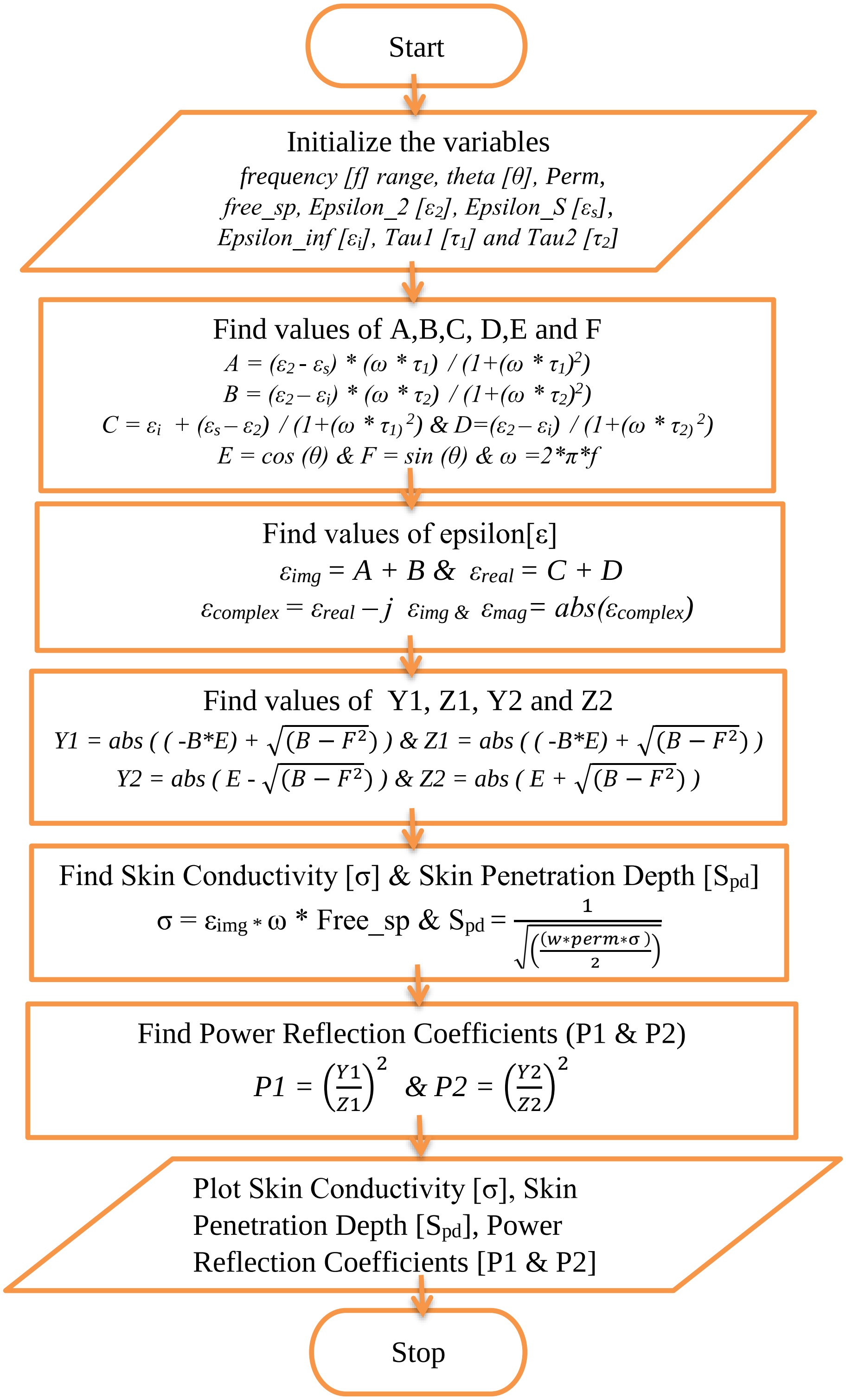
Start
Initialize the variables
frequency [f] range, theta [θ], Perm,
free_sp, Epsilon_2 [$\varepsilon_2$], Epsilon_S [$\varepsilon_s$],
Epsilon_inf [$\varepsilon_i$], Tau1 [$\tau_1$] and Tau2 [$\tau_2$]
Find values of A,B,C, D,E and F
$A = (\varepsilon_2 - \varepsilon_s) * (\omega * \tau_1) / (1+(\omega * \tau_1)^2)$
$B = (\varepsilon_2 - \varepsilon_i) * (\omega * \tau_2) / (1+(\omega * \tau_2)^2)$
$C = \varepsilon_i + (\varepsilon_s - \varepsilon_2) / (1+(\omega * \tau_1)^2)$ & $D=(\varepsilon_2 - \varepsilon_i) / (1+(\omega * \tau_2)^2)$
$E = \cos(\theta)$ & $F = \sin(\theta)$ & $\omega = 2*\pi*f$
Find values of epsilon[ε]
$\varepsilon_{img} = A + B$ & $\varepsilon_{real} = C + D$
$\varepsilon_{complex} = \varepsilon_{real} - j\ \varepsilon_{img}$ & $\varepsilon_{mag} = abs(\varepsilon_{complex})$
Find values of Y1, Z1, Y2 and Z2
$Y1 = abs((-B*E) + \sqrt{(B - F^2)})$ & $Z1 = abs((-B*E) + \sqrt{(B - F^2)})$
$Y2 = abs(E - \sqrt{(B - F^2)})$ & $Z2 = abs(E + \sqrt{(B - F^2)})$
Find Skin Conductivity [σ] & Skin Penetration Depth [$S_{pd}$]
$\sigma = \varepsilon_{img} * \omega * Free_sp$ & $S_{pd} = \frac{1}{\sqrt{\left(\frac{(w*perm*\sigma)}{2}\right)}}$
Find Power Reflection Coefficients (P1 & P2)
$P1 = \left(\frac{Y1}{Z1}\right)^2$ & $P2 = \left(\frac{Y2}{Z2}\right)^2$
Plot Skin Conductivity [σ], Skin Penetration Depth [$S_{pd}$], Power Reflection Coefficients [P1 & P2]
Stop

## Code illustrates Path loss Parameter for different constituent substances:

```
function prog_main()
clc;
close all;
clear all;

f=0.2e12:0.01e12:1e12;
w=2.*pi.*f;
perm=1.25e-6;
c=3e8;
lambda_THz=c./f;
free_sp=8.85e-12;
theta=0:pi./160:pi./2;
d=0e-3:0.5e-3:3e-3 ;      % distance range

%% Whole Blood
epsilon_inf=2.1;
epsilon_1=130;
epsilon_2=3.8;
tau1=14.4e-12;
tau2=0.1e-12;
[n1,k1,sigma1,epsilon_img1,epsilon_real1 ] = function1 (epsilon_inf,
epsilon_1,epsilon_2, tau1, tau2, f,w, free_sp);  % find conductivity
epsilon_c1=epsilon_real1-i.*(epsilon_img1);
[reflection_par1, reflection_per1]= function2 (epsilon_c1,theta);
% find power reflection coefficients
[PL1,spr_coe1,abs_coe1]= function3(f,c,d,n1,k1);  % find pathloss[PL]

%% Blood Plasma
epsilon_inf=1.7;
epsilon_1=78;
epsilon_2=3.6;
tau1=8e-12;
tau2=0.1e-12;
[n2,k2,sigma2,epsilon_img2,epsilon_real2 ] = function1 (epsilon_inf,
epsilon_1,epsilon_2, tau1, tau2, f,w, free_sp);  % find conductivity
epsilon_c2=epsilon_real2-1i.*(epsilon_img2);
[reflection_par2, reflection_per2]= function2 (epsilon_c2,theta);
% find power reflection coefficients
[PL2,spr_coe2,abs_coe2]= function3(f,c,d,n2,k2);  % find pathloss[PL]

%% Water
epsilon_inf=3.3;
epsilon_1=87.8;
epsilon_2=4.5;
tau1=8.4e-12;
tau2=0.5e-12;
[n3,k3,sigma3,epsilon_img3,epsilon_real3 ] = function1 (epsilon_inf,
epsilon_1,epsilon_2, tau1, tau2, f,w, free_sp);  % find conductivity
epsilon_c3=epsilon_real3-i.*(epsilon_img3);
[reflection_par3, reflection_per3]= function2 (epsilon_c3,theta);
% find power reflection coefficients
[PL3,spr_coe3,abs_coe3]= function3(f,c,d,n3,k3);  % find pathloss[PL]
```

```
%% Skin Dermis
epsilon_inf=3.6;
epsilon_1=60;
epsilon_2=3;
tau1=10e-12;
tau2=0.2e-12;
[n4,k4,sigma4,epsilon_img4,epsilon_real4] = function1 (epsilon_inf,
epsilon_1,epsilon_2, tau1, tau2, f,w, free_sp);  % find conductivity
epsilon_c4=epsilon_real4-i.*(epsilon_img4);
[reflection_par4, reflection_per4]= function2 (epsilon_c4,theta);
% find power reflection coefficients
[PL4,spr_coe4,abs_coe4]= function3(f,c,d,n4,k4);  % find pathloss[PL]

%% Skin Epidermis
epsilon_inf=3.6;
epsilon_1=58;
epsilon_2=3;
tau1=9.4e-12;
tau2=0.18e-12;
[n5,k5,sigma5,epsilon_img5,epsilon_real5 ] = function1 (epsilon_inf,
epsilon_1,epsilon_2, tau1, tau2, f,w, free_sp);  % find conductivity
epsilon_c5=epsilon_real5-i.*(epsilon_img5);
[reflection_par5, reflection_per5]= function2 (epsilon_c5,theta);
% find power reflection coefficients
[PL5,spr_coe5,abs_coe5]= function3(f,c,d,n5,k5);  % find pathloss[PL]


%% Plot Epsilon Imaginary
figure;
plot(f.*(1e-12),epsilon_img1,'bo-',f.*(1e-12),epsilon_img2,'k-*', f.*(1e-
12),epsilon_img3,'r-s',f.*(1e-12), epsilon_img4,'g-d',f.*(1e-12),
epsilon_img5,'m-v')
xlabel('Frequency (THz)')
ylabel('Imaginary part of relative permitivity');
legend('Whole','Plasma', 'Water', 'Dermis', 'Epidermis');
title('Epsilon Imaginary ');
grid on;

%% Plot Epsilon Real
figure;
plot(f.*(1e-12),epsilon_real1,'bo-',f.*(1e-12),epsilon_real2,'k-*', f.*(1e-
12),epsilon_real3,'r-s',f.*(1e-12), epsilon_real4,'g-d',f.*(1e-12),
epsilon_real5,'m-v')
xlabel('Frequency (THz)')
ylabel('Real part of relative permitivity');
legend('Whole','Plasma', 'Water', 'Dermis', 'Epidermis');
title('Epsilon Real');
grid on;



%% Plot Conductivity
figure;
plot(f.*(1e-12),sigma1,'bo-',f.*(1e-12),sigma2,'k-*', f.*(1e-12),sigma3,'r-
s',f.*(1e-12), sigma4,'g-d',f.*(1e-12), sigma5,'m-v')
```

```
xlabel('Frequency (THz)')
ylabel('Sigma Conductivity (S/m)');
legend('Whole','Plasma', 'Water', 'Dermis', 'Epidermis');
title('Conductivity');
grid on;


%% Plot Parallel Power Reflection Coefficient
figure;
plot(theta.*(180./pi),reflection_par1, 'bo-
',theta.*(180./pi),reflection_par2, 'k-*',theta.*(180./pi),reflection_par3,
'r-s',theta.*(180./pi),reflection_par4, 'g-d',
theta.*(180./pi),reflection_par5, 'm-v');
xlabel('Angle of Incidence')
ylabel('Power Reflection Coefficient ')
legend('Whole','Plasma', 'Water', 'Dermis', 'Epidermis');
title('Parallel Polarization')
grid;

%% Plot Perpendicular Power Reflection Coefficient
figure;
plot(theta.*(180./pi),reflection_per1, 'bo-
',theta.*(180./pi),reflection_per2, 'k-*',theta.*(180./pi),reflection_per3,
'r-s', theta.*(180./pi),reflection_par4, 'g-d',
theta.*(180./pi),reflection_par5, 'm-v');
xlabel('Angle of Incidence')
ylabel('Power Reflection Coefficient ')
legend('Whole','Plasma', 'Water', 'Dermis', 'Epidermis');
title('Perpendicular Polarization')
grid;

%% Plot abs_coe
figure;
plot(d,abs_coe1,'bo-',d, abs_coe2,'k-*', d,abs_coe3,'r-s',d,abs_coe4,'g-
d',d,abs_coe5,'m-v')
xlabel('Distance in mm')
ylabel('abs-coe');
legend('Whole','Plasma', 'Water', 'Dermis', 'Epidermis');
title('abs-coe ');
grid on;

%% Plot spr_coe
figure;
plot(d,spr_coe1,'bo-',d, spr_coe2,'k-*', d,spr_coe3,'r-s',d,spr_coe4,'g-
d',d,spr_coe5,'m-v')
xlabel('Distance in mm')
ylabel('spr-coe');
legend('Whole','Plasma', 'Water', 'Dermis', 'Epidermis');
title('spr-coe ');
grid on;


%% Plot Pathloss [PL]
figure;
plot(d,PL1,'bo-',d,PL2,'k-*', d,PL3,'r-s',d,PL4,'g-d',d,PL5,'m-v')
```

```
xlabel('Distance in mm')
ylabel('Pathloss PL(db)');
legend('Whole','Plasma', 'Water', 'Dermis', 'Epidermis');
title('Pathloss ');
grid on;


% Function to Find Conductivity
function [n,k,sigma,epsilon_img, epsilon_real] = function1 (epsilon_inf,
epsilon_1,epsilon_2, tau1, tau2, f,w,free_sp)
a=(epsilon_1-epsilon_2).*(w.*tau1)./(1+((w.*(tau1)).^2));
b=(epsilon_2-epsilon_inf).*(w.*tau2)./(1+(w.*(tau2).^2));
c=(epsilon_inf)+((epsilon_1-epsilon_2)./(1+((w.*(tau1)).^2)));
d=(epsilon_2-epsilon_inf)./(1+(w.*(tau2).^2));
epsilon_img=a+b;
epsilon_real=c+d;
epsilon_complex=epsilon_real-i.*(epsilon_img);
one_1=sqrt((epsilon_real.^2)+(epsilon_img.^2));
k=sqrt((one_1-epsilon_real)./2);
n=sqrt((one_1+epsilon_real)./2);
sigma=epsilon_img.*2.*pi.*f.*free_sp;

% Function to find Power Reflection Ceofficients
function [power_reflection_parallel, power_reflection_perpendicular]=
function2 (epsilon_complex,theta)
% Parallel Power Reflection Coefficient
y=abs(-(epsilon_complex).*cos(theta)+sqrt((epsilon_complex)-
(sin(theta)).^2));
z=abs(epsilon_complex.*cos(theta)+sqrt((epsilon_complex)-(sin(theta)).^2));
R_parallel=y./z;
power_reflection_parallel=(R_parallel).^2;

% Perpendicular Power Reflection Coefficient
y1=abs(cos(theta)-sqrt((epsilon_complex)-(sin(theta)).^2));
z1=abs(cos(theta)+sqrt((epsilon_complex)-(sin(theta)).^2));
R_perpendicular=y1./z1;
power_reflection_perpendicular=(R_perpendicular).^2;

%% Find Pathloss (PL)
function [PL,spr_coe,abs_coe]= function3(f,c,d,n,k)
index=1;
%f=0.2e12:0.01e12:1e12;
%c=3e8;
lamda0=c./f;
lamda1=lamda0./n;
%d=0e-3:0.5e-3:3e-3 ;      % distance range
atten1=(4*pi*k)./lamda0;  % attenuation [atten]
spr_coe=(lamda1(index)./(4*pi.*d)).^2;
abs_coe=exp(-atten1(index).*d);
%Pathloss, PL=PL_spread+PL_absorbtion
PL_spread=-10*log10(spr_coe);
PL_absorbtion=-10*log10(abs_coe);
PL=PL_spread+PL_absorbtion;
```

## Code illustrates Reflection occurred when transmitting From implanted sensor to wearable sensor:

```
function prog_plot_()

clc;
close all;
clear all;

theta=90;
theta1=0:theta;

% Case 1
n2=1;
n1=1.73;
[rte, rtm]=func_rte_rtm(n1, n2, theta);

figure;
plot(theta1, rte);
grid on;
xlabel( 'Angle of Incidence');
ylabel('Magnitude of TE Reflection Coefficient, |TE|');
legend('Skin-Air Interface');

figure;
plot(theta1, rtm);
grid on;
xlabel( 'Angle of Incidence');
ylabel('Magnitude of TM Reflection Coefficient, |TM|');
legend('Skin-Air Interface');

% Case 2
n2=1.73;
n1=1.58;
[rte, rtm]=func_rte_rtm(n1, n2, theta);

figure;
plot(theta1, rte);
grid on;
xlabel( 'Angle of Incidence');
ylabel('Magnitude of TE Reflection Coefficient, |TE|');
legend('Fat-Skin Interface');
figure;
plot(theta1, rtm);
grid on;
xlabel( 'Angle of Incidence');
ylabel('Magnitude of TM Reflection Coefficient, |TM|');
legend('Fat-Skin Interface');
```

```
% Case 3
n2=1.58;
n1=1.97;
[rte, rtm]=func_rte_rtm(n1, n2, theta);

figure;
plot(theta1, rte);
grid on;
xlabel( 'Angle of Incidence');
ylabel('Magnitude of TE Reflection Coefficient, |TE|');
legend('Blood-Fat Interface');
figure;
plot(theta1, rtm);
grid on;
xlabel( 'Angle of Incidence');
ylabel('Magnitude of TM Reflection Coefficient, |TM|');
legend('Blood-Fat Interface');

function [rte, rtm] = func_rte_rtm(n1, n2, theta1)
    n=n2/n1;
    for i=0:theta1
        theta=i;
        rte(i+1)=abs((cosd(theta) - sqrt(n^2 -sind(theta)^2))/ (cosd(theta) + 
sqrt(n^2 -sind(theta)^2)));
        rtm(i+1)=abs(((n^2*cosd(theta)) - sqrt(n^2 -sind(theta)^2))/ 
((n^2*cosd(theta)) + sqrt(n^2 -sind(theta)^2)));
    end
```

## Code illustrates Reflection occurred when transmitting From wearable sensor to implanted sensor:

```
function prog_plot_()

clc;
close all;
clear all;

theta=90;
theta1=0:theta;

% Case 1
n2=1.73;
n1=1;
[rte, rtm]=func_rte_rtm(n1, n2, theta);

figure;
plot(theta1, rte);
grid on;
xlabel( 'Angle of Incidence');
ylabel('Magnitude of TE Reflection Coefficient, |TE|');
legend('Air-Skin Interface');
```

```
figure;
plot(theta1, rtm);
grid on;
xlabel( 'Angle of Incidence');
ylabel('Magnitude of TM Reflection Coefficient, |TM|');
legend('Air-Skin Interface');

% Case 2
n2=1.58;
n1=1.73;
[rte, rtm]=func_rte_rtm(n1, n2, theta);

figure;
plot(theta1, rte);
grid on;
xlabel( 'Angle of Incidence');
ylabel('Magnitude of TE Reflection Coefficient, |TE|');
legend('Skin-Fat Interface');
figure;
plot(theta1, rtm);
grid on;
xlabel( 'Angle of Incidence');
ylabel('Magnitude of TM Reflection Coefficient, |TM|');
legend('Skin-Fat Interface');

% Case 3
n2=1.97;
n1=1.58;
[rte, rtm]=func_rte_rtm(n1, n2, theta);

figure;
plot(theta1, rte);
grid on;
xlabel( 'Angle of Incidence');
ylabel('Magnitude of TE Reflection Coefficient, |TE|');
legend('Fat-Blood Interface');
figure;
plot(theta1, rtm);
grid on;
xlabel( 'Angle of Incidence');
ylabel('Magnitude of TM Reflection Coefficient, |TM|');
legend('Fat-Blood Interface');

function [rte, rtm] = func_rte_rtm(n1, n2, theta1)
    n=n2/n1;
    for i=0:theta1
        theta=i;
        rte(i+1)=abs((cosd(theta) - sqrt(n^2 -sind(theta)^2))/ (cosd(theta) +
sqrt(n^2 -sind(theta)^2)));
        rtm(i+1)=abs(((n^2*cosd(theta)) - sqrt(n^2 -sind(theta)^2))/
((n^2*cosd(theta)) + sqrt(n^2 -sind(theta)^2)));
    end
```